\documentclass{aastex}
\usepackage{rotating}
\usepackage{natbib}
\usepackage{url}
\usepackage[usenames]{color}
\usepackage{hyperref}
\usepackage{float}
\pdfoutput=1
\begin{document}
\title{Observations of Transiting Exoplanets with the James Webb Space Telescope (JWST)}

\author{Charles Beichman}
\affil{\it NASA Exoplanet Science Institute, California Institute of Technology, Jet Propulsion Laboratory}
\author{ Bjoern	Benneke, Heather	Knutson, Roger	Smith}
\affil{\it California Institute of Technology}
\author{ Pierre-Olivier	Lagage}
\affil{\it CEA, Saclay}
\author{ Courtney	Dressing, David	Latham}
\affil{\it Center for Astrophysics, Harvard University}
\author{ Jonathan	Lunine}
\affil{\it Cornell University}
\author{Stephan Birkmann, Pierre Ferruit, Giovanna Giardino}
\affil{\it European Space Agency}
\author{Eliza	Kempton}
\affil{\it Grinnell College}
\author{Sean	Carey, Jessica	Krick}
\affil{\it Infrared Procesing and Analysis Center, California Institute of Technology}
\author{Pieter D.	Deroo, Avi	Mandell, Michael E.	Ressler, Avi	Shporer, Mark	Swain, Gautam Vasisht}
\affil{\it Jet Propulsion Laboratory, California Institute of Technology}
\author{George	Ricker}
\affil{\it Massachusetts Institute of Technology}
\author{Jeroen	Bouwman, Ian	 Crossfield	}
\affil{\it Max Planck Institute of Astrophysics}
\author{Tom	Greene, Steve	 Howell	}
\affil{\it NASA Ames Research Center}
\author{Jessie	Christiansen, David	Ciardi}
\affil{\it NASA Exoplanet Science Institute, California Institute of Technology}
\author{Mark Clampin, Matt Greenhouse}
\affil{\it NASA Goddard Spaceflight Center}
\author{Alessandro	Sozzetti}
\affil{\it Osservatorio Astronomico Di Torino}
\author{Paul	Goudfrooij,	 Dean	Hines,	 Tony	Keyes,	 Janice	Lee,	 Peter	McCullough,	 Massimo	Robberto,	 John	Stansberry, Jeff	Valenti}
\affil{\it Space Telescope Science Institute}
\author{Marcia	 Rieke, George Rieke}
\affil{\it University of Arizona}
\author{Jonathan Fortney}
\affil{\it University of California, Santa Cruz}
\author{Jacob	Bean, Laura	 Kreidberg}
\affil{\it University of Chicago}
\author{ Drake	Deming}
\affil{\it University of Maryland}
\author{Lo\"{i}c	Albert, Ren\'e	 Doyon}
\affil{\it Universit\'e de Montr\'eal}
\author{David	Sing}	
\affil{\it University of Exeter}

\clearpage
\tableofcontents
\clearpage
\section{\href{http://nexsci.caltech.edu/committees/JWST/Goals\_for\_Transit\_Meeting2.pptx}{Introduction}}

The study of exoplanets is called out explicitly in NASA's Strategic Plan\footnote{http://www.nasa.gov/sites/default/files/files/FY2014\_NASA\_SP\_508c.pdf.} as one of NASA's high level science goals: ``{\it Objective 1.6:} Discover how the universe works, explore how it began and evolved, and search for life on planets around other stars." The Strategic Plan calls out the James Webb Space Telescope ( {\it JWST} ) explicitly as a critical facility for studying exoplanets: ``JWST will allow us to \ldots study in detail planets around other stars." The  {\it JWST}  project also cites exoplanet research among its most important goals\footnote{http://jwst.nasa.gov/science.html}: ``{\it \href{http://jwst.nasa.gov/science.html}{The Birth of Stars and Protoplanetary systems}} focuses on the birth and early development of stars and the formation of planets...{\it Planetary Systems and the Origins of Life} studies the physical and chemical properties of solar systems (including our own) and where the building blocks of life may be present." Now, with the construction of  {\it JWST}  well underway, the exoplanet community is starting to think in detail about how it will use  {\it JWST}  to advance specific scientific cases.

Thus, a workshop on exoplanet transit opportunities with  {\it JWST}  is timely. The study of exoplanets by transits has become a mature subfield of exoplanetary science, and arguably the most exciting. Two developments have occurred almost simultaneously to make this so. First, the  {\it Kepler}  spacecraft completed a four-year transit survey mission which has revolutionized the study of exoplanets by obtaining high-precision, high-cadence, continuous light curves of $\sim$150,000 stars, providing the first statistically significant determination of planet frequency down to objects the size of the Earth \citep{Burke_et_al2014}. Second, transit spectroscopy of individual nearby transiting planets using primarily the Hubble Space Telescope ({\it HST}) and the Spitzer Space Telescope have provided proof-of-principle that useful data on the composition of exoplanet atmospheres can be derived in this way, even down to ``super-Earths" with diameters less than three times that of the Earth \citep{Kreidberg_et_al2014}.

 {\it JWST}  will revolutionize our knowledge of the physical properties of dozens to possibly hundreds of exoplanets by making a variety of different types of observations. Here we focus on transits and phase curves; direct detection via coronagraphy was considered in a 2007 white paper (for all  {\it JWST}  white papers see http://www.stsci.edu/jwst/doc-archive/white-papers) and will be revisited in the near future. 

 {\it JWST} 's unique combination of high sensitivity and broad wavelength coverage enables the accurate measurement of transit and orbital parameters with high signal-to-noise (SNR). Most importantly,  {\it JWST}  will investigate planetary atmospheres, determine atomic and molecular compositions, probe vertical and horizontal structure, and follow  dynamical evolution (i.e. exoplanet weather). It will do this for a diverse population of planets of varying masses and densities, in a wide variety of environments characterized by a range of host star masses and metallicities, orbital semi-major axes and eccentricities. \footnote{
Of particular interest for  {\it JWST}  will be small planets (R$<2-4$R$_\oplus$) located at a distance from their host stars such that their equilibrium temperatures could be comparable to that of our Earth. The range of the so-called ``Habitable Zone'' has been argued over by many authors since its original definition \citep{kasting1993}. We take an agnostic approach to this question, referring loosely to planets whose stellar insolation  is comparable to that of our own.}

The sensitivity of  {\it JWST}  over its wavelength range of 0.6 to 28 microns compared to other missions and ground-based facilities has been amply documented (http://www.stsci.edu/jwst/science/sensitivity) and  {\it JWST} 's halo orbit around the Earth-Sun L2 point provides long, highly stable, uninterrupted observing sequences compared with the ground or {\it HST}.  {\it JWST} 's detectors are capable of much better than 100 parts per million (ppm) precision over time periods from hours to days. Its suite of four instruments and multiple operating modes provides a large range of choices in trading off spectral resolution (R between 4 -3000), photometric sensitivity, and observing time. Taken together, these characteristics will make JWST's transit and eclipse observations  the   best method for characterizing exoplanet atmospheres in the foreseeable future. 

While JWST's instrument suite will be described in detail below ($\S$7),  we present a brief introduction here:
\begin{itemize}

\item  \href{http://www.stsci.edu/jwst/instruments/nirspec}{NIRSpec} is a highly versatile spectrometer covering the spectral range from 0.7 to  5 $\mu$m with spectral resolution options ranging from a R$\sim$100 prism to gratings for R$\sim$1000 and 2700 ($\S7.1$). In addition to its multi-object capability and integral field unit, NIRSpec has a number of fixed slits, including one optimized for transit spectroscopy (1.6\arcsec$\times$ 1.6\arcsec). 

\item \href{http://www.stsci.edu/jwst/instruments/niriss}{NIRISS} has a variety of imaging and spectroscopy modes. Of particular interest for transit science is a grism mode ($\S7.2$) covering wavelengths 0.6 to 2.5 $\mu$m with a spectral resolution of R$\sim$300-800. This grism has been optimized for transit spectroscopy using a cylindrical lens to broaden the spectrum to a width of 20-30 pixels to reduce inter-pixel noise and to improve the bright-star limit (J$\sim$7-8 mag).

\item \href{http://www.stsci.edu/jwst/instruments/nircam}{NIRCam} has a suite of broad-, medium-, and narrow-band filters covering wavelengths from 0.7 to 5 $\mu$m ($\S7.3$). By combining a fast sub-array readout mode with a defocusing lens, NIRCam will be capable of imaging transit host stars as bright as K$\sim$ 6 mag. NIRCam also has a grism mode covering the  2.4-5 $\mu$m range at R$\sim$1700 which, with a fast sub-array readout, can  observe stars as bright at K$\sim$ 4 mag.

\item \href{http://www.stsci.edu/jwst/instruments/miri}{MIRI} provides photometric, coronagraphic  and spectroscopic capabilities between 5 and 28 $\mu$m ($\S7.4$). In addition to broad-band filters, MIRI has  medium (R$\sim$70) and high (R$\sim$1550-3250) resolution spectroscopic modes covering this entire range. The saturation limit is K$\sim$ 6 mag for imaging at 8 $\mu$m and K$\sim$3-4 mag for spectroscopy.

\end{itemize}

This white paper is based on a meeting held at Caltech in March of 2014, bringing together observers, modelers and  {\it JWST}  project personnel with the following goals: 

\begin{itemize}
 \item Identify key science cases for  {\it JWST}  transit observations.
 \item Improve the understanding of instrument capabilities among the scientists attending the meeting and, through this white paper, among the broader exoplanet community.
 \item Inform instrument teams, the Project and Space Telescope Science Institute about the particular requirements for maximum precision, e.g. cadence, astrometry, observation duration.
 \item Identify lessons learned from other facilities such as {\it HST}, {\it Kepler} and {\it Spitzer}.
 \item Identify precursor activities needed for observational planning and interpretation.
 \item Develop plans for near-term work to ensure optimum on-orbit performance, and to engage the wider exoplanet community to plan transit observations with  {\it JWST} . 
\end{itemize}

The white paper is organized along the structure of the meeting, with key science opportunities ($\S$\ref{opportunities}) and target selection ($\S$\ref{targets}) followed by sections on best practices for transits learned from other missions ($\S$\ref{sec:bestpractices}) and  {\it JWST}  operational considerations specific to transits ($\S$\ref{operations}). Detector issues and features and the modes of each {\it JWST} instrument relevant to transits are discussed in $\S$\ref{detectors} and $\S$\ref{modes}. Section $\S$\ref{science} presents some illustrative science programs with an emphasis on cross-instrument capabilities. The challenges of data processing and community engagement are then considered ($\S$\ref{engage}), and the paper concludes with a summary of the most important findings ($\S$\ref{conclusions}). Individual presentations at the Workshop go into much greater detail on these topics than can be given in this White Paper. These presentations are available on-line at the \href{http://nexsci.caltech.edu/committees/JWST/agenda.shtml}{NASA Exoplanet Science Institute}. Many of the section headings link directly to these presentations. The meeting agenda with live links (\ref{tab:Agenda}) and a list of attendees (\ref{tab:Attendees}) are given in the Appendix.

\begin{figure}[t!]
\begin{center} 
\includegraphics[width=0.7\textwidth]{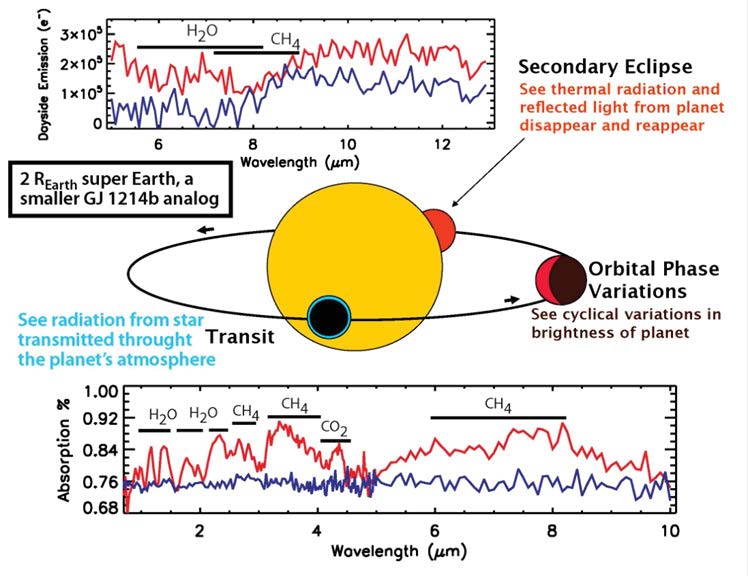} 
\caption{\small\it Illustrative simulated spectra are shown for a smaller (2.0 R$_\oplus$) planet similar to GJ 1214b at that planet's orbital location with the same host star. Primary (bottom) and secondary (top) transits are shown, and the planet model spectra were based on  \citet{Fortney2013}.
\label{summary}} 
\end{center} 
\end{figure}

\section{Key Science Opportunities\label{opportunities}}


As a planet passes either in front of or behind its host star we detect either a ``primary transit'' or a ``secondary eclipse" (Figure~\ref{summary}). In the former case we see stellar light filtered through a thin annulus of the upper atmosphere of the planet. While primary transit signals are relatively insensitive to the vertical structure of the atmosphere, the measurements sample only a few scale heights in the upper atmosphere which may not be representative of the atmosphere as a whole. In the case of a secondary eclipse we directly detect radiation emitted by the star-facing side of the planet from which we can retrieve both atmospheric temperature structure and more fully representative chemical abundances. The figure also shows that it is possible to follow a planet throughout its orbit to measure its phase curve from which one can infer important information on the planet's albedo (in reflected light) and its horizontal temperature structure (in thermal emission). In this section we address science opportunities for spectroscopy of gas and ice giant planets ($\sim 0.05-5 M_{Jup}$, $\S$\ref{GiantSpectra}),  Super Earths ($\sim5-15 M_\oplus$, $\S$\ref{SuperSpectra}) and even terrestrial-sized planets   ($\sim1-5 M_\oplus$, $\S$\ref{EarthSpectra} and \ref{EarthPhotometry}). {\it JWST} will also investigate the dynamical processes of weather through the study of phase curves $\S$\ref{Phase}.

\subsection{Transit and Secondary Eclipse Spectroscopy}

\subsubsection{\href{http://nexsci.caltech.edu/committees/JWST/Fortney\_JWST\_giantplanets_2014.ppt}{Giant Planet Spectroscopy}\label{GiantSpectra}}

\begin{figure}[b!]
\begin{center} 
\begin{tabular}{c}
\includegraphics[width=0.8\textwidth]{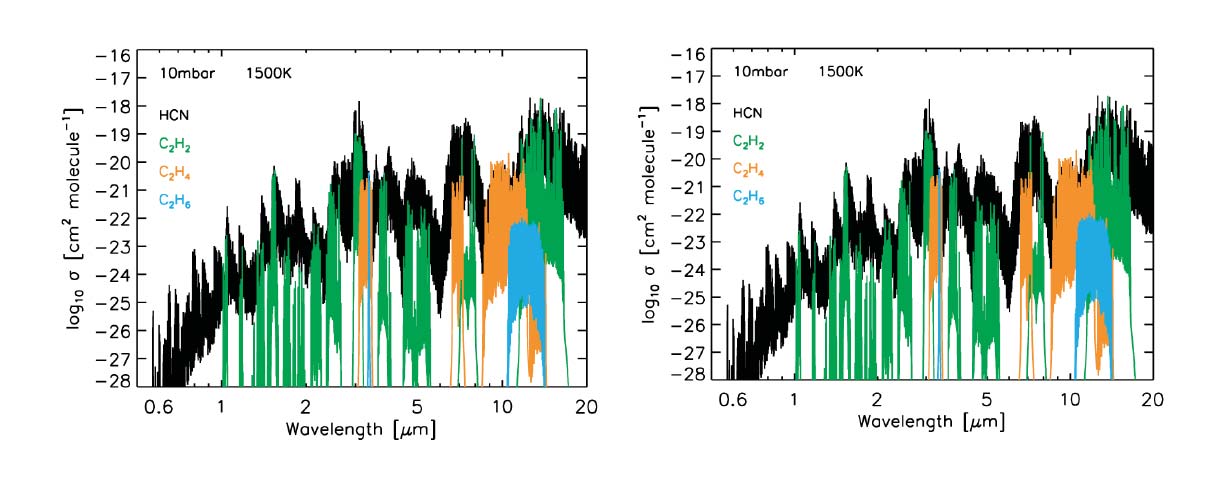} \\
\includegraphics[width=0.5\textwidth]{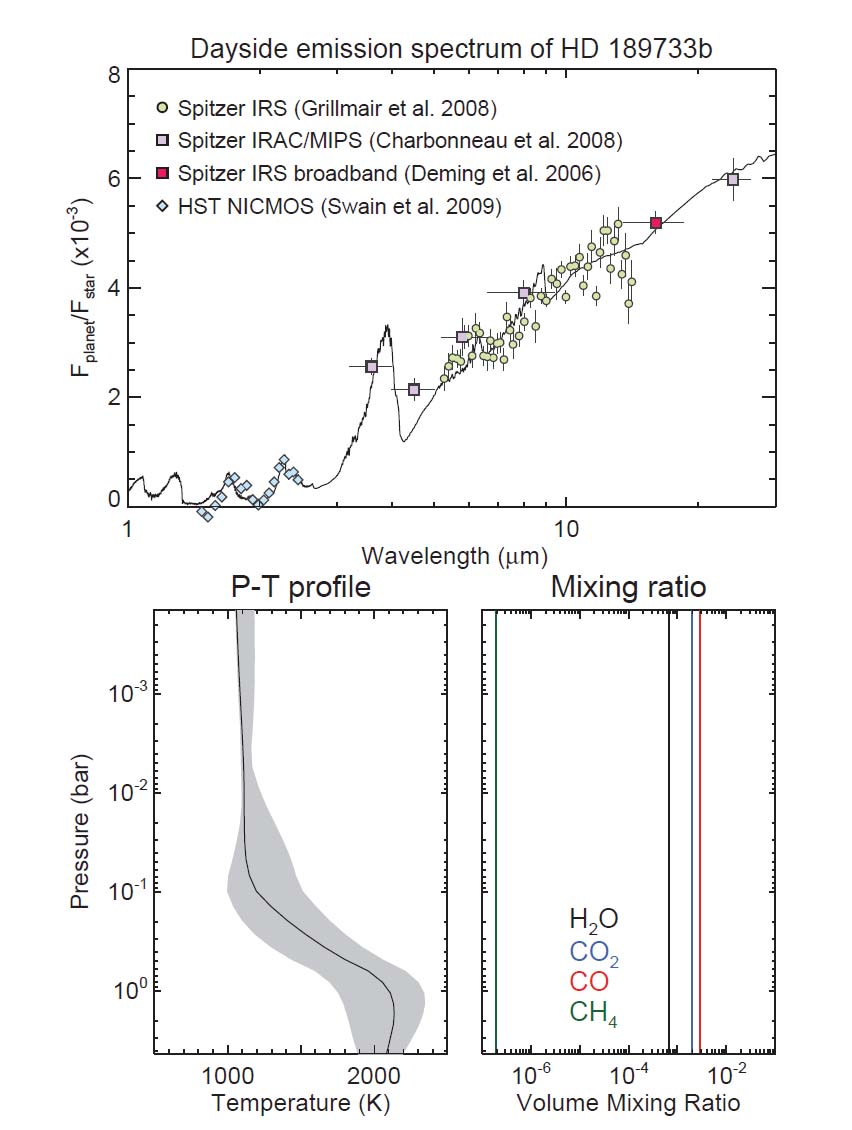} \\
\end{tabular}
\caption{\small\it top) Model abundances for a 1 M$_{Jup}$ planet at 1500 K \citep{Shabram2011}; bottom) Early observations of the transmission and emission spectra of HD189733b from {\it Spitzer} and {\it HST} have been inverted to yield Pressure-Temperature profile of the atmosphere along with mixing ratios of H$_2$O, CO$_2$, CO, and CH$_4$ \citep{Lee2012}. \href{http://nexsci.caltech.edu/committees/JWST/Fortney\_JWST\_giantplanets\_2014.ppt}{(Fortney (2014), this workshop)}.
\label{Shabram}} 
\end{center} 
\end{figure}

With  {\it JWST}  transit spectroscopy we will address fundamental questions about gas and ice giant planets, putting their present-day characteristics into the context of theories of their formation and evolution. For example, to what extent is metal enrichment   a hallmark of giant planets? How does a planet's composition vary as a function of planet mass and migration history, stellar type and metallicity and, in particular, stellar C/O ratio \citep{Oberg2011}. By determining the abundances of key molecules (Figure~\ref{Shabram}, \citet{Shabram2011}a) we can begin to address these and other issues.

Today's knowledge of the spectroscopy of exoplanets is roughly at the same level as we had about the planets in our solar systems 50 years ago when, for example, it was determined that Jupiter's atmosphere had a strong temperature inversion with CH$_4$ as a primary absorber \citep{Gillett1969}. As spectroscopic observations of exoplanets become available from {\it HST} and {\it Spitzer}, we are beginning  to gain a comparable level of knowledge (Figure~\ref{Shabram}b; \citet{Lee2012}). However, with  {\it JWST} 's high signal-to-noise ratio (SNR) at multiple wavelengths and high resolution we will obtain a much better characterization of the atmospheres of exoplanets, including vertical structure, elemental and molecular abundances, surface gravity, and the effects of non-equilibrium and/or photochemistry \citep{Madhusudhan2014,Line2013}. 

\begin{figure}[t!]
\begin{center} 
\includegraphics[width=0.5\textwidth]{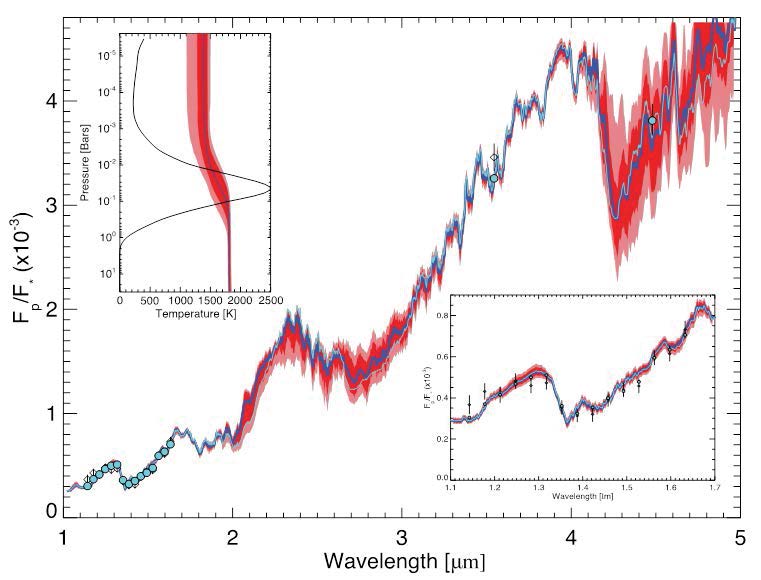} 
\caption{\small\it Recent, high precision observations of the emission spectrum of WASP-43b from {\it Spitzer} and {\it HST} have been inverted to yield Pressure-Temperature profile of the atmosphere.  \href{http://nexsci.caltech.edu/committees/JWST/Fortney\_JWST\_giantplanets\_2014.ppt}{(Fortney (2014), this workshop)}. \citep{Kreidberg2014}.
\label{KreidbergTransit}} 
\end{center} 
\end{figure}

By examining elemental abundances through a broad survey of gas and ice giants covering a wide range of planet mass (0.05$<M<5 M_{Jup}$), stellar metallicity (-0.5$<[Fe/H]<0.5$) and stellar spectral type (F5V-M5V), it will be possible to improve our understanding of the primary factors in the formation of massive planets. For example, an enhanced C/O ratio compared to  the host star's might be indicative of a core accretion formatino mechanism\citep{Madhusudhan2014, Konopacky2013}.  Lowering the mass range to 0.03 M$_{Jup}$ (10 M$_\oplus$) would reach into the realm of mini-Neptunes and Super-Earths ($\S$\ref{SuperSpectra}). Extending the mass range up to 10 M$_{Jup}$ would provide an interesting comparison with low mass brown dwarfs to investigate whether there are compositional fingerprints of a different formation scenario. Recent progress with observations taken with {\it HST}/WFC3 is encouraging about what  {\it JWST}  will be able to accomplish with its greater collecting area and broader range of wavelengths. Figure \ref{KreidbergTransit} shows how the combination of {\it HST} transmission spectra and {\it Spitzer} emission photometry constrain the temperature profile of WASP-43b \citep{Kreidberg2014}.

\subsubsection{\href{http://nexsci.caltech.edu/committees/JWST/JWST\_Kempton.pdf}{Spectroscopy of super-Earths and mini-Neptunes} \label{SuperSpectra}}

\begin{figure}[t!]
\begin{center} 
\begin{tabular}{c}
\includegraphics[width=0.5\textwidth]{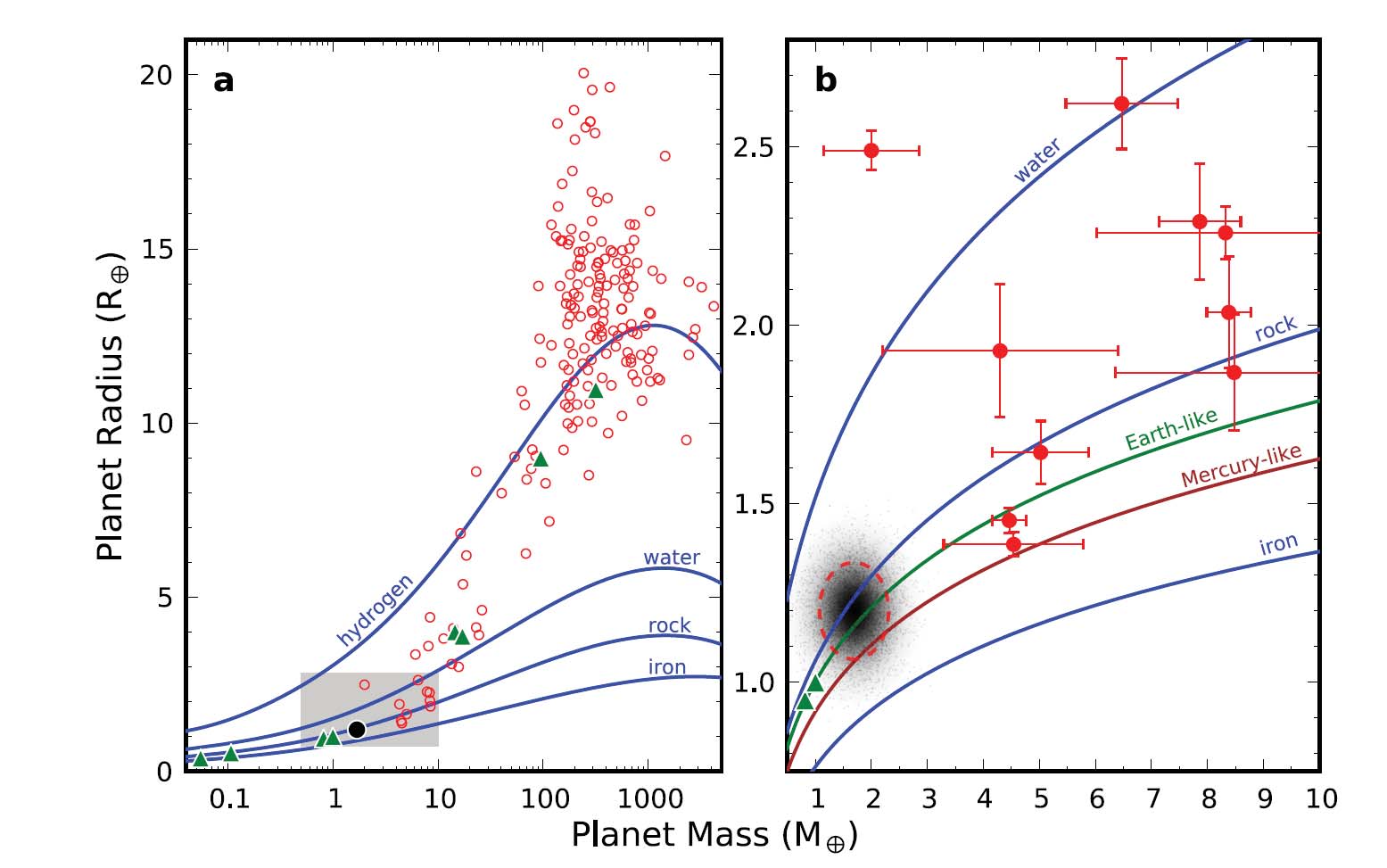} \\
\includegraphics[width=0.5\textwidth]{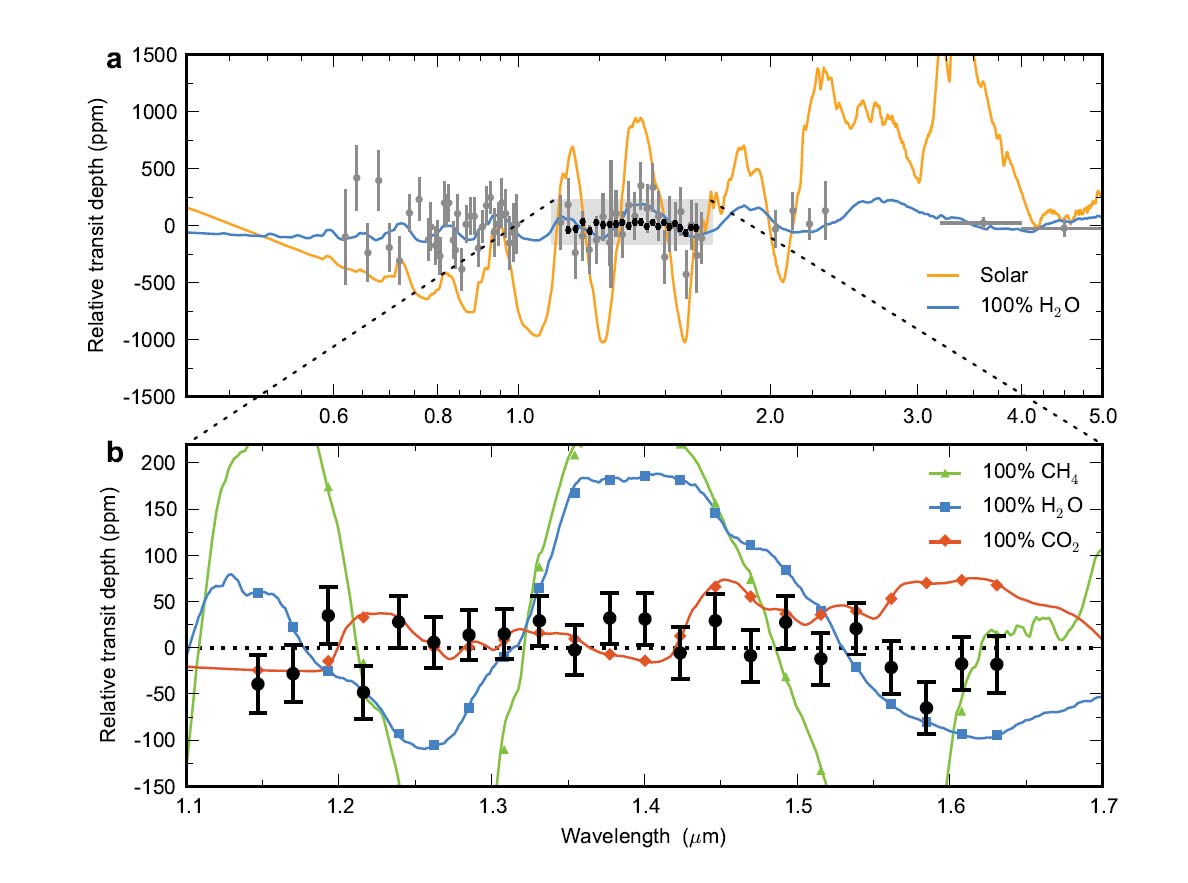} \\
\end{tabular}
\caption{\small\it Top,left) The distribution of exoplanets across the full range of mass and radius. Solar system planets are denoted as green triangles. top,right) A close-up of planets with masses less than 10 $M_{Jup}$ showing a range of bulk densities ranging from predominantly water and volatile rich to dense, rocky-iron bodies. The shaded area corresponds to the likely properties of Kepler-78b \citep{Howard2013}. bottom) {\it HST} observations of the spectrum of GJ 1214b with $\sim$25 ppm precision \citep{Kreidberg2014} along with various theoretical models none of which provide a particularly good fit to the data. One possible interpretation for the relatively flat spectrum is the presence of clouds or photochemical haze. \href{http://nexsci.caltech.edu/committees/JWST/JWST\_Kempton.pdf}{(Kempton (2014), this workshop}; \citep{Kreidberg2014}).
\label{density}} 
\end{center} 
\end{figure}

``Super Earths'' or ``mini-Neptunes'' can be defined loosely as planets with a range of 5-15 M$_\oplus$. Results from  {\it Kepler}  suggest that such planets are very common \citep{Petigura2013, Dressing2013}, especially around late  K and early M stars. On-going surveys with Kepler/K2 will investigate this trend for still later spectral types. It is, however, striking that no ``Super-Earth''  exists in our own solar system, making their study in other planetary systems all the more exciting. The bulk properties of ``Super-Earths'' are quite diverse with some objects having large radii and low masses whereas others have comparable masses but much smaller radii and thus higher densities (Figure~\ref{density}a). Large low-density super-Earths with large atmospheric scale heights will typically be easier to observe, both in transit and secondary eclipse phase.  However, the highly interesting small, high-density super-Earths with potentially Earth-like properties require a substantial investment of {\it JWST}  time in order to build up reasonable SNR \citep{Batalha2014}.  This trade-off in terms of time investment vs. scientific return will need to be confronted when planning observing strategies aimed at characterizing this group of planets.

The prototype for this class of object is GJ1214b, a planet with a mass of 6.6 M$_\oplus$, radius of 2.7 R$_\oplus$ and density of 1.9 g cm$^{-3}$ orbiting an M3 star at a distance where its  equilibrium temperature is approximately 550 K \citep{Charbonneau2009}. This object has been the target of numerous spectroscopic investigations but there remains fundamental uncertainty as to its nature, either a ``mini-Neptune" composed primarily of a rock/ice interior with a large hydrogen dominated atmosphere and traces of water and methane, or a true ``water world'' with a predominantly icy interior and an H$_2$O atmosphere \citep{Rogers2010}. {\it HST}/WFC3 spectroscopy of GJ1214b does not constrain its atmospheric  composition, showing  definitive evidence for a high-altitude, optically thick cloud or haze layer \citep{Kreidberg2014}. A similarly puzzling result holds for a second Super-Earth, HD 97658b, which also has a flat spectrum across the 1-1.8 $\mu$m region. In both cases, the presence of high altitude clouds or photochemical haze has been suggested as the origin for the relatively featureless spectra at these wavelengths. Alternatively, a high mean molecular weight atmosphere with low hydrogen abundance remains a viable alternative interpretation for HD 97658b.  {\it JWST}  observations spanning at least the 1-10 $\mu$m region may offer our first glimpse through the clouds as they become transparent at longer wavelengths.

\begin{figure}[b!]
\begin{center} 
\includegraphics[width=0.8\textwidth]{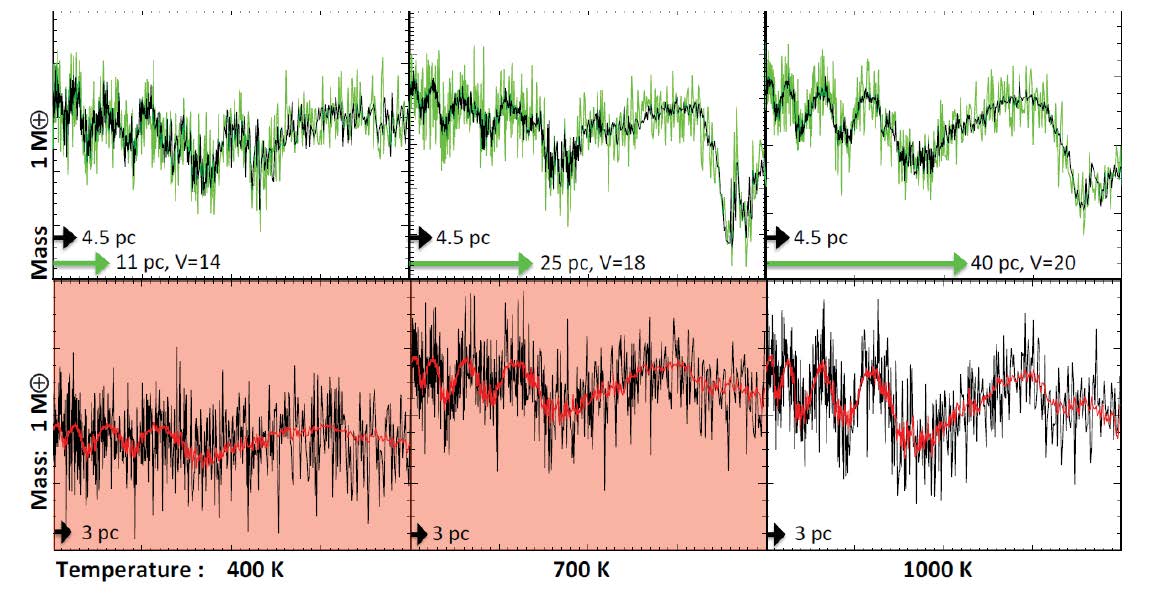} 
\caption{\small\it Predictions for the detectability of the 1-5 $\mu$m spectra of a 1 M$_\oplus$ planet using  {\it JWST}  to observe 25 primary transits. The planet is assumed to orbit an M star like GJ1214 and to have any of three equilibrium temperatures and two different atmospheric types (top, Hydrogen-rich; bottom, Hydrogen-poor). The simulations include a wide variety of noise sources \citep{Batalha2014}. In the H-rich case, the black lines indicate the simulated observations for a star at 4.5 pc while the green lines indicate the simulated spectra for different limiting distances. In the H-poor case, the red lines denote the model and the black lines the observations for a star at 3 pc. Temperature/mass combinations shaded in red have low SNR  ($<$15) and may be challenging to interpret . \href{http://nexsci.caltech.edu/committees/JWST/JWST\_Kempton.pdf}{(Kempton (2014), this workshop)}.
\label{JWSTBatalha}} 
\end{center} 
\end{figure}

\subsubsection{Terrestrial Planet Spectroscopy\label{EarthSpectra}}

Terrestrial-sized planets can be defined as objects with a range of 1-5 M$_\oplus$. \citet{Batalha2014} have made simulations of transit spectra obtained after observing 25 transits with NIRSpec for 1,4 and 10 M$_\oplus$ planets at a variety of equilibrium temperatures (Figure~\ref{JWSTBatalha}). The Hydrogen-rich atmospheres have a large scale height and are thus detectable over a broad range of temperatures and distances whereas the Hydrogen-poor atmospheres are barely detectable at the closest distances and hottest temperatures. These simulations suggest that it will be nearly impossible to measure spectra of terrestrial analogs (H-poor, low scale height atmospheres) for all but the closest systems without a very large investment of dedicated telescope time.

\subsection{\href{http://nexsci.caltech.edu/committees/JWST/knutson\_jwst\_workshop1.pptx}{Atmospheric Dynamics and Weather}\label{Phase}}
\subsubsection{Giant Planet Phase Curves}
\begin{figure}[t!]
\begin{center} 
\plottwo{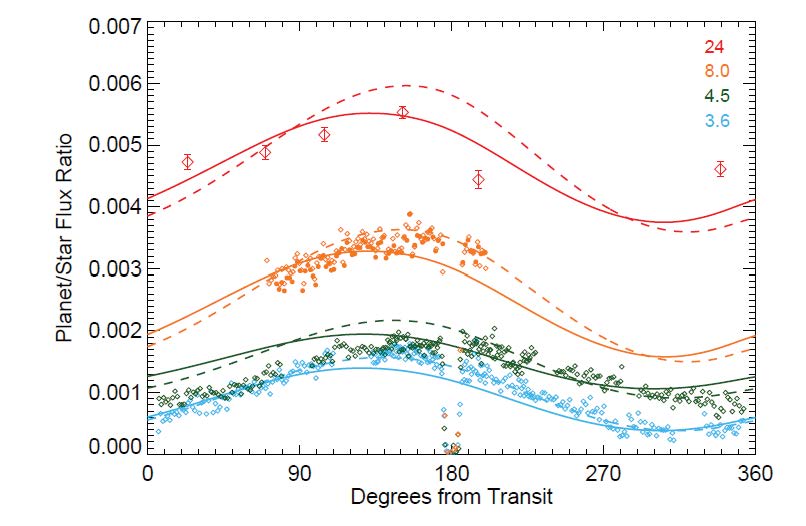}{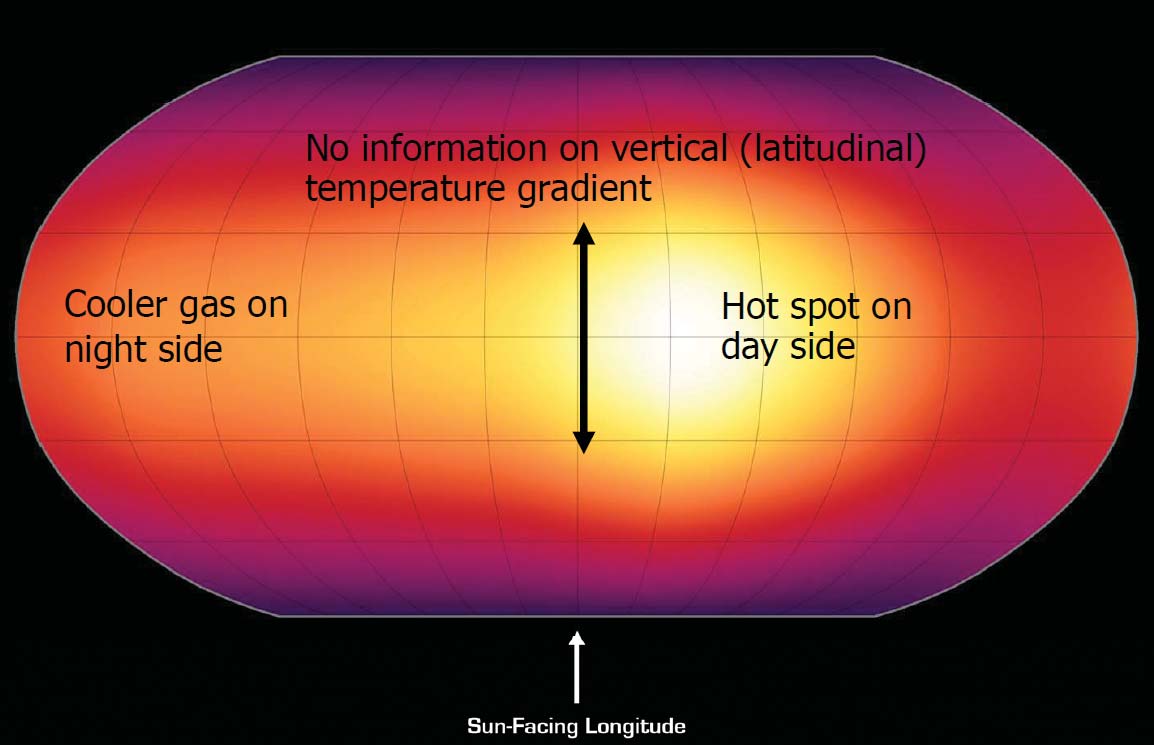} 
\caption{\small\it Multi-wavelength {\it Spitzer} observations through an entire orbit of the planet HD 189733b yield a thermal phase curve that can be used to infer properties of the planet's atmosphere.  left) Data at different wavelengths are plotted along with atmospheric circulation models using  different amounts of metallicity \citep{Knutson2012}, solar (solid)  and 5$\times$ solar (dashed). right) A reconstructed map of the distribution of temperature across the surface of the planet suggests  that supersonic winds shift the hottest gas away from the sub-stellar point \citep{knutson_et_al2007}.  \href{http://nexsci.caltech.edu/committees/JWST/knutson\_jwst\_workshop1.pptx}{(Knutson (2014), this workshop)}.\label{knutson}} 
\end{center} 
\end{figure}

Transiting planets are usually found close to their host stars, because the probability of a transit is inversely proportional to the semi-major axis of the planetary orbit. The strong stellar tidal force experienced by planets in close orbits drives the planetary rotation to be synchronous with orbital revolution \citep{guillot_et_al1996}. Synchronous rotation implies that, for a circular orbit, the star-facing hemisphere of a planet is constantly irradiated, while the anti-star hemisphere is perpetually dark. Strong zonal winds re-distribute energy longitudinally, mediated by radiative cooling of the atmosphere, leading to longitudinal infrared brightness gradients. Using infrared photometry, the most basic aspect of zonal thermal structure that we can detect is the day-night temperature difference. The existence and phase offset of localized hot spots is also inferred from the phase curve (Figure~\ref{knutson}; \citet{knutson_et_al2007}). Using rotational inversion techniques, the spatial resolution that can be obtained in longitude is intrinsically limited to about 5 cycles per planetary circumference \citep{cowan_agol2008}. Nevertheless, significant insight can be obtained into the chemistry and dynamics of giant exoplanet atmospheres at that resolution, especially if spectroscopy can be obtained as a function of orbital phase. Photometric observations of hot Jupiter phase curves have been made with {\it Spitzer}, but  {\it JWST}  will be able to acquire spectroscopy across a broad range of wavelengths simultaneously. Only  {\it JWST}  will have the capability to acquire spectroscopy of multiple molecules at good SNR over the full exoplanet orbit. Time-resolved, multi-molecule  spectroscopy will be informative concerning non-equilibrium chemical and thermal processes that can potentially play a major role in the atmospheric dynamics of close-in giant planets.

\subsubsection{Super-Earth Phase Curves}
Phase curve observations of super-Earths and mini-Neptunes are of particular scientific interest. A tidally-locked rocky planet that lacks an atmosphere will have a high contrast phase curve, with minimal heat transfer from the star-facing to anti-stellar hemisphere. The amplitude of the phase curve can be compared to the depth of the secondary eclipse in order to determine how much heat transfer is occurring \citep{knutson_et_al2007}. Significant longitudinal heat transfer is not expected in the absence of an atmosphere, hence rocky planet phase curves can be used to establish that an atmosphere exists \citep{seager_deming2009}. {\it Spitzer} has been able to place some rough limits on rocky planet phase curves \citep{seager_deming2009}, but there are no definitive  phase curve measurements from {\it Spitzer} for super-Earths. With JWST's 6$\times$ increased SNR due to its larger mirror diameter, we estimate that   {\it JWST}  will be able to measure the broad-band  phase curves of hot super-Earths and mini-Neptunes down to $\leq 4$ R$_\oplus$.

\subsubsection{Secondary Eclipse Mapping}
Beyond phase curves,  {\it JWST}  will be able to use a technique now being pioneered by {\it Spitzer}, wherein the photometric structure of secondary eclipse light curves at ingress and egress can be inverted to map the disk of the planet. The scale height of stellar atmospheres is about two orders of magnitude smaller than the radii of hot Jupiter exoplanets, hence the stellar limb resolves the star-facing hemisphere of the planet during ingress and egress at secondary eclipse. When the impact parameter is near zero, the limb of the star scans the planet approximately parallel to circles of longitude. In that case, secondary eclipse mapping gives the longitudinal temperature structure of the exoplanetary disk, but little to no latitude information \citep{williams_et_al2006}. Longitudinal structure includes the phase offset of the hottest point due to advection by zonal winds \citep{knutson_et_al2007}. In the case where the impact parameter differs significantly from zero, the stellar limb scans the planet at two oblique angles, that differ at ingress and egress. In this case, secondary eclipse mapping \citep{majeau_et_al2012, dewit_et_al2012} can also provide latitude information. The slow rotation of tidally-locked hot Jupiters causes their zonal jets to be broad in latitude compared to Jupiter in our Solar System \citep{menou_et_al2003}. The broadness of the jets facilitates spatial mapping, by making it easier to resolve that structure on the exoplanetary disk.

\subsubsection{\href{http://nexsci.caltech.edu/committees/JWST/birkmann\_simulations\_challenges.pptx}{Modeling Considerations}}

Modeling will have to make significant advances to keep up with improvements in the observations. For example, we need to verify that different atmosphere models, which seek to fit physical and chemical parameters, e.g. pressure-temperature profiles and chemical abundances, yield consistent results when presented with identical physical assumptions and data. A valuable test of the community's analysis methods  would be to verify that multiple groups can retrieve  the same atmospheric  parameters from benchmark simulated datasets. Additionally, we must ensure that our conclusions are statistically robust (in a Bayesian sense) when compared with other physical  interpretations \citep{Hansen2014}. 

We must  also examine model assumptions. Perhaps the most important thing is to consider how realistic it is to use one dimensional models to represent hemisphere-averaged conditions. A related issue is how significant patchy clouds are in affecting observations during transit and secondary eclipse. Clouds are now understood to be very important in the appearance of the brown dwarf siblings of exoplanets and probably in the spectra of Super-Earths as well (Figure\ref{density}b; \citet{Morley2014}). Further explorations of the conditions under which clouds and hazes emerge in the giant planet and super-Earth contexts are definitely needed. Progress can be made on several of these model validation issues by making use of observed spectra of brown dwarfs and directly imaged planets that sample temperature ranges similar to transiting planets.  

Another modeling challenge results from the limited availability of opacity data in the temperature and pressure range appropriate for exoplanet atmospheres.  To properly interpret spectroscopic observations, an accurate understanding of the relevant molecular and atomic opacities is required.  While there are publicly available databases for this purpose, e.g. HITRAN and ExoMol,  there are known deficiencies in these databases such as  a reliable formulation of near-IR and optical methane opacities.

\subsection{\href{http://nexsci.caltech.edu/committees/JWST/Deming.pptx}{Transit Photometry}}

\subsubsection{\href{http://nexsci.caltech.edu/committees/JWST/smallest.pptx}{Thermal Emission of Nearby Earths}\label{EarthPhotometry}}
Simple calculations suggest that broadband MIRI photometry at 10-15 $\mu$m would be able to detect the secondary eclipse of a 300-350 K, 1 R$_\oplus$ planet orbiting a nearby late M dwarf like GJ1214. In a single 45 minute transit the SNR at 15 $\mu$m would be $\sim$ 0.5-1, taking into account both stellar and zodiacal noise and a noise floor of 50 ppm. Observing 25 transits would result in a positive detection at the 3 to 5 $\sigma$ level. In the more favorable cases of a system at 10 pc or closer, a target star of even later spectral type, or a lower noise floor of 25 ppm, it might even be possible to detect the broad spectral line of CO$_2$ at 15 $\mu$m. But this would be a very challenging project at the limit of  {\it JWST} 's prowess and would require the detection by {\it TESS}, Kepler/K2 or other transit survey of an appropriate target ($\S$\ref{subsec:newtargets}).

\subsubsection{Transit Timing Variations}
Gravitational perturbations between planets in a multi-planet system can alter individual transit times by readily observable amounts. These deviations from simple orbits enable mass determinations for planets whose stellar Doppler reflex is unmeasurable and has been demonstrated in multi-planet {\it Kepler} systems \citep{lissauer_et_al2011}.  {\it JWST}  could extend the temporal baseline for {\it Kepler} systems to over a decade for greatly improved parameter estimation and could make new measurements for {\it TESS} planets. Simple photometric observations would suffice to accomplish this goal, but the nature of the planetary system would have to be scientifically compelling in order to justify the  {\it JWST}  time.

\subsubsection{Known Transit Validation and Characterization}
Simple photometry of transits (as opposed to secondary eclipses) can be valuable because the atmospheric opacity of the planet may vary with wavelength, allowing multi-wavelength observations to probe the composition of the atmosphere. {\it Spitzer} photometry has played an  important role in the initial characterization of bright transiting systems and  {\it JWST}  may play a similar role especially for systems too faint for more detailed spectroscopy.  

A second reason to observe transits in the infrared is to validate transits as being due to genuine planets and not false-positive signals such as an eclipsing binary composed of small stars or a triple system with a much brighter star. While most systems observed with {\it JWST}  will already be well characterized, there may be compelling faint M star targets observed with TESS, for example, needing  a simple validation step showing that the transit signal is indeed achromatic. Infrared transit observations using {\it Spitzer} have been a valuable tool to eliminate false-positives in the  {\it Kepler}  data \citep{ballard_et_al2011}. For scientifically compelling exoplanetary systems such as a habitable super-Earth transiting a nearby M-dwarf star,  {\it JWST}  could potentially help to eliminate false-positives while constraining the composition of the atmosphere.

\subsubsection{Searches for New Planets in Known Systems}
The photometric precision that can be achieved using space-borne platforms has numerous applications in exoplanetary science. Among these is the search for transits of additional planets in systems that are already known to host transiting planets. Space-borne searches of this type are warranted when the detection would be of great scientific importance and the transits that are being sought are beyond the sensitivity of ground-based photometry. Such searches are most efficient when conducted using an observatory that has the capability to observe a given exoplanetary system for long uninterrupted periods of time, i.e. not occulted by the Earth. For example, {\it Spitzer} was used to search for a habitable planet in the GJ\,1214 system \citep{fraine_et_al2013, gillon_et_al2014} and also detected two sub-Earth-sized planet candidates around GJ 436b \citep{Stevenson2012}.  {\it JWST}  would be able to conduct a similar search for Earths and super-Earths in the habitable zones of nearby M-dwarf stars, provided that the specific case is scientifically compelling.

\section{\href{http://nexsci.caltech.edu/committees/JWST/agenda.shtml\#targets}{Targets for  {\it JWST} }:
\href{http://nexsci.caltech.edu/committees/JWST/bean.pptx}{Ground Surveys},
\href{http://nexsci.caltech.edu/committees/JWST/Howell\_JWST\_Workshop.pptx}{Kepler}, 
\href{http://nexsci.caltech.edu/committees/JWST/Ricker\_TESS\_Pasadena\_JWST\_v2.pdf}{TESS}, 
\href{http://nexsci.caltech.edu/committees/JWST/dressing\_jwst\_mdwarfs.pptx}{M Star Surveys},
\href{http://nexsci.caltech.edu/committees/JWST/CHEOPS\_JWST\_transits.pdf}{CHEOPS} \&
\href{http://nexsci.caltech.edu/committees/JWST/JWST\_Gaia\_Sozzetti.pdf}{GAIA}
\label{targets}}

In order to maximize the scientific yield from  {\it JWST}  planet observations, the community must plan ahead to select a well-vetted sample of planets. In this section we discuss desirable target system properties, sources of targets for  {\it JWST} , the importance of measuring host star properties, initial vetting procedures for likely targets, and observation timing.

\subsection{Desirable Target Characteristics}
\label{subsec:goodtargets}

Observations of transits, eclipses and phase curves with  {\it JWST}  will yield the highest signal-to-noise for targets with these characteristics: bright, photometrically stable host stars; large ratio of planet to star radius; large atmospheric absorbing areas, driven by high planet temperature, low planet surface gravity, and low mean molecular weight of planetary atmospheres for transmission spectroscopy; deep spectral absorption features; relatively high planet to star temperatures for emission observations; and short orbital periods  for phase curves (less than a few days for reasonable observation times).

These observational considerations are often times at odds with each other (e.g., bright, small, and photometrically stable host stars) and are frequently inconsistent with the most scientifically interesting systems (e.g., small, cooler planets). As explained in the following section, new survey facilities are expected to produce significant numbers of interesting new targets: both small and large planets around mostly small, nearby stars. Many are expected to be in or near habitable zones, with semi-major axes large enough to reduce the effect of stellar insolation on the planetary atmospheres.

\subsection{Surveys for New Targets \label{subsec:newtargets}}

Many of the hot Jupiters that have already been detected with ground-based surveys and studied with {\it HST} are prime targets for  {\it JWST} , but the smallest planets observable with  {\it JWST}  have most likely not yet been discovered. Small planets are interesting but challenging targets due to their smaller transit depths, but the challenge can be reduced by targeting smaller stars. The transit depth scales as the inverse square of the stellar radius, so the signal of a 2 R$_\oplus$ planet around a 0.2 R$_\odot$ late M dwarf is comparable to the transit depth of a gas giant around a Sun-like star. While the final SNR for such a system depends on the stellar radius, the  orbital period, and the number of transits observed, the SNR advantage of studying planets orbiting small stars is very significant.

Several ground-based transit surveys are now targeting bright, low-mass stars to find planets with reasonable transit depths. MEarth \citep{nutzman+charbonneau2008, berta_et_al2012} began monitoring the northern sky with eight 40-cm telescopes in 2008 and recently expanded to the southern sky with an additional eight telescopes. The KELT \citep{Siverd2009} and HATNET \citep{Bakos2011} surveys are continuing to find interesting systems. The APACHE Project \citep{sozzetti_et_al2013} has started a survey from the Alps in 2012 and SPECULOOS \citep{gillon_et_al2013} will begin operations in the Atacama Desert in 2015. The Next Generation Transit Survey project (NGTS)\footnote{http://www.ngtransits.org/index.shtml} will build on the success of its predecessor Super-WASP  to search stars brighter than V=13 mag looking for Neptune-sized objects. The MASCARA survey will find several dozen planets around stars with $V<8$ \citep{Snellen2012}.

\begin{figure}[t!]
\begin{center} 
\begin{tabular}{c}
\plottwo{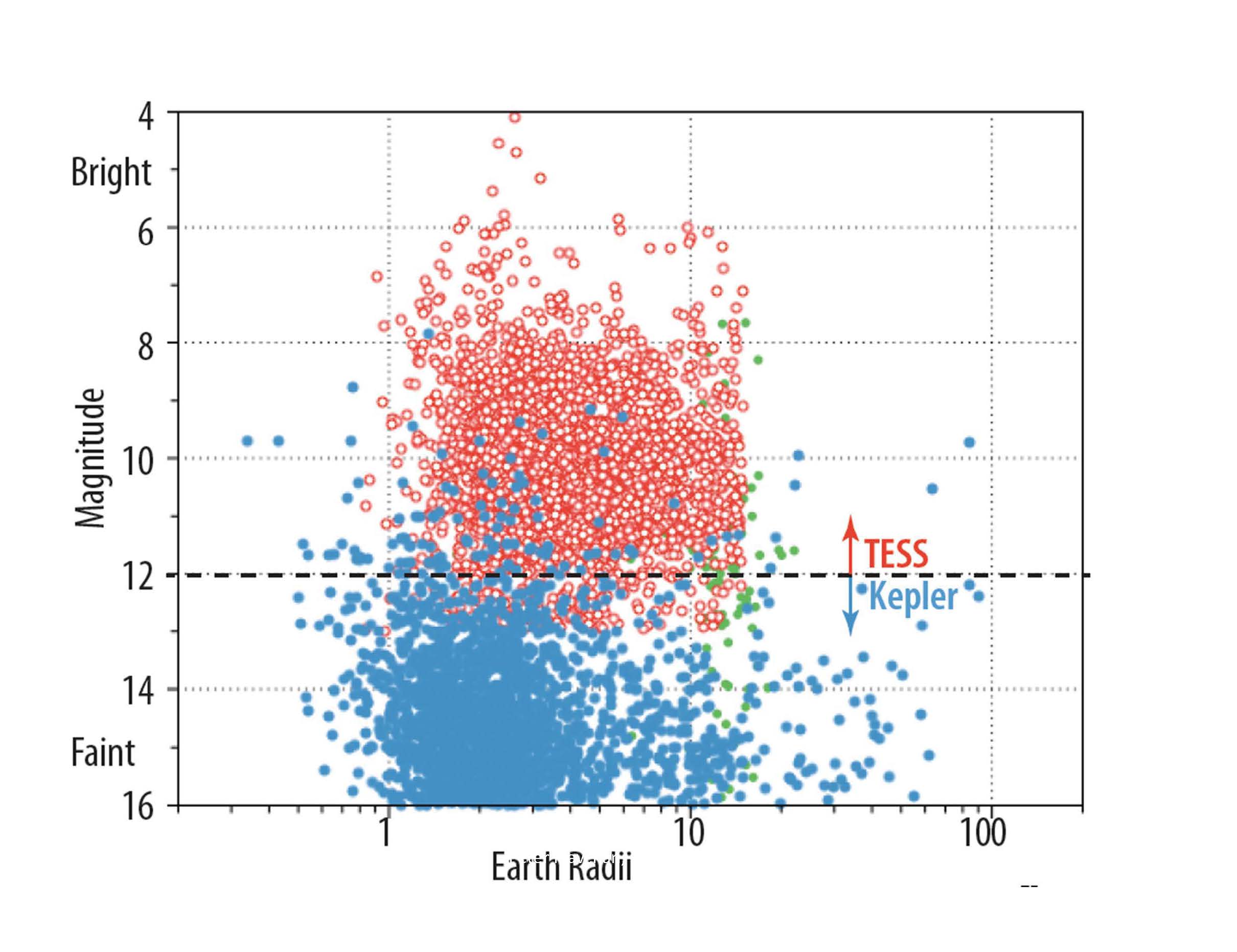}{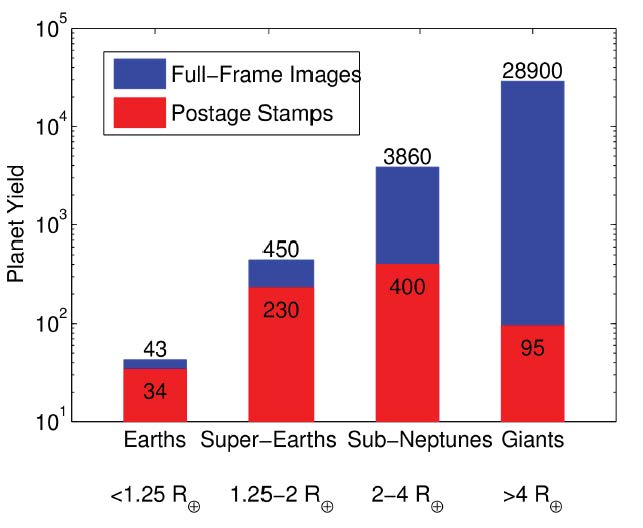}\\
\includegraphics[width=0.8\textwidth]{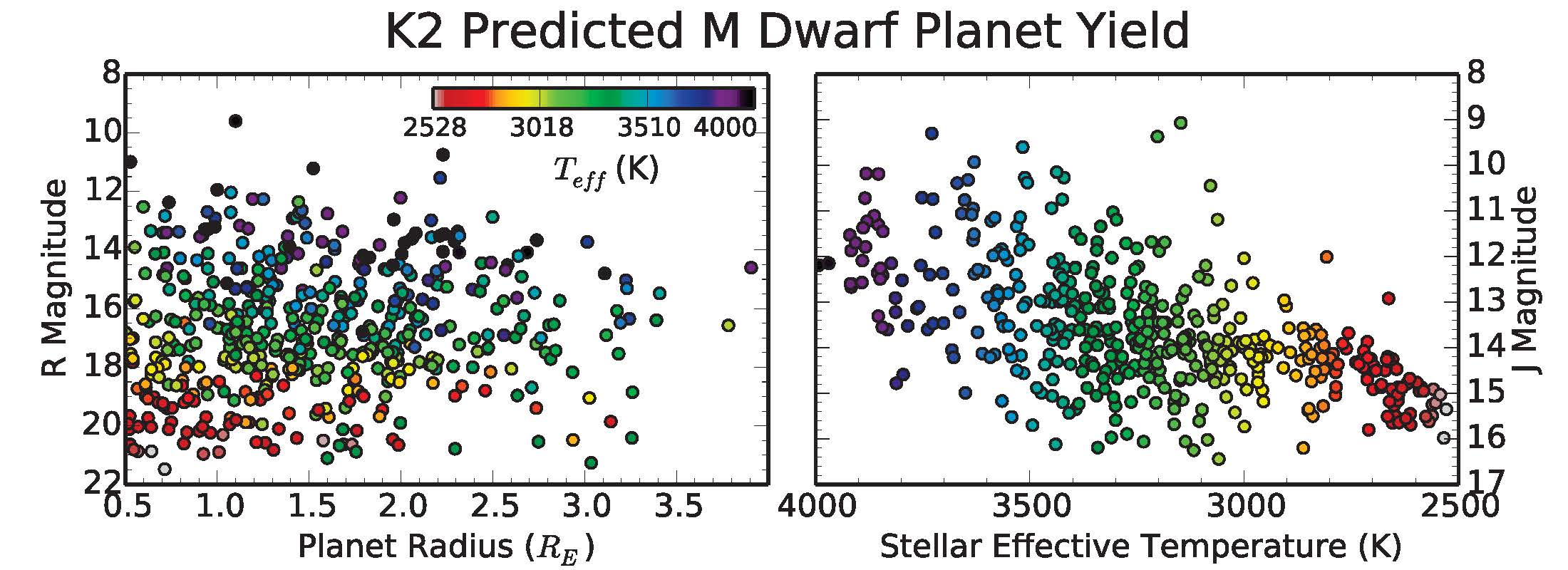} \\
\end{tabular}
\caption{\small\it top, left) The planets detected by  {\it TESS}  (red) will typically be 5 magnitudes brighter than those found by {\it Kepler} (blue) thus making them excellent targets for  {\it JWST}  \citep{Ricker2014}. top, right) A histogram showing the distribution of planet sizes expected to be found by TESS, either in the full frame images or in postage stamps selected for closer examination \citep{Sullivan2014}. \href{http://nexsci.caltech.edu/committees/JWST/Ricker\_TESS\_Pasadena\_JWST\_v2.pdf}{(Ricker (2014), this workshop)}. bottom) Predicted R- and J-band magnitudes of planet-bearing M dwarf stars found by the Kepler/K2 survey as a function of planet radius (left) and stellar effective temperature (right; Crossfield, private comm.).  \label{TESS}} 
\end{center} 
\end{figure}

From space, K2 \citep{howell_et_al2014} is  likely to detect several hundred small planets orbiting bright stars. Although most of the stars observed during the prime  {\it Kepler}  mission were fainter than $V=13$ mag, a large fraction of K2 targets will be brighter than $V=12$. Based on the initial testing for the K2 mission, K2 should find roughly 50 $\leq R_\oplus$ planets orbiting M dwarfs per year. After completion of its roughly 2.5-yr mission, K2 will have observed 10 times as much area on the sky as Kepler. For a uniform population of M stars, this corresponds to an average brightness increase of over 1.5 mag compared with Kepler. And since  both the brightest M stars and the latest known M stars in each field are being targeted, one can always get lucky. 

Larger K2 planets will also be interesting. K2's 75 day observing period will find planets with much less insolation than the currently known hot Jupiters orbiting, on average,  stars brighter than the average  {\it Kepler}   star. These large planets will produce strong signals when they transit their host stars, enabling high signal-to-noise characterization observations with  {\it JWST} . It is possible that dozens of these candidates will have measured RV masses or sensitive mass upper limits by the time of the  {\it JWST}  launch.{\it Because  K2 targets are being observed, validated and characterized as this document is being written, K2 will be an important  source for both large and small new targets for  observation by   {\it JWST}  early in the mission lifetime}.

In later {\it JWST} observing cycles , the most important source of targets for  {\it JWST}  will be the all-sky  {\it TESS} survey \citep{Ricker2014}.  The majority of {\it TESS} targets will be bright stars, $V < 12$ mag (Figure~\ref{TESS}). {\it TESS} yield simulations (Sullivan et al. in prep) suggest that {\it TESS} will detect approximately 40 Earth-size planets and 340 Super-Earths during the course of the two-year survey. Roughly 20 of these small planets are expected to lie in or near the habitable zones of their host stars and a quarter of the small, habitable zone planets are likely to lie within the  {\it JWST}  continuous viewing zone.  {\it TESS} will have good sensitivity to Super-Earths  around early- to late-type M  stars which  {\it JWST}  will be able to  characterize in detail. A few verified {\it TESS} planets may be known at the time of the  {\it JWST}  launch, with significant numbers confirmed with mass measurements after the first couple years of  {\it JWST}  operations. {\it The timely validation and characterization of K2,  {\it TESS} and other survey targets will require a significant investment in precision radial velocity measurements and adaptive optics imaging.}

\subsection{Expected Number of Targets}

The yield predictions for {\it TESS} and K2 are based on planet occurrence rates estimated from the results of the  {\it Kepler}  mission \citep{fressin_et_al2013, dressing+charbonneau2013, petigura_et_al2013} and the census of nearby stars. The RECONS survey \citep{henry_et_al2006} has cataloged the population of stars within 10~pc and determined that 75\% of nearby stars are M~dwarfs. Supplementing the sample of stars within 10~pc with additional catalogs of stars with known parallaxes \citep{van_altena_et_al1995, perryman_et_al1997, lepine2005, dittmann_et_al2014} and assuming planet occurrence rates from \citet{dressing+charbonneau2013} reveals that there should be approximately 80~transiting planets within 20~pc. Roughly ten of those planets are likely to be potentially rocky planets orbiting within the habitable zone with the closest system located within 9$\pm$3 pc and orbiting a star as bright as J$\sim$ 5.5$\pm$0.9 mag (Dressing et al., \emph{in prep}). 

It must be noted that these estimates are based on as yet incomplete processing of all of the  {\it Kepler}  data and may be optimistic \citep{foreman2014}, particular for the latest spectral types where  {\it Kepler}  data are sparse. Improved  {\it Kepler}  analysis followed by  results from K2 and   finally  {\it TESS} will be critical to refining these estimates and, of course, identifying the best targets. To the extent that one of these planets is orbiting a late type M star with a correspondingly deep transit signal, it may be possible to characterize this planet in some detail, although at the expense of a large amount of  {\it JWST}  time to cover 25 or more transits (Figure~\ref{JWSTBatalha}).

\subsection{\href{http://nexsci.caltech.edu/committees/JWST/precursor\_JWST\_ciardi\_20140312a.pptx}{Characterizing Host Stars and Screening Candidates}}

Planetary mass and radius estimates are directly tied to stellar properties. Accordingly, our knowledge of planetary properties is constrained by our knowledge of the host star. Prior to devoting  {\it JWST}  time (or even ground-based follow-up resources) to observing a planet it is crucial to characterize the host star. The parallaxes provided by Gaia \citep{perryman_et_al2001} will be useful for constraining the distances and absolute magnitudes of host stars. Coupled with ground-based spectra, those distance measurements should constrain the metallicities, temperatures, radii, and surface gravities of potential target stars.

In addition to obtaining spectra to refine the properties of the host star, we will also require high-resolution images in order to search for nearby stars and properly account for transit signal contamination. In some cases, a signal that initially appears to be a small planet transiting a single star may actually be caused by a background eclipsing binary, an eclipsing binary physically bound to the target star, a larger planet transiting a star in a binary star system or a larger planet transiting a background star \citep{brown2003}. Obtaining high-resolution imagery to rule out such blended systems is particularly important for determining the radii of planets detected by satellites with large pixels such as {\it Kepler}, {\it K2}, and {\it TESS}.

\subsection{\href{http://nexsci.caltech.edu/committees/JWST/McCullough\_JWST\_transits.pptx}{Effects of Stellar Variability}\label{variable}}

For targets that pass the initial screening tests, we will also need to examine the variability of the host star. Stars that exhibit large amplitude brightness variations due to spots may be challenging targets for transit or phase curve measurements. Secondary eclipse measurements are less affected by variability which evolves on a much slower timescale than the duration of the eclipse. 

Starspots can greatly complicate the interpretation of transit data as a planet scans across  a mottled stellar surface, changing the apparent depth of a transit from its true value or producing a flux change if the planet itself occults a starspot. \citep{Pont2013, Huitson2013}. JWST's multi-wavelength capabilities can help to  disentangle these effects as the contrast between a starspot and its star decreases to longer wavelengths.

\subsection{Refining the Properties of Confirmed Planets\label{PRV}}

Planetary masses and radii must be determined before assessing a system's suitability for  {\it JWST}  observations. Precision radial velocity (RV)  measurements (or transit timing variations) confirm that an object is real and yield critical information needed for interpreting  {\it JWST}  spectra. The amplitude of the expected atmospheric signal depends the surface gravity of the planet, making knowledge of planetary mass and radius crucial when selecting which targets to observe. 

Visible light spectrometers,  including Keck/HIRES, HARPS, HARPS-N, the Lick Automated Planet Finder (APF) and  VLT/ESPRESSO (on-line in 2016), will provide critical RV data. An advantage of TESS's targets is that they will 3-5 magnitudes brighter than Kepler's which will allow spectrometers on smaller telescopes to play a key role in follow-up programs (Figure~\ref{TESS}). {\it NASA should make it a high priority to ensure ready access by the US community to appropriate RV capabilities}.

As discussed in $\S$\ref{subsec:goodtargets} and \ref{subsec:newtargets}, many of the best planets for  {\it JWST}  will orbit M dwarfs. Due to their red colors, M dwarfs are faint in the optical bands traditionally used in radial velocity observations. However, M~dwarf spectra contain considerable velocity information in the red optical and near-infrared. For instance, \citet{bean_et_al2010} have achieved 5~m s$^{-1}$ precision on late M dwarfs using CRIRES with an ammonia cell on the VLT. While promising, that precision is not sufficient for measuring the masses of potentially habitable Earths. However, the next generation of fiber-fed red optical and near-infrared spectrographs is expected to reach 1~m s$^{-1}$ precision. CARMENES \citep{quirrenbach_et_al2012}, HPF \citep{mahadevan_et_al2010}, Spirou \citep{artigau_et_al2011}, and Maroon-X\footnote{\url{http://astro.uchicago.edu/~jbean/spectrograph.html}} are expected to begin operations in 2017 and should be able to provide mass estimates for the population of small planets that will be detected by transit surveys of bright stars.

The most promising  {\it JWST}  targets could be observed by the targeted (non-survey) CHEOPS satellite \citep{broeg_et_al2013} for better transit data to refine radius and bulk density determinations. CHEOPS is scheduled to be launched in 2017 and is expected to achieve a precision of 20~ppm over 6~hours for G-type stars with $V<9$. The well-vetted planets with precisely measured masses and radii could even be could even be observed with HST to obtain lower S/N spectra, to prescreen for relatively featureless transmission spectra obscured by clouds.

\subsection{Determining When to Observe}

\begin{figure}[t!]
\centering
 \includegraphics[width=6in]{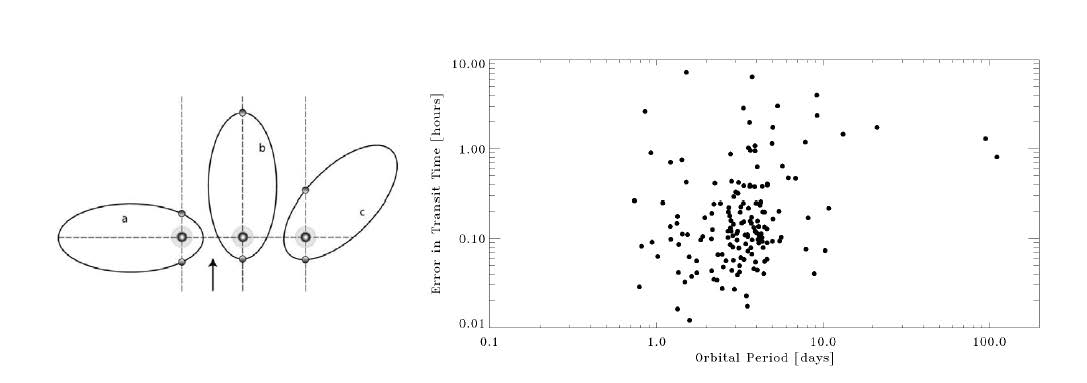}
 \caption{\small\it left) The geometry of primary and secondary transits is shown for different orbital configurations. Determining the exact timing for a secondary transit requires detailed radial velocity observations. When the planet's eccentricity and/or longitude of periastron are unknown or poorly known, the uncertainty in the secondary eclipse can be hours or even days. right) The uncertainty in future transit times in 2018 for a sample of transiting systems based on  extrapolations from present day uncertainties in the parameters of simple Keplerian orbits. A few systems will have uncertainties as large as one hour or more by 2018, highlighting the need for continuing monitoring to update  orbital information through the lifetime of  {\it JWST}  \href{http://nexsci.caltech.edu/committees/JWST/precursor\_JWST\_ciardi\_20140312a.pptx}{(Ciardi (2014), this workshop)} \label{fig:Secondary} }
\end{figure}

One of the most basic requirements for conducting transit and secondary eclipse measurements is knowing when to expect events. For systems without observed transit timing variations, the typical precision  of measured transit times and periods is  sufficient to estimate the timing of primary transits to within $<<$ 1 hour out to mid 2018 (Figure~\ref{fig:Secondary}). But for many systems and for later observing epochs it will be important to update orbital information before investing large amounts of  {\it JWST}  time.

Predicting secondary eclipses is challenging due to the uncertainty in the planet's eccentricity and longitude of periastron. For planets originally detected via transit, a significant number of radial velocity measurements will be required to constrain the eccentricity and orientation of the system. In many cases, the secondary eclipse might not even be detected in the original photometry. While careful analysis of high SNR transit data may  constrain these orbital parameters \citep{dawson2014}, important  K2 and  {\it TESS}   targets observed over only a few periods will have poorly determined  orbital parameters without ground-based follow-up. {\it A continuing program of precision radial velocity monitoring to ensure precise timing of both primary and secondary transits is essential}.

If its operations continue beyond 2014, {\it Spitzer} would be the best observatory for searching for secondary eclipses for existing and newly-discovered planets.  In addition,  the success of ground-based studies of the secondary eclipses of hot Jupiters \citep{Croll2011}, suggests that such observations will be able to track the ephemerides of   {\it TESS} and K2 planets with  transit depths as shallow as $\sim 10^{-3}$.

\begin{sidewaystable}
\tiny
\centering
\begin{tabular}{|lllll|}
\hline
Systematic &Impact on photometry &Timescale &Frequency &Suggested mitigation \\
\hline
\multicolumn{5}{|l|}{\bf HST}\\
Thermal breathing & $<$2\% ramp in flux time series, & 30 minutes & Every orbit & Detrend in processing\\
 & \hspace{1em} and offset first orbit & & & \\
Electronics & Offset in first exposure per orbit & Length of exposure & Every orbit & Take a short first exposure and remove\\
 & & & & \\
\multicolumn{5}{|l|}{\bf Kepler}\\
Pointing tweaks & Discontinuities in flux time series& $<$30 minutes & Sporadically & Isolate pointing management timescales\\
 & & & & \hspace{1em} from scientific timescales of interest\\
Seasonal thermal changes & Ramps in flux time series & 90 days & $\sim$Yearly & Detrend in processing\\
S/c pointing thermal changes & Ramps in flux time series & $\sim$2 days & $\sim$Monthly & Detrend in processing\\
Heater cycle thermal changes & Correlated noise in flux time series & 3--6 hourly & Continuously & Isolate heater management timescales\\
& & & & \hspace{1em} from scientific timescales of interest\\
Electronics & Correlated noise in flux time series & Various & Continuously & Robust detector characterisation\\
 & \hspace{1em} and cross-contamination & & & \hspace{1em} before launch \\
Reaction wheel zero crossings & Additional noise ($<$2\%) in flux time series & 0.01--0.5 days & Sporadically & Avoid zero crossings during times of \\
 & & & & \hspace{1em} scientific interest (ingress/egress/etc)\\
\multicolumn{5}{|l|}{\bf Spitzer IRAC}\\
Pointing wobble & Correlated noise in flux time series & 0.08$^{\prime\prime}$ in 30--60 minutes, & Continuously & Build well understood gain maps\\
\hspace{1em} and pointing drift & & \hspace{1em} and 0.3$^{\prime\prime}$/day & & \\
Electronics & 1--10\% ramp up (charge trapping?), & 1-24 hours & Per target & Robust detector characterisation \\
 & \hspace{1em} and $\sim$10\% ramp down (persistence?) & \hspace{1em} (flux-dependent) & & \hspace{1em} before launch \\
\multicolumn{5}{|l|}{\bf Spitzer MIPS/IRS}\\
Pointing absolute offsets & 2\% discontinuity in flux & Each frame & Each frame & Avoid dithering\\
Flatfield errors & 0.2\% correlated noise in flux & 1--3 hours & Continuously & \\
Electronics & 2\% ramp up in flux time series, & 2--10 hours, & Start of observations, & Robust detector characterisation \\
 & \hspace{1em} 0.2\% fallback in flux time series, & \hspace{1em} 10--30 hours, & \hspace{1em} After ramp up saturates, & \hspace{1em} before launch \\
 & \hspace{1em} and $<$2\% latents (bright/dark) & \hspace{1em} Hours--days & \hspace{1em} Sporadically & \\
Artificial background variations & 0.2\% discontinuity in flux & Constant within AOR & Each AOR & \\
\hline
\end{tabular}

\caption[Summary of systematic effects by spacecraft]{Summary of systematic effects, their impacts and suggested mitigation strategies by spacecraft.}
\label{tab:systematicsummary} 
\end{sidewaystable}

\section{\href{http://nexsci.caltech.edu/committees/JWST/agenda.shtml\#transits}{Transit Best Practices}\label{sec:bestpractices}}

Many of the transit science opportunities described $\S$\ref{opportunities} will be challenging even for  {\it JWST} . Over the last decade we have amassed a wealth of experience with space-based transit observations from {\it HST}, {\it Spitzer} and {\it Kepler}. All of these have achieved photometric precision of a 20-30 ppm with the careful design and execution of the observations and substantial post-processing. However, with  {\it JWST} , we will not have the same baseline (17 years for STIS on {\it HST} and 10 years for {\it Spitzer}) for developing and refining the ideal observing strategies. Therefore it behooves us to examine the best practices and lessons learned from these missions, in order to begin the first cycle of  {\it JWST}  as well prepared as possible. Here we briefly describe the systematic sources of error which have been identified as contributing significantly to the quality of the photometric precision; the three recurring themes are thermal stability, pointing stability and electronic artifacts. The systematics and the suggested mitigation strategies, where available, are summarised in Table \ref{tab:systematicsummary}.

\subsection{\href{http://nexsci.caltech.edu/committees/JWST/JWST\_meeting\_Sing.pdf}{The Hubble Space Telescope} ({\it HST})}

{\it HST} is a 2.4-m telescope in a low-Earth orbit. Launched in 1990, it has hosted several instruments spanning from the near-UV to the near-IR. For transit observations, some of the notable instruments have been STIS, the Near Infrared Camera and Multi-Object Spectrometer (NICMOS), the Advanced Camera for Surveys (ACS) and the Wide Field Camera 3 (WFC3).

Due to its 96-minute orbit, the thermal environment of the telescope is highly variable. The thermal breathing of the telescope, due largely to day/night temperature changes, causes changes in the focus over the duration of the observations. When generating a flux time series on a given target, this results in a ramp in the photometry which is highly repeatable from orbit to orbit. These ramps are common across multiple instruments in the focal plane. The first orbit in a set of observations on a given target typically experiences a different form of the ramp, and the standard observing procedure to offset this is to schedule an additional orbit at the start of the observations which is then discarded or deweighted. Due to the repeatable nature of the ramp in subsequent orbits, a standard common-mode systematic removal model has been developed to remove the thermal breathing.

An additional systematic that has been noted is the trend for the first exposure in each orbit to be offset from the remainder of the exposures. This is potentially attributable to electronics, and the current strategy is to take a short exposure at the beginning of each orbit which is then discarded.

Recent efforts to improve the photometric precision in transit observations have concentrated on the new scanning mode with the WFC3 instrument (Figure~\ref{KreidbergTransit}). By scanning the target Point Spread Function (PSF) across the CCD during the exposure, the total exposure time can be increased without saturation, improving the observing efficiency. The improvement in signal-to-noise from the additional photons arguably outweighs the errors introduced by the varying inter- and intra-pixel sensitivities sampled during the exposure, but this is still being actively investigated.

Data analysis strategies for removal of systematic trends in HST transit and eclipse spectroscopic data have focused on removing repeatable and common-mode trends through self-calibration.  The common trends in each orbit, usually attributed to thermal breathing of the telescope (see Table 1), can be removed by using data from before and after the transit/eclipse event to create a template which is then be divided out from all the orbits.  This method is commonly called {\it divide-oot} \citep{Berta2012}, but has been used in similar applications as well.  Additionally, trends that are common in either the temporal dimension (such as the stellar spectrum and the sensitivity function) or the wavelength dimension (e.g. common-mode flux variations) can be removed by creating a template spectrum or a template light curve and then subtracting it \citep{Deming2013, Mandell2013}.

\subsection{\href{http://nexsci.caltech.edu/committees/JWST/keplerbestpractices.pptx}{Kepler}}

The {\it Kepler} mission launched a 0.95-m optical telescope in 2009 into an Earth-trailing orbit, with the goal of obtaining very high precision photometry on 150,000 stars in a 100 square degree field of view in the constellation Cygnus. The prime mission ended in May, 2013 after the failure of its second reaction wheel, but NASA has recently approved plans to continue using the telescope for high precision time domain science in the ecliptic plane \citep{Howell2014}.

{\it Kepler's} orbit is much more thermally stable than {\it HST}'s. A yearly cycle as the spacecraft orbits the Sun produces a common mode effect for all stars across the focal plane and is removed in detrending. Additionally, the spacecraft pointed toward Earth every month to download the accumulated data, and it took approximately two days after the spacecraft returned to the science field for the temperature to settle back down to the seasonal average. This effect resulted in ramps of up to a few percent in the flux time series over those two days. In the early months of the mission, the reaction wheel heaters cycled every 3 to 6 hours, manifesting itself as a sinusoidal variation in the flux time series. While the mechanism of the coupling between the reaction wheel heaters and the focal plane were not understood, changes were made to the flight software to decrease the cycle period and amplitude to below a single observation ($\sim$30 minutes), which largely eliminated this effect. {\it Ideally, thermal management timescales should be well isolated from scientific timescales of interest } (1--30 minutes for transit ingress and egress, 1--15 hours for transit durations, 1--10 days for phase curves). Additionally, as much thermal ancillary data as possible should be gathered, with sufficient resolution and precision for use in post-processing.

Another source of noise in the {\it Kepler} flux time series is due to the pointing of the spacecraft.  {\it Kepler}  met its pre-launch requirement of 3 mas ($\sim$1 millipixel) pointing jitter in 15 minutes and had a measured month-to-month drift of 10--15 millipixels. In general this very precise pointing largely eliminated the requirement for absolute flat-fielding. Problems arose, however, when there were sudden offsets in the pointing, such as pointing adjustments (``tweaks") to adjust the spacecraft attitude. These typically generated discontinuities in the flux time series that were substantially more difficult to detrend than slow drifts. In the early part of the mission, {\it Kepler} was forced to make several pointing tweaks, which significantly degraded the photometric precision of those data. The Attitude Determination Control System (ADCS) was modified to eliminate the drifts necessitating the tweaks. {\it Pointing changes during observations should be avoided to the greatest extent possible}.

Several electronic artifacts impact {\it Kepler} photometry, including undershoot (an apparent reduction in pixel sensitivity downstream of saturated pixels for up to 20 pixels), Moire pattern noise (due to a temperature- and time-dependent resonance in the Local Detector Electronics (LDE) amplifier circuit aliased to near-Nyquist spatial frequency), rolling bands (due to the same resonance aliased to near-zero spatial frequency), and both FGS and electronic cross-talk. {\it Early detector calibration tests allowed identification and significant mitigation of the undershoot effect and demonstrated the power and indeed, the necessity, of robust, early detector characterisation}. 

Finally, not only the temperature but the speed of the reaction wheels was found to impact {\it Kepler}'s photometry. When any of the wheels were going through zero RPM, degraded pointing precision resulted in increased photometric noise. If there is the ability to manage the times at which the reaction wheels have zero crossings, care should be taken to avoid critical times of interest, e.g. transit ingress or egress. In addition, after the failure of the first reaction wheel, the remaining three wheels were routinely operated at significantly higher RPM than before, which resulted in improved pointing precision and decreased noise in the photometry. Of course, longevity of the reaction wheels should be considered in any decision about operating RPM, but this was a happy outcome of the increased RPM.

It is encouraging to note that initial reductions of K2 data are showing  nearly the same level of precision as {\it Kepler's} primary dataset.  The effects of modestly  degraded pointing performance have been mitigated by standard decorrelation techniques.\footnote{http://keplerscience.arc.nasa.gov/K2/Performance.shtml}

\subsection{\href{http://nexsci.caltech.edu/committees/JWST/combined\_IRAC-MIPS\_talk\_updated.pptx} {The Spitzer Space Telescope}}

{\it Spitzer } is a 0.85-m infrared telescope that was launched in 2003. On-board it hosted the Infrared Array Camera (IRAC; 3--8 $\mu$m), the Infrared Spectrograph (IRS, 5.3--40 $\mu$m), and the Multiband Imaging Photometer for {\it Spitzer} (MIPS; 24, 70 and 160 $\mu$m) instruments. In May, 2009 its supply of liquid helium was exhausted and the longest wavelengths were no longer usable. However, in its Warm Mission phase {\it Spitzer} has continued to obtain high precision infrared photometry with the shortest two bands of the IRAC instrument.

\subsubsection{IRAC\label{IRAC}}

\begin{figure}[t!]
\centering
 \includegraphics[width=3in]{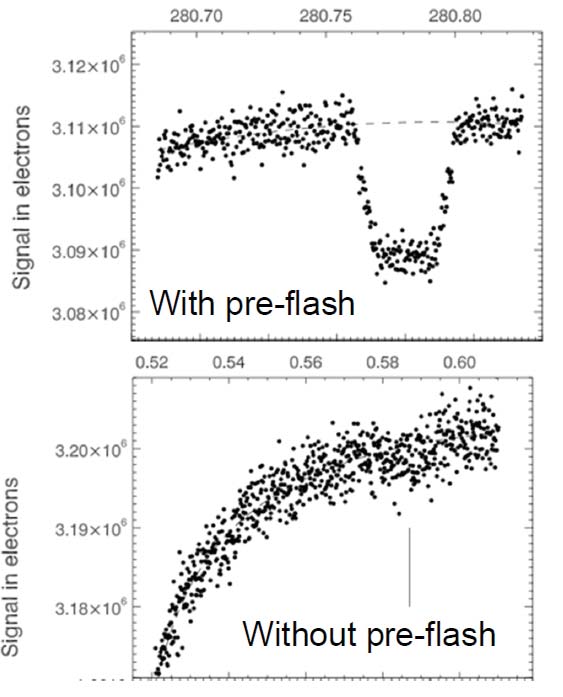}
 \caption{\small\it An example of the ameliorating effect of pre-flashing the IRAC 8 $\mu$m detector is shown in the comparison between two observations of GJ436 \citep{seager_deming2009}. The bottom panel shows the steep rise in signal without pre-flashing, the top panel shows the improved performance with flashing \href{http://nexsci.caltech.edu/committees/JWST/combined\_IRAC-MIPS\_talk\_updated.pptx}{(Carey (2014), this workshop)}. \label{fig:IRACFlash} }
\end{figure}

Despite numerous challenges, the IRAC instrument has been able to achieve a precision level of 25-30 ppm for transit measurements with both its In:Sb and Si:As detectors \citep{Knutson2009, Fraine2013}. The noise level continues to improve as the square root of the number of epochs for a non-variable host star \citep{Demory2012} so that stacking multiple transits is an effective technique to improve SNR. This section describes some of these challenges and the mitigation in either spacecraft operations or post processing needed to achieve these levels of precision.

One prominent systematic in the flux time series generated with IRAC is a substantial ramp in the photometry in the 8 $\mu$m channel. This was attributed to charge-trapping in the detector material, changing the effective gain in those pixels. The timescale for filling the traps in a given pixel depends on the flux falling on that pixel, ranging from 1--24 hours, with the amplitude of the ramp ranging from 1--10\%. One strategy that was developed to mitigate this effect was by `pre-flashing' the detector---observing a bright target and filling the charge traps before moving to the target of interest---and correcting the results on a pixel-by-pixel basis (Figure~\ref{fig:IRACFlash}; \citet{seager_deming2009}).

The same Si:As detector in the 5.8 $\mu$m channel exhibited the opposite behaviour---an anti-ramp that was possibly caused by a persistence effect in the read-out multiplexers. This effect occurs over a timescale of hours and is flux-dependent as it is not observed at low flux levels. The typical post-processing strategy was to use the data themselves to fit the ramp, with the caveat that the effect may be over-fit in this fashion.

\begin{figure}[t!]
\begin{center}
\begin{tabular}{c}
\includegraphics[width=0.5\textwidth]{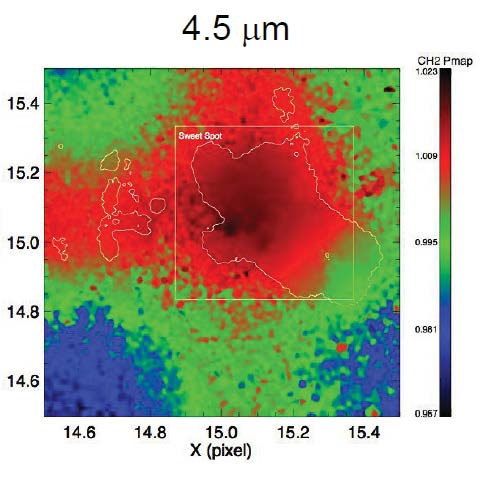} \\
\includegraphics[width=0.9\textwidth]{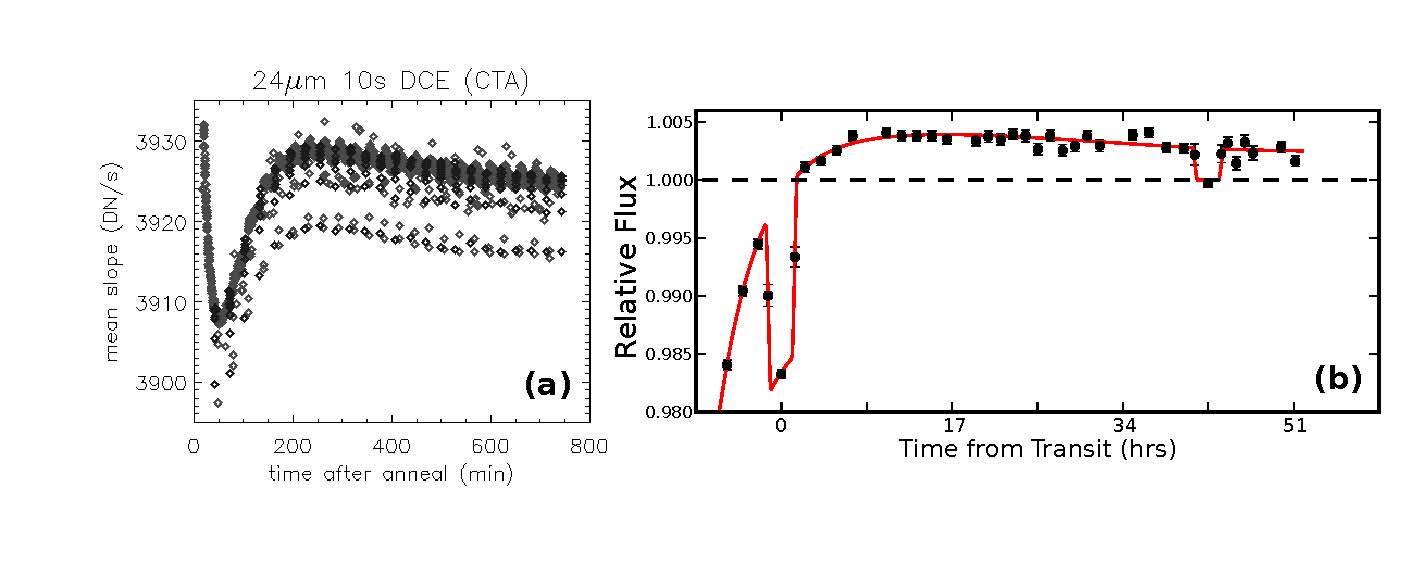} \\
\end{tabular}
\end{center}
 \caption{\small\it top) The pixel gain map used in Spitzer transit observations at 4.5 $\mu$m. Many calibration observations were used to generate the map of the responsivity of this specific pixel. Transit targets were then placed accurately on the marked square known as the ``sweet spot".   Bottom) Examples of detector systematics: MIPS 24\,$\mu m$ detector ramp and subsequent fallback as seen in pre-launch lab tests \citep[(a);][]{young_et_al_2003} and on orbit in observations of HD~209458b \citep[panel (b);][]{crossfield_et_al_2012b}. Without sufficient lab tests and documentation, the decrease in flux seen in (b) could be mistakenly interpreted as a planetary phase curve. \href{http://nexsci.caltech.edu/committees/JWST/combined\_IRAC-MIPS\_talk\_updated.pptx}{(Carey (2014), this workshop)}. \label{fig:mips} }
\end{figure}

As with {\it Kepler}, the spacecraft pointing is a significant source of correlated noise in the photometry. For {\it Spitzer}, the pointing variations consist of (1) semi-regular pointing wobble, with an amplitude of $\sim$0.08$^{\prime\prime}$, and a period of 36--60 minutes, (2) pointing drift of 0.3$^{\prime\prime}$ per day in 80\% of observations, and pointing jitter of $\sim0.03^{\prime\prime}$. The IRAC pixels are 1.2$^{\prime\prime}$ in size and are thus undersampled, which increases the noise introduced to the photometry due to intra-pixel variations. The most successful strategy for removing these variations has been to build a high fidelity intra-pixel gain map for well-behaved pixels in the two remaining IRAC channels. The observations are then placed onto these pixels, and the gain map is used to detrend the data in post-processing (Figure~\ref{fig:mips}a).

\subsubsection{IRS \&\ MIPS Photometry\label{sec:MIPS}}
{\em Spitzer}'s Si:As cameras, used for 16 and 24\,$\mu m$ photometry with IRS and MIPS \citep{rieke_et_al_2004}, provide yet another set of lessons for optimal, highly stable, photometry. These derive from the various instrumental effects: temporal variations in detector sensitivity, and nonuniform detector response coupled to variations in telescope pointing.

The MIPS 24\,$\mu m$ instrumental response varied across the field of view, even after calibration including flat-field corrections. In the MIPS instrument, every time series was obtained by dithering across 14 different dither positions. The sensitivity at each position varied by $\sim$2\%, introducing unnecessary systematic offsets into the photometry. Fourteen positions is an excessive number, but this effect is not wholly undesirable -- data acquired at multiple, independent positions can provide an internal consistency check. More serious was pointing-dependent sensitivity variations which varied at different dither positions, possibly attributable to flat-field errors \citep{crossfield_et_al_2010}. Because this effect did not correlate with pixel phase, residual flat-field errors seem a more likely source than intra-pixel variations; applying an empirical flat field generated from the data themselves may slightly reduce the impact of this effect \citep{crossfield_et_al_2012b}. Unfortunately, the flat-fielding explanation means that pixel-mapping does not effectively mitigate the effect and so a large number of polynomial terms (one set per dither position) must be employed. The primary lessons from these effect are: {\em (1) do not dither across $\gtrsim2$ positions; (2) minimize pointing variations during long time series; (3) pay closer attention to accurate and precise flat-field corrections.}

Several electronic effects were also observed in the MIPS data. These include a ramp in the flux time series at the start of some observations ($\sim$2\%, lasting 2--10 hours), similar to that seen in IRAC 8\,$\mu m$ data \citep{crossfield_et_al_2012b}. Unlike the IRAC ramp, the MIPS ramp occurred only occasionally (perhaps related to recent thermal anneals of the detector). There is some evidence that even over
60~hr the instrumental response had not stabilized, instead transitioning from ramp to a $\sim$0.2\% ``fall-back'' in sensitivity after 10--30 hours (see Fig.~\ref{fig:mips}; this effect was characterized rather poorly, to the extent that it precluded any definitive statement about HD~209458b's thermal phase curve \citep{young_et_al_2003,crossfield_et_al_2012b}. For transits and eclipses the ramp can be removed by simultaneous fitting of a transit light curve and ramp function. Many functional forms must be tested, and the final choice of function must be statistically justified \citep[e.g.,][]{stevenson_et_al_2012a}. Failure to follow this approach can result in significantly underestimated uncertainties. 

 Bright and dark latents are also apparent in the MIPS observations, which can persist for days, and are a potential source of noise in photometry even when recognised. One suggested strategy is to take several frames at one position, then offset the telescope for the main science observations. This approach allows
identification of latent regions and hot \&\ bad pixels. {\it As was suggested for {\it Kepler}, pre-flight calibration tests for detailed detector characterisation are extremely useful in identifying the source and form of the signals for subsequent mitigation.}

\section{\href{http://nexsci.caltech.edu/committees/JWST/agenda.shtml\#jwstops}{ {\it JWST}  Operations for Transits}\label{operations}}

\subsection{Observatory Overview\label{sec:overview}}

Although it  was not optimized for transit operations, many aspects of the observatory will make  {\it JWST}  a superb platform for high precision observations. The instruments and detectors are  described in $\S$\ref{detectors} and  $\S$\ref{modes}. In this section we describe  aspects of the observatory relevant to transit observations.

JWST's optical quality is excellent with diffraction limited imaging at 2 $\mu$µm. The wavefront error budget that determines the image quality comprises static and dynamic allocations for the telescope topics, the ISIM and the science instruments, combined with pointing jitter.  The imaging performance at wavelengths less than 2 $\mu$m is not diffraction-limited, but still provides excellent image quality for photometric and spectroscopic observations. Performance limitations imposed by the stellar image will arise from the sampling and the intra- and inter-pixel variations of the detectors within individual instruments. 

While HST showed significant orbit-to-orbit changes in PSF due to thermal effects in low earth orbit, JWST's highly stable L2 orbit should result in very low levels of wavefront drift. JWST's requirement (worst case) is 57 nm over 14 days, after slewing from its hottest to its coldest pointing configuration. It is expected that more typical wavefront errors should be $<$ 10 nm over a single observation compared with $\sim$30 nm over an HST orbit \citep{gersch2014}. More significant changes might arise in phase curve observations lasting a few days. These effects will be studied early in the commissioning of JWST. Careful scheduling of transit measurements to avoid large slews before settling on a transit target might further mitigate the effects of thermal drifts \citep{gersch2014}.

{\it Kepler} and {\it Spitzer}  demonstrated the critical importance of first  minimizing  position drifts relative to  target pixels and second of having sufficient knowledge  after the fact to enable   decorrleation techniques to remove the effects of drifts on  photometry or spectroscopy. The nominal  {\it JWST}  requirement on image motion is $\sim$6.6-6.8 mas for the near-IR instruments and 7.4 mas for MIRI (per axis on any 15 second interval over a 10$^4$ second observation). For NIRCam this motion corresponds to an rms jitter of 0.1-0.2 pixel in the long or short  wavelength cameras, or 0.05-0.1$\times$ the 2 pixel height of the NIRCam grism. For NIRSpec with its poorly sampled PSF (the core size is $\sim$ 1 detector pixel) the jitter effects will be worse whereas for NIRISS with its $\sim$20 pixel high spectra drift effects should  be greatly reduced. Finally, for  MIRI with its 110 mas pixels, the relative jitter will be small, 0.06 pixel, although responsivity variations in the less well behaved Si:As detectors may amplify the jitter effects at longer wavelengths ($\S$\ref{sec:MIPS}). After the fact decorrelation using accurate position determinations can greatly reduce the effects of jitter as has been demonstrated with {\it Kepler} and {\it Spitzer}.

{\it Spitzer}'s pixel-mapping technique  (Figure~\ref{fig:mips}a) proved  to be  very powerful in improving photometric precision. On  {\it JWST}  it  will be possible to use  the Fast Steering Mirror (FSM) to map out intra-pixel variations within ``sweet spots" on the detectors. For JWST, FSM offsets with a precision of few mas will allow an image to be stepped around individual pixels to map out pixel response functions.  

A technique recently demonstrated on HST of scanning the stellar image across a large number of pixels is being considered for JWST. Fast, large scale motions across many pixels could  improve the saturation limit and reduce flat-fielding errors by averaging over many pixels, but flight software and new operational modes would have to be developed to support this technique. The use of moving target capability would provide spatial scanning but at much slower rates up to $\sim$60 mas/sec.

The requirement for long, uninterrupted transit observations under extremely stable conditions over 6-12 hours (and even up to a few  days for phase curve measurements) presents a variety of challenges. Some  challenges have already been identified and resolved, some have been  identified and should be solvable with appropriate attention paid to them, some will require  significant resources and a high level commitment to resolve,  and finally some may be so fundamental to the operation of the observatory or the  instruments that they may  never be resolved, only mitigated.

The four basic limits to long, uninterrupted observations include the following:

\begin{itemize}

  \item {\it Re-pointing of the High Gain Antenna} nominally occurs every $\sim$10000 sec. Even though the observatory remains in fine-lock, an HGA re-pointing will cause small ($<$70mas, or 1-2 NIRCam pixels!), short ($<$ 1 min) pointing disturbances. The project is working to provide the option of deferring  re-pointings around transit observations or, at a minimum, to allow transit observations to observe through a re-pointing without forcing a break in the instrument cadence. Either approach is, at least, preferable to stopping and restarting exposures which would likely result in data gaps and/or response drifts. 

  \item {\it Momentum wheel unloads} are planned to occur every 25 days, but could occur as frequently as every 5 days. While this is unlikely to represent a significant problem for most observations, STScI will have to investigate balancing the scheduling of transit observations to avoid interruptions with the competing need  to minimize the buildup of momentum and thus minimize the consumption of propellant, a limited resource. 
  
  \item{\it Station-Keeping manuevers} occur every 20 days to maintain the orbit around L2. This is unlikely to be an issue for the majority of transit observations. 
  
  \item {\it Wavefront Sensing} to maintain image quality is planned to occur every 2 days with the possibility of shifting an event by $\pm$1 day. In principle, it should be possible to  schedule these activities to avoid transit observations. In practice this will depend upon wavefront error stability observing in orbit over the 14 day WFSC cadence. 
  
   \item {Maximum exposure duration} is limited at the detector and/or instrument level, as described in Sections \ref{sec:exposures} and \ref{modes}.
   
\end{itemize}


\subsection{Observatory Operations for Transit Science}
\subsubsection{Exposures Overview\label{sec:exposures}}

Exposures are executed in a completely regular, structured manner. Frame times, reset patterns, and the intervals between them are deterministic and invariant during an exposure. Clocking, reading and resetting of the detector pixels occurs at a constant cadence throughout the exposure. There are no irregularites in the cadence: data flows through the ASIC to the spacecraft without interruption. There is no difference, from the perspective of the detectors, between frames that are stored and frames that are discarded: the ASIC amplifies, digitizes, and packetizes \emph{all} frames, and then simply marks the data appropriately afterwards. This exceptionally regular mode of running the detectors should result in exceptional photometric stability, particularly within exposures and after any (possible) transients that may occur at the beginning of exposures.

Details of frame times in full-frame and subarray modes for both H2RG and MIRI detectors are provided in Appendix~\ref{App:AppendixC}.

For the 3 instruments using H2RG detectors (NIRCam, NIRISS and NIRSpec), the maximum number of integrations within a single exposure, NINT, is 65535. For bright host stars, where short integrations will be needed in order to avoid saturating the detectors, this limit may result in transits to be observed using multiple \emph{exposures.} In other words, after 65535 short integrations have been collected in the manner described above, the detector would return to idle mode, the end of that exposure would be detected by the scripts, and a new exposure command would be issued. The effect of such exposure breaks on photometry, at the precision relevant for transit science, is unknown and exceptionally hard to determine on the ground. Unfortunately the NINT limit is inherent to the design of the ASIC and can not be circumvented. Table \ref{tab:MaxExp} gives maximum exposure durations for a few relevant subarray configurations.
  
\begin{deluxetable}{lccc}
\tablewidth{5in}
\tablecaption{Maximum Exposure Duration, H2RG Detectors\tablenotemark{a} \label{tab:MaxExp} }
\tablehead{
\colhead{Subarray~~~~~~~~~} & \colhead{Frame}      & \colhead{Integration}    & \colhead{Exposure} \\ 
\colhead{Configuration~~~}    & \colhead{Time (sec)} & \colhead{Duration (sec)} & \colhead{Duration (hr)\tablenotemark{b}} }
\startdata
64 $\times$ 64\tablenotemark{c}    &  0.0494 &	 0.148	&   2.7 \\
160 $\times$ 160\tablenotemark{c}  &  0.277  &   0.831  &  15.1  \\ 
2048 $\times$ 64\tablenotemark{c}  &  1.34   &	 4.02   &  73.1  \\
2048 $\times$ 64\tablenotemark{d}  &  0.341  &	 1.02 	&  18.6  \\
2048 $\times$ 2048\tablenotemark{e}& 10.74   &	 32.2   &  586  \\
\enddata 
\tablenotetext{a}{For integrations with two NFRAMES=1 groups.}
\tablenotetext{b}{Each exposure is limited to a maximum of 65,535 integrations.}
\tablenotetext{c}{Normal subarray, using a single output channel.}
\tablenotetext{d}{Stripe-mode subarray, using 4 parallel output channels.}
\tablenotetext{e}{Full-frame mode, using 4 parallel output channels.}
\end{deluxetable}

\subsubsection{Data Volume and Data Rates}

All science and engineering data produced by  {\it JWST}  and intended for downlink is temporarily stored on the solid state recorder (SSR). One half of the SSR capacity is available daily, the other half retaining a copy of older data as insurance against potential downlink failures. For science data, the available daily data volume is 57.5 GB. Transit observations, with their high photon-collection duty cycle, integrations that will typically use NFRAMES = 1 (i.e. no coadding, and very lengthy observations, could significantly strain the data volume allocation. 

The data rate from the H2RG detectors depends on whether the detector is configured in full-array (or subarray stripe) mode, or in subarray mode. As mentioned earlier, in normal subarray mode the all the pixels are read out through a single output channel; the data rate in this mode (neglecting small overheads for telemetry values) is approximately 16 bits every 10 $\mu$sec, or 1.6 Mb/sec. Running continuously for a day, this equates to 17.3 GB/day. In full-array and stripe modes the pixels are read out through 4 output channels in parallel (each output handling pixels in a 512-column wide region), and the data rate is 4 times larger: 6.4 Mb/sec or 69.2 GB/day. These data rates neglect the fact that no data is generated during reset frames -- for short integration ramps the resets can reduce the data rate by as much as 50\% (for NGROUPS=1, NFRAMES=1). However, comparing these \emph{single-detector} data rates to the SSR allocation given above, it becomes clear that there is the potential for transit observations to overfill the SSR. This is particularly true for NIRCam because of the large number of detectors. To avoid exceeding the daily SSR allocation NIRCam data will only be taken using two detectors (one each in the short-wave and long-wave channels). NIRSpec transit observations will use a single output (window mode) and  should not present   data rate problems. Other minor restrictions in the observing templates for transits will guarantee that the data volume limit is not violated.

\subsubsection{Event Driven Operations}

JWST will use an event-driven observation schedule, rather than one where observations are all initiated at a pre-planned absolute time. The event-driven approach should provide higher observatory efficiency because any failed observations can (usually) be immediately followed by the next observation in the queue. The event-driven observation plan \emph{is} a strict queue: observations will execute in the order specified in the observation plan. Observations consist of one or more visits to a target or pointing. Observations may have to be composed of multiple visits, for example, if more than one guide-star is required in order to complete the observation. Photometry of a point source, including small dithers, could be accomplished using a single visit. Mapping of a region or an extended source, where offsets larger than about 20\arcsec\ are required, would be executed as multiple visits. Exoplanet transit/eclipse observations will be executed as single visits.

While most observations will begin when event-driven operations allow them to be, \emph{observers are allowed to specify fixed-time constraints.} Observatory requirements dictate that such observations will begin within $\pm 2.5$ minutes of the observer-specified start time. Some additional flexibility will be available for planning and scheduling recurring events such as transits and eclipses. Observers will be able to specify a desired PHASE for the start of their observation, providing schedulers with multiple opportunities to include their observation into the observing plan. Once the schedulers put the observation into a plan, the PHASE constraint will be converted into the fixed-time constraint appropriate to the planning period. 

When the observation plan includes a fixed-time observation (such as a transit or eclipse), schedulers will estimate the time required to execute all prior observations in the plan and include enough margin such that the observatory will be prepared to begin observing at the appointed moment. If a prior observation fails, the observatory will remain idle until the next observation (whether it is unconstrained or has a fixed-time constraint) can begin. All observations have validity windows during which they can begin. These windows, for all targets, depend on guide-star availability and any other observer-specified constraints. Guide stars must usable for the entire duration of a visit to a target. The observation plan includes multiple guide stars for every target or pointing, and includes the visibility windows for each guide star. The guide star is chosen from that list by the on-board scripts at the time the observation is actually ready to start.

The event-driven operations planning system for JWST is designed from the outset to take into account the possibility that individual observations may fail.  Current modeling of the scheduling process indicates that this possibility can  be accommodated by including several observations with execution windows  that open at (or soon after) the transit observation timing constraint. The  schedule can then adjust to the failure of even quite long observations. The  active observation plan {\bf would} end earlier than previously predicted in such  a case, but there would be adequate time to react to the failure and append to  the observation plan to avoid any periods in which the observatory became idle. 

\subsubsection{Observation Planning}
Observation planning for  {\it JWST}  will make use of the Astronomers Proposal Tool (APT), familiar to many who have observed with  \emph{HST}. Transit observations will have their own user interface because of the unique requirements they will place on the observatory, and to provide a clean, intuitive interface for science users. As mentioned above, the PHASE constraint will allow observers to specify that they want an observation starting at a particular orbital phase for a transiting planet without having to specify exactly which transit event they wish to observe. 

Saturation will be a significant issue that must be addressed during observation planning for transit science. Details regarding the saturation limits for the various instruments are summarized in Section~\ref{modes}, and can be used with a reasonable degree of confidence now. By late 2016 APT will include direct access to an exposure time calculator (ETC). The ETC will provide saturation estimates for sources, and APT will issue warnings if an observer specifies exposure parameters that would result in saturated science images. 

Another key aspect of transit observations will be balancing requirements for avoiding saturation of the detectors (accomplished through choice of subarray size and number of groups and frames in each integration ramp) with the need to take very long exposures. For bright stars it may be impossible to do both without breaking the observation into multiple exposures, as discussed earlier. Given a target star or flux estimate, the proposal tool will provide information on saturation limits in the various available modes (see Section \ref{modes}), and suggest possible exposure parameters. Once the user picks an observing mode and set of exposure parameters, the system will indicate the total exposure duration and allow the user to select multiple exposures if that is required in order to get a time-series spanning the event and out-of-event baselines. The exposures, if more than one is necessary, would be executed with the minimum possible gaps, and with no interruptions for other engineering or science activities. 

The current plan is to include a target acquisition (TA) step as an integral part of the transits observing template. For the slitless grism modes of NIRISS and NIRCam, TA is required in order to know the wavelength calibration of the data. For NIRSpec, TA is required to accurately place the target into the dedicated slit, thereby minimizing slit losses and pointing-dependent photometric variations. The need for TA in NIRCam imaging modes is less clear given that the NIRCam pixels sample the PSF rather well, and intrapixel sensitivity variations for the H2RG detectors appear to be quite small. (This is unlike the situation with, e.g., Spitzer/IRAC, where pixel variations are a significant effect that is magnified by the under-sampling of the PSF.) Nevertheless, by repeatedly observing transits on the same real-estate on the NIRCam detectors, it may be possible to derive extremely precise methods of calibration of the data. 

\subsubsection{Transit Timing}

The timestamp for transit events is important for easy inter-comparison of transits from multiple observatories to look for small timing transit variations (TTVs) due to planet-planet interactions. The community has identified $BJD\_TDB$, which uses the barycenter of the solar system is its reference point.  $TDB$ is a barycentric dynamical time and it is distinct from $UTC$ in that it doesn't have leap seconds \citep{Eastman2010}. Spitzer adopted the approach of providing timing information in a variety of ways. Table~\ref{tab:SpitzerTiming} gives  an extract from a Spitzer FITS header. 

\begin{deluxetable}{lcl}
\tablecaption{Extract of Spitzer FITS Header Illustrating Multiple Timestamps\label{tab:SpitzerTiming} }
\tabletypesize{\scriptsize}
\tablehead{Quantity& Value&Comments}
\startdata
DATE\_OBS&  2009-01-28T08:16:46.745  & Date \& time ($UTC$) at DCE start   \\         
UTCS\_OBS&        286402606.745 & [sec] DCE start time from noon, Jan 1, 2000 \\
UTCMJD\_OBS &   54859.3449855 & [days] Mod. Julian Date ($MJD$) in $UTC$ at DCE start (=JD-2400000.5) \\
HMJD\_OBS&        54859.3462516 & [days] Corresponding Helioc. Mod. Julian Date \\
BMJD\_OBS&        54859.3462694 & [days] Solar System Barycenter Mod. Julian Date\\
ET\_OBS  &        286402672.929 & [sec] DCE start time ($TDB$ seconds past J2000)  \\
SCLK\_OBS&        917598037.993 & [sec] SCLK time (since 1/1/1980) at DCE start \\
\enddata
\end{deluxetable}

The precision needed for  absolute timing is driven by TTVs requirements. The uncertainties in  the measured midpoint of a well sampled, relatively high SNR transit measured with  {\it Kepler}  can be as low as a few seconds, particularly with short cadence data. Thus, to avoid increasing the uncertainties in a TTV measurement, the  after-the-fact,  absolute timing uncertainty in  {\it JWST}  measurements should be $<<$ 1 minute. These issues are under active study by the  {\it JWST}  project.

\section{\href{http://nexsci.caltech.edu/committees/JWST/agenda.shtml\#detectors}{Detector Issues and Features}\label{detectors}}

\subsection{\href{http://nexsci.caltech.edu/committees/JWST/NIRCam\_Detectors\_transit\_mtg.pptx}{HgCdTe Detectors for NIRCam, NIRISS and NIRSpec}}

Three of  {\it JWST} 's instruments share a common detector technology, using 2048$\times$2048 pixel, $H2RG$ detectors from Teledyne Imaging Systems with either short- or long-wavelength cutoff material. For many of JWST's programs the critical parameters will be high quantum efficiency (QE) and low read noise, and the newly installed devices in NIRCam exceed project requirements with an average QE of 0.90$\pm0.05$ at 2 $\mu$m and 0.85 at 3.5 $\mu$m. Total noise in a 1000 sec observation from read noise and dark current is 6 $e^-$ in the short wavelength arrays and 8 $e^-$ in the long wavelength arrays. But for transit observations which stare for long periods of time at bright sources, a different set of parameters become important, including long term stability, inter- and intra-pixel variation of QE, persistence, readout speed and well-depth before the onset of non-linear effects. As described in the discussion of the individual instruments ($\S$\ref{modes}), the subarray capability of the H2RGs allows  {\it JWST}  to observe very bright sources, particularly when combined with a high spectral resolution setting. 

NIRCam tests have revealed latent images, i.e. persistent charge that bleeds off slowly after a bright sources is removed from the array. The persistence has the level of a few $e^- s^{-1}$ after illumination at 80\% of full well and shows at least two times constants, $\sim$ 60 and $\sim$ 1000 seconds. A converse effect is slowly increasing responsivity at the start of an observation of a bright source may affect the first few hundred seconds of an  {\it JWST}  observation at the few 10s of ppm level. 

Overall, the performance of  {\it JWST} 's H2RGs should be excellent in the transit application with photometric precision of $\leq$25 ppm. This conclusion is based on laboratory tests ($\S$\ref{testbed}) and, most importantly, on existing space observations.  {\it HST}/WFC3  uses a 1.7 $\mu$m cutoff H2RG and has achieved 70 ppm precision per band-integrated exposure and a final precision on spectroscopic transit depths of ~25 ppm for 5-pixel-wide binning. \citep{Kreidberg2014}.

\subsection{\href{http://nexsci.caltech.edu/committees/JWST/Ressler\_MIRI\_Detectors\_for\_exoplanets.pdf}{MIRI Detectors}}

The detector arrays used in MIRI are the direct descendants of the long wavelength Si:As arrays used on {\it Spitzer}/IRAC. They have the same four science outputs, channel interleaving and readout procession, and overall noise performance. The chief differences are the larger format, 1024$\times$1024 vs 256$\times$256, and a slightly smaller pixel size, 25 $\mu$m vs. 30 $\mu$m. As discussed in $\S$\ref{sec:mirimodes}, the detectors can be operated in a variety of subarray modes which will enable observations of bright sources for transits.

Just as with IRAC ($\S$\ref{IRAC}), there are numerous detector pathologies that will have to be mitigated against in observation planning and in post-processing. Some of the bad habits affecting transit observations are listed below:
\begin{itemize}
\item Response drifts during exposure -- a serious problem for transit observations as identified in Spitzer observations ($\S$\ref{IRAC}).
\item The Reset anomaly represents ``left-over" signal from previous resets which contaminate the beginning of subsequent integrations. Fortunately, the effect is not flux dependent and can be corrected with information extracted from darks.
\item Latent images on timescales from a few seconds to a few hours. The largest component has an 8 sec time constant, a 2 min time constant term has an 800$\times$ smaller amplitude, and there is a 10 min effect with even smaller amplitude (1$\times$ –- 4$\times$). The effect manifests itself as latent images in subsequent exposures.
\item Settling when changing operational mode. Switching in or out of subarray mode can unsettle the detectors for up to 20 minutes and observations should be planned to avoid unnecessary switching.
\end{itemize}

\begin{figure} [t!]
\begin{center} 
\includegraphics[width=4in]{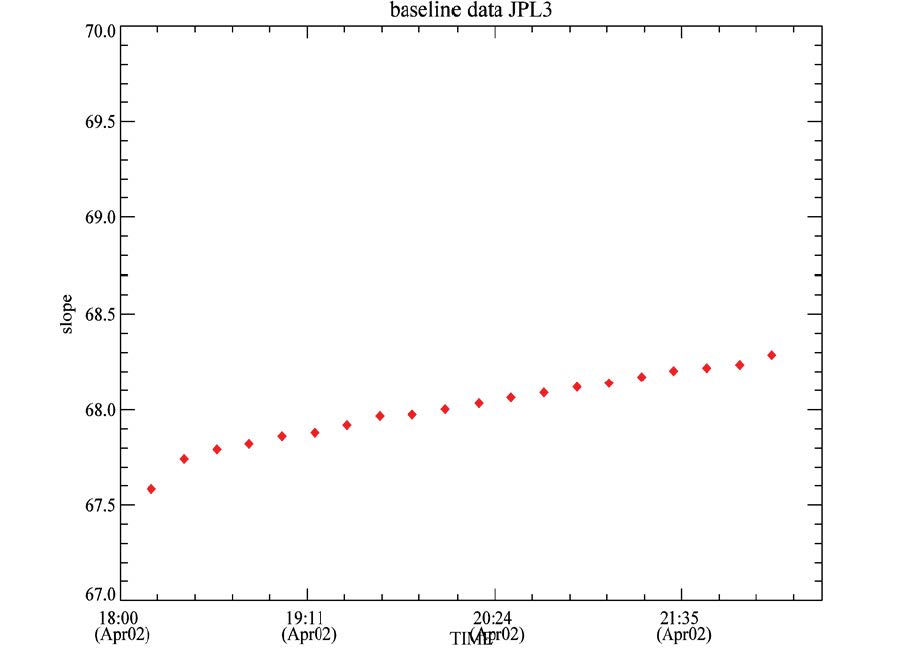} 
\caption{\small\it A long 4 hour exposure shows a long term drift of $\sim$ 1\% in the responsivity of a MIRI Si:As detector under constant illumination. \href{http://nexsci.caltech.edu/committees/JWST/Ressler\_MIRI_Detectors\_for\_exoplanets.pdf}{(Ressler (2014), this workshop)}. \label{fig:MIRIDrift}} 
\end{center} 
\end{figure}

Figure~\ref{fig:MIRIDrift} shows an example of a $\sim$1\% change in response over a four hour exposure. Effects like this are seen in IRAC and will have to be corrected for in post-processing. Figure~\ref{fig:mips} shows similar effects in {\it Spitzer}/MIPS detectors.

Overall, however, the MIRI detectors are very sensitive and very good cosmetically. Extensive calibration efforts are ongoing and will continue throughout flight operations. {\it Spitzer}/IRAC experience should be an excellent guide to planning for  {\it JWST} /MIRI observations.

\begin{figure}[b!]
\begin{center}  
\includegraphics[width=0.5\textwidth]{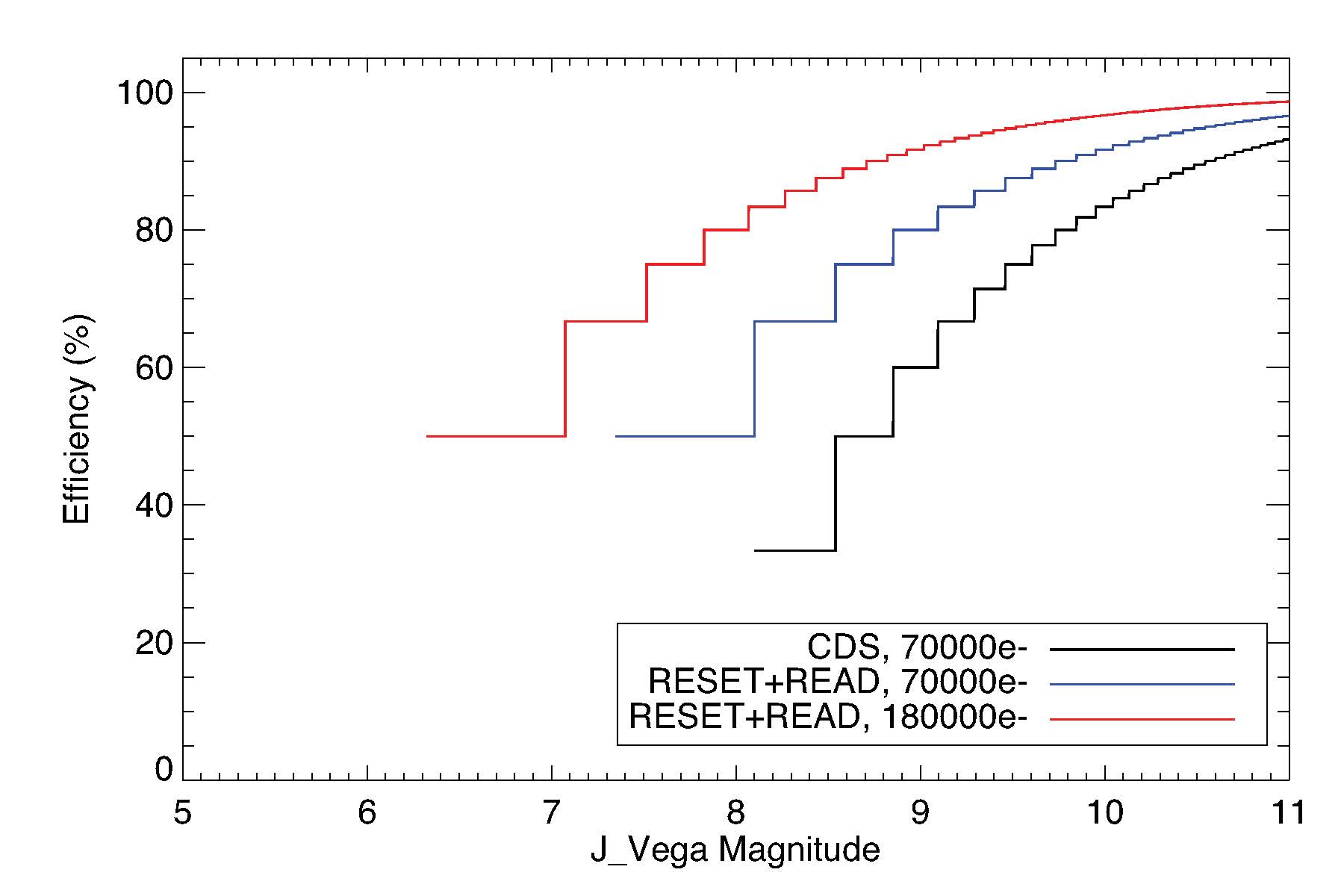}  
\caption{\small\it NIRISS On-sky efficiency during integration as a function of target J-band magnitude for 3 read-out schemes, the standard correlated double sampling (CDS) mode, the reset+read mode and a boosted detector gain mode. This is for the Standard mode with a sub-array size of 256$\times$2048 pixels. The curve ends at the onset of saturation. The curves would shift to the left by 1.2 magnitudes when observing with the Bright mode (80$\times$2048 pixels). The pixel-wise ``$READ-RESET-READ$" would approach 100\% efficiency up to a saturation limit set by the frame readout time.
\label{fig_niriss_readout}}  
\end{center}  
\end{figure} 

\subsection{\href{http://nexsci.caltech.edu/committees/JWST/smith\_bright\_objects.pdf}{Efficient Detector Readout Scenarios and Bright Star Limits}\label{efficient}}

Most of the  {\it JWST}  detectors have a readout arrangement optimized for lowest possible noise for long integrations. Such arrangements can be inefficient for the short observations of bright sources required for transit measurements. The region of interest must be readout three times in a frame-wise ``$RESET-READ-READ$" pattern with each step requiring a read time, $T_r$: 1) reset, 2) read frame (initial value) and 3) read frame (final value). The measured signal then equals the final read (\#3) minus the initial read (\#2). Although the total exposure time is 3$\times T_r$, the time over which useful photons are collected is only $T_r$ for a duty cycle of 33\%. As much signal accumulates prior to the first read as during the actual integration so bright limit is halved. Furthermore, more photons accumulate after the final read, resulting in more charge trapping and image persistence. A variant of the frame-wise ($RESET-READ-READ$) pattern uses only a ($RESET-READ$) to improve efficiency to 50\% and reduce overall charge accumulation (Figure~\ref{fig_niriss_readout}). The penalty of this variation is uncorrected $kTC$ noise $\sim 35e^-$, which would, however, be small compared to the stellar shot noise for bright sources, $\sim245 e^-$ at 80\% of full well. A larger concern for  the $RESET-READ$ mode  is the long term stability of the per-pixel DC offset which is no an issue with  $RESET-READ-READ$.

An alternative approach for improved efficiency uses a per pixel-``$READ-RESET-READ$" pattern: 1) address a pixel, 2) read final value of previous exposure, 3) reset that pixel, 4) read initial value for next exposure and 5) move to next pixel. The advantages of this approach is near 100\% duty cycle and a reduction in the charge on a pixel past the final read so persistence effects will be reduced.

Finally, a hardware approach to improving  the saturation limit and overall efficiency would  increase the bias voltage on the HgCdTe detectors to increase the well depth to $\sim$180,000 e$^-$. This change  would yield a  $\sim$ 1 mag improvement in limiting magnitude and an increased efficiency (Figure~\ref{fig_niriss_readout}), but is as yet  untested in terms of noise performance or an increase in the number of hot pixels and other detector artifacts.

Incorporating new readout schemes into the flight software for the instruments will require significant effort including testing to ensure that subtle side-effects do not otherwise degrade the performance of the detector. It is not yet clear that the scientific return is sufficiently compelling to warrant such changes.

\begin{figure} [t!]
\begin{center} 
\includegraphics[width=5in]{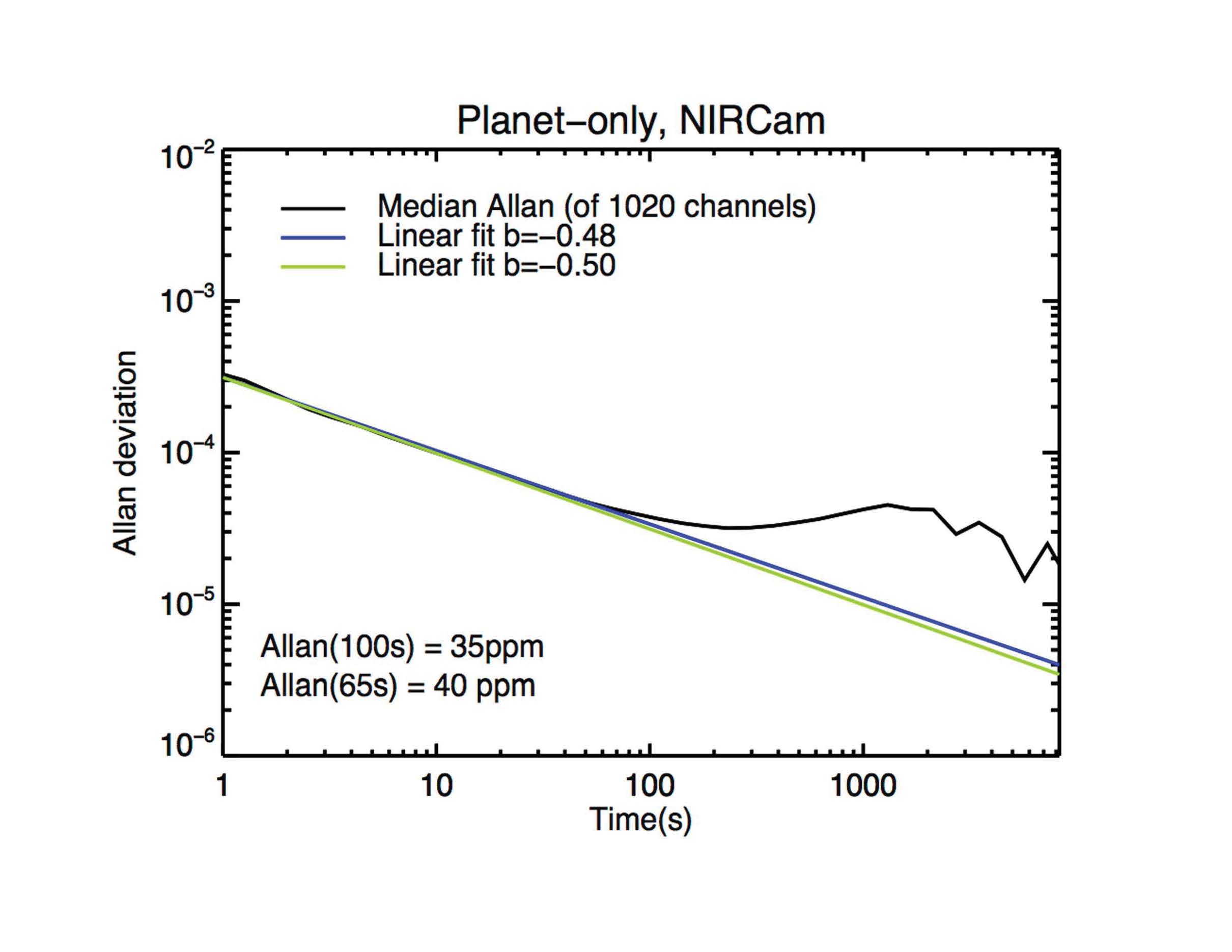} 
\caption{\small\it An Allan Deviation plot showing shot noise limited performance of a NIRCAM-like simulated spectrum at a level of 35 ppm before leveling off.  \href{http://nexsci.caltech.edu/committees/JWST/csb\_TransitExptJWSTMtg3.pptx}{(Beichman (2014), this workshop)}\label{fig:NIRCamTestbed}} 
\end{center} 
\end{figure} 

\subsection{\href{http://nexsci.caltech.edu/committees/JWST/csb\_TransitExptJWSTMtg3.pptx}{Laboratory Tests of HgCdTe detectors for Transit Observations}\label{testbed}}

A laboratory testbed has been established at Caltech with the express purpose of testing HgCdTe detectors in transit applications. Initial experiments were aimed at demonstrating levels of precision suitable for an infrared all-sky transit survey \citep{Clanton2012} using a 1.7 $\mu$m H2RG detector similar to that used on {\it HST}/WFC3. With suitable decorrelation of pointing and lamp drifts it was possible to achieve a precision level for point sources of $<$ 50 ppm. The testbed has been modified to project images of simulated spectra using monochromatic light. The simulated spectra have an extent in the dispersion direction of 1020 pixels and 1, 2 or 10 pixels in the spatial direction. These sizes are similar to the spectra produced by NIRSpec, NIRCam's grism, and NIRISS, respectively (although the NIRISS has an extent of 20 pixels not 10, ($\S$\ref{niriss}). Figure~\ref{fig:NIRCamTestbed} is an Allan Deviation plot for a NIRCam-like spectrum and shows that after decorrelation for lamp drifts and pointing drifts up to 0.5 pixels, the data average down as $t^{-1/2}$ as expected for photon-noise limited observations until reaching a floor of $<$35 ppm. In at least one experiment, observations with a simulated NIRISS spectrum achieved a noise floor of 20 ppm. Work with the Caltech testbed is continuing and will eventually use a flight-spare, short wavelength H2RG for higher fidelity testing. The NIRISS team is developing its own testbed capable of projecting a true polychromatic spectrum using a flight-like instrument configuration. These testbeds will be valuable for assessing the limits of  {\it JWST}  measurements, developing on-orbit procedures, and optimizing post-processing algorithms.

\begin{figure}[b!]
\begin{center} 
\includegraphics[width=5in]{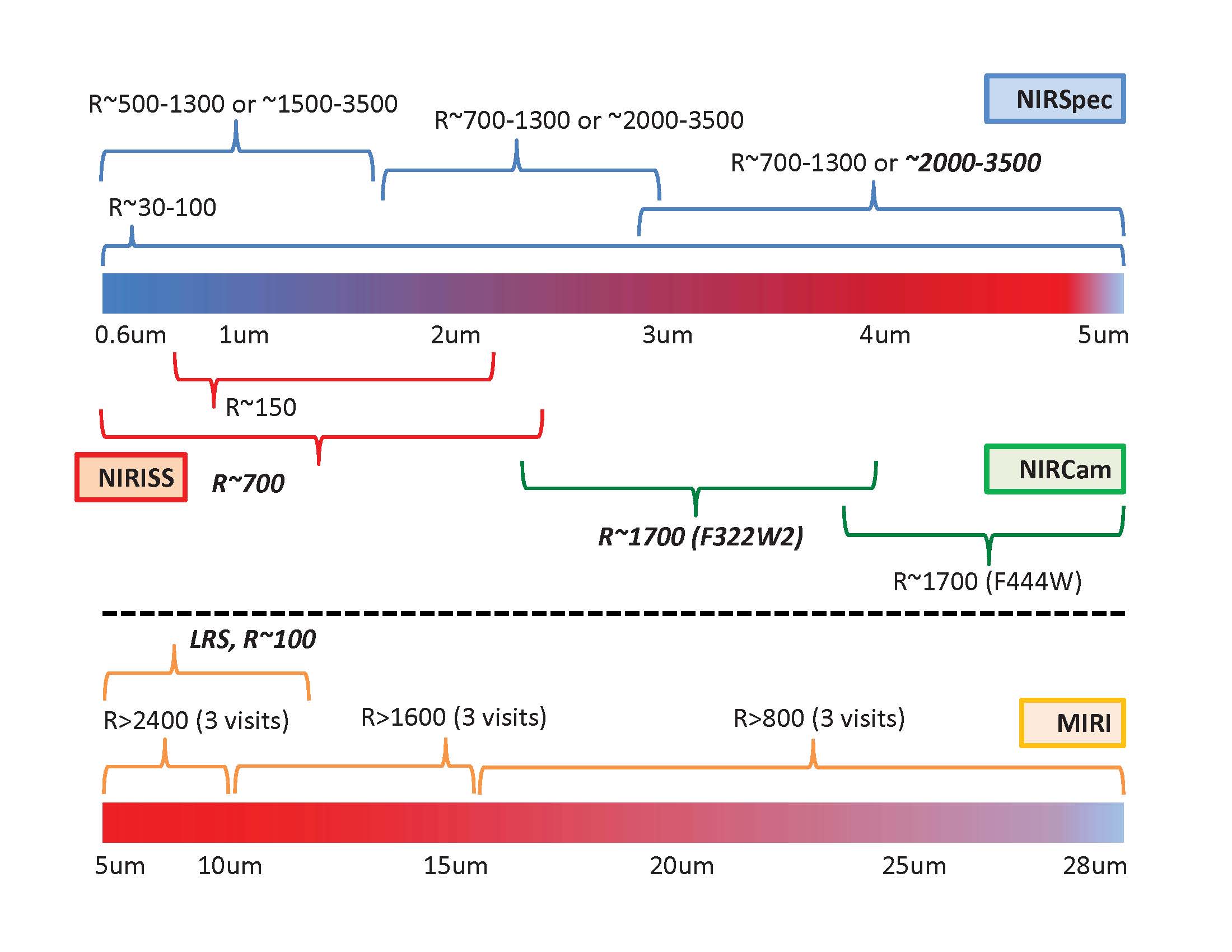}
\caption{\small\it As described in subsequent sections, a wide variety of instrument modes can be used for transit observations. In general, observations of at least 4 separate transits will be required for complete spectral coverage from $\sim$1 to $>$10 $\mu$m. In some cases, for fainter sources $J<11$ mag and at low spectral resolution, the NIRSpec prism mode can cover the entire 1-5 $\mu$m range. 
\label{fig:JWSTmodes}} 
\end{center} 
\end{figure}

\section{\href{http://nexsci.caltech.edu/committees/JWST/agenda.shtml\#modes}{Instrument Modes for Transits}\label{modes}}

Transit observations require  {\it JWST}  instruments to operate in conditions rather different from those that have driven their design (with the exception of NIRISS). In order to detect light variations of the order of a few 10s of ppm over time scales of several hours, an accurate control of systematic effects such as intra-pixel sensitivity variations, linearity and image persistence becomes critical. The summary of instrument capabilities presented in this section will therefore focus on the characteristics most relevant for this type of observations. Further information, including a  {\it JWST}  Exposure Time Calculator, can be found at {\rm http://www.stsci.edu/jwst/instruments}.

\begin{figure}[b!]
\begin{center} 
\includegraphics[width=5in]{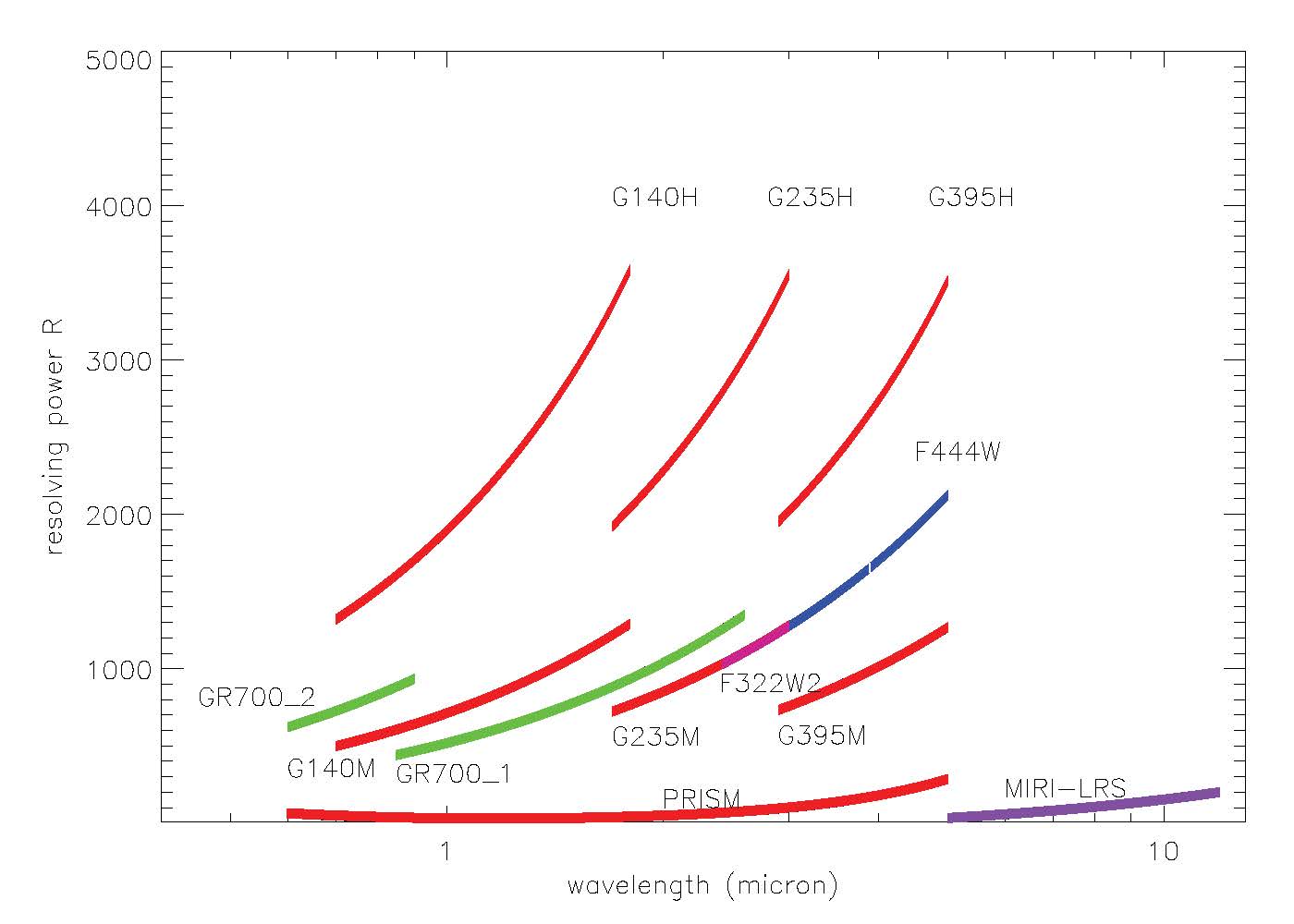} 
\caption{\small\it The resolving power vs. wavelength is shown for all instrument modes relevant to transits. The instruments are coded by color: red=NIRSPEC; green=NIRISS; BLUE=NIRCam; PURPLE=MIRI. There is an overlap between NIRSPEC-G235M and NIRCAM-F322W2, indicated with a violet area. The thin white line in the NIRCam blue field separates F322W2 from F444W. \label{fig:Rlambda}} 
\end{center} 
\end{figure}
\begin{figure}[t!]
\begin{center} 
\includegraphics[width=5in]{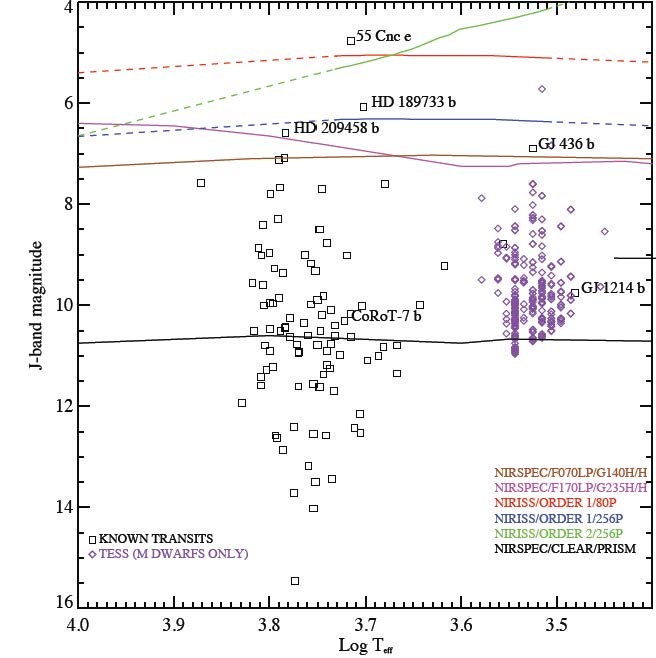}
\caption{\small\it The figure shows the saturation limits for various NIRSpec and NIRISS modes relative to the brightness of known transiting planet host stars. Stars below the lines can be safely observed in a particular mode. Note in particular the limit around J=11 mag for the prism. With its fast readout mode the NIRCam grism can obtain spectra for sources as bright as K$\sim$3 mag. \href{http://nexsci.caltech.edu/committees/JWST/20140311-NIRSpec-transit-spectroscopy\_v4.pdf}{(adapted from Ferruit and Birkmann (2014), this workshop)}. \label{Saturation}} 
\end{center} 
\end{figure}

Figure~\ref{fig:JWSTmodes} lays out the relevant modes of the four  {\it JWST}  instruments which will be discussed in the following sections. Figure~\ref{fig:Rlambda} shows the same information in a different from, plotting spectral resolution as a function of wavelength. The saturation levels for different modes are given in Table~\ref{tab:brightlimits} and shown graphically in Figure~\ref{Saturation}.

\begin{deluxetable}{llr}
\tablecaption{Brightness Limits for Various Instrument Modes\label{tab:brightlimits}}
\tablehead{Instrument & Mode & Brightness Limit (mag)\tablenotemark{a}}
\tabletypesize{\scriptsize}
\startdata
{\bf NIRSpec} & & \\
& Low-resolution spectroscopy & $J>11$ \\
& Medium-resolution spectroscopy & $J\gtrsim 6$ (best-case) \\
{\bf NIRISS} & & \\
& Standard spectroscopy & $J>8.1$\\
& Subarray spectroscopy & $J>6.9$\\
{\bf NIRCam} & L (mag) limits & \\
& Long-wavelength spectroscopy & 32$\times2048$ Subarray \\
&F277W	&	3.7	\\    
&F322W2	&	3.7	\\    
&F356W	&	3.8	\\    
&F410M	&	3.8	\\    
&F444W	&	3.8	\\    
& Photometry & 64$\times64$ Subarray\\  
&F277W	&	9.2	\\    
&F322W2	&  10.0	\\    
&F356W	&	9.3	\\    
&F410M	&	8.2	\\    
&F444W	&	9.2	\\    
{\bf MIRI} & & \\
& 8\,\micron\ imaging (F770W) & $K>6$ \\
& 15\,\micron\ imaging (F1500W) & $K>3-4$ \\
& LRS Spectroscopy & $K>3-4$ \\ 
\enddata
\tablenotetext{a}{For integrations with NFRAMES=1, NGROUPS=2.}
\end{deluxetable}

\subsection{\href{http://nexsci.caltech.edu/committees/JWST/20140311-NIRSpec-transit-spectroscopy\_v4.pdf}{NIRSpec}}
NIRSpec is a near-infrared spectrograph capable of low- (R$\sim100$), medium- (R$\sim1000$ and high- (R$\sim2700$) resolution spectroscopy between 0.6 and 5.3 $\mu$m \citep{ferruit_et_al2012}. The combination of wavelength coverage and spectral resolution provided by the instrument will make it possible to resolve spectral features of many molecules expected to be found in exoplanet atmospheres, including H$_2$O, CO$_2$, CO, CH$_4$, and NH$_3$. Although current space-based facilities can deliver spectroscopy of H$_2$O features ({\it HST}/WFC3) and photometry sensitive to CO, CO$_2$, and H$_2$O ({\it Spitzer} 3.6- and 4.5-$\mu$m channels), NIRSpec will provide improved resolution, sensitivity, and wavelength coverage, enabling new constraints on the elemental abundances of exoplanet atmospheres and their thermal structures.

The instrument can be operated in three different modes: multi-object spectroscopy (MOS mode) over an area of 9 square arcminutes with micro-shutter arrays for the selection of the sources; integral field spectroscopy (IFU mode) over a field of view of 3"$\times$3" and with a sampling of 0.1"; slit spectroscopy (SLIT mode) using five high-contrast slits. {\it One of these five slits is a square 1.6"$\times$1.6" aperture designed specifically for exoplanet transit spectroscopy and it will be the primary mode for observing exoplanets with NIRSpec.}

\begin{table}[b!]
\centering
\small
\begin{tabular*}{\textwidth}{|l|l|}
\hline
\multicolumn{2}{l}{NIRSpec transit spectroscopy}\\
\hline
Instrument mode & SLIT/A1600 \\
Aperture size (projected on the sky) & 1.6"$\times$1.6" \\
Aperture size (projected on the detectors) & $\sim$16$\times$16 pixels \\
\hline
\end{tabular*}
\begin{tabular*}{\textwidth}{|l|r|r|r|}
\multicolumn{4}{r}{Low spectral resolution spectroscopy (R$\sim$100)} \\
\hline
Wavelength range: & \multicolumn{3}{r}{from 0.6 $\mu$m to 5.3 $\mu$m in a single band}\\
Spectral resolution: & \multicolumn{3}{r}{30-300}\\
Instrument configuration: & \multicolumn{3}{r}{CLEAR/PRISM}\\
Readout window size: & \multicolumn{3}{r}{512$\times$32 pixels (spectral $\times$ spatial)}\\
\hline
\multicolumn{4}{r}{Medium spectral resolution spectroscopy (R$\sim$1000)} \\
\hline
Instrument configuration: & Range & Resolution & Readout window size\\
F070LP/G140M & [0.7$\mu$m, 1.2$\mu$m] & 500-850 & 2048$\times$32 pixels\\
F100LP/G140M & [1.0$\mu$m, 1.8$\mu$m] & 700-1300 & 2048$\times$32 pixels\\
F170LP/G235M & [1.7$\mu$m, 3.1$\mu$m] & 700-1300 & 2048$\times$32 pixels\\
F290LP/G395M & [2.9$\mu$m, 5.2$\mu$m] & 700-1300 & 2048$\times$32 pixels\\
\hline
\multicolumn{4}{r}{High spectral resolution spectroscopy (R$\sim$2700)} \\
\hline
Instrument configuration: & Range & Resolution & Readout window size\\
F070LP/G140H & [0.7$\mu$m, 1.2$\mu$m] & 1300-2300 & 2048$\times$32 pixels\\
F100LP/G140H & [1.0$\mu$m, 1.8$\mu$m] & 1900-3600 & 2048$\times$32 pixels\\
F170LP/G235M & [1.7$\mu$m, 3.1$\mu$m] & 1900-3600 & 2048$\times$32 pixels\\
F290LP/G395H & [2.9$\mu$m, 5.2$\mu$m] & 1900-3600 & 2048$\times$32 pixels\\
\hline
\end{tabular*}
\caption{Characteristics of the NIRSpec/A1600 mode for transit spectroscopy.
\label{tab:transitNIRSpec} }
\end{table} 

A summary of the characteristics of the spectral configurations of NIRSpec available for transit spectroscopy is provided in Table~\ref{tab:transitNIRSpec} with a plot of the associated spectral resolution curves found in Figure~\ref{fig:JWSTmodes}.

For science cases that do not require high spectral resolution, the low spectral resolution mode of NIRSpec (CLEAR/PRISM, Table~\ref{tab:transitNIRSpec}) provides very broad spectral coverage from 0.6 to 5.3 $\mu$m in a single exposure. This configuration has no equivalent in the other  {\it JWST}  near-infrared instruments. It also provides some wavelength overlap with the MIRI LRS configuration, opening the possibility to ``stitch" NIRSpec and MIRI spectra together. Besides the low spectral resolution, the main limitation of this configuration is its relatively high saturation limit of J$<$ 11 mag (Table~\ref{tab:brightlimits} as well as {\rm http://www.cosmos.esa.int/web/jwst/exoplanets}). For reference, the typical host star for the {\it TESS} planet detections is expected to have a K-band brightness $<10$ mag (Dressing et al., in prep). This corresponds to J-band magnitudes of $<10.5-11.0$ using a typical M-dwarf J-K$_S$ color index of 0.5 to 1.0 \citep{kirkpatrick_et_al1999}. The saturation limit is worse at short wavelengths due to the combination of the typical stellar spectrum and the lower resolution of the prism at shorter wavelengths. This brightness limit becomes much less stringent for higher spectral resolution configurations. The worst case saturation limits correspond to J-band magnitudes of 8.5 and 7.5 for the medium and high spectral resolution cases, respectively (see {\rm http://www.cosmos.esa.int/web/jwst/exoplanets} for more details). 

\begin{figure}[b!]
\begin{center} 
\includegraphics[width=0.8\textwidth]{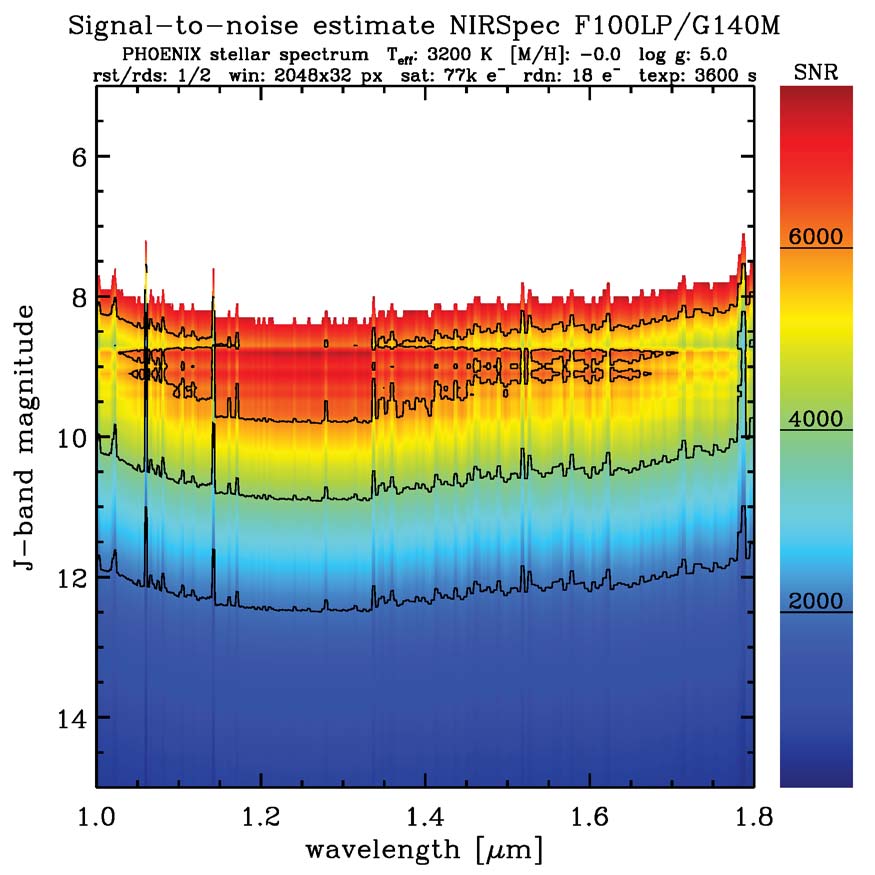} 
\caption{\small\it A plot showing the signal-to-noise ratio (detector and shot noise; no systematics) per spectral pixel for a transit spectroscopy observation at medium spectral resolution with NIRSpec (F100LP/G140M configuration) as a function of wavelength and of the host star J-band magnitude. The computation has been performed for a star with an effective temperature of 3200~K (M3-M4 typically) and an observation duration of 1 hour only.  \href{http://nexsci.caltech.edu/committees/JWST/20140311-NIRSpec-transit-spectroscopy\_v4.pdf}{(Ferruit and Birkmann (2014), this workshop)}. \label{fig:NIRSpecSN}} 
\end{center} 
\end{figure} 

The excellent sensitivity of NIRSpec means that the so-called ``noise-floor" corresponding to an ideal observation limited only by the detector and shot noises will be low. As an example (Figure~\ref{fig:NIRSpecSN}), only one hour of transit observation is necessary to reach a noise floor of 200 parts per million (ppm) for a 9$^{th}$ magnitude host star (J-band). This means that NIRspec transit spectroscopy programs will routinely have photon-noise limited noise floors of a few tens of ppm. It is therefore extremely important to be able to eliminate or calibrate out sources of systematic noise that could prevent us from approaching this theoretical limit.

\begin{table}[t!]
\centering
\small
\begin{tabular*}{\textwidth}{|l|l|l|}
\hline
Noise contribution & Description & Comment \\
\hline
\begin{minipage}[t]{0.2\textwidth}
Detector and shot noises
\end{minipage} &
\begin{minipage}[t]{0.4\textwidth}
These contribution are included in the baseline simulations and used to estimate the "noise floor".\\
\end{minipage} & 
\begin{minipage}[t]{0.3\textwidth}
Used as a benchmark.
\end{minipage} \\
\hline

\begin{minipage}[t]{0.2\textwidth}
Variable aperture losses
\end{minipage} &
\begin{minipage}[t]{0.4\textwidth}
The level of aperture losses can change when the source drifts during a transit observation.\\
\end{minipage} & 
\begin{minipage}[t]{0.3\textwidth}
\textbf{Raw} contribution of typically 40 ppm (B. Dorner, PhD thesis, 2012).
\end{minipage} \\
\hline

\begin{minipage}[t]{0.2\textwidth}
Intra-pixel sensitivity changes
\end{minipage} &
\begin{minipage}[t]{0.4\textwidth}
The response of the detectors can change when the source drifts during a transit observation.\\
\end{minipage} & 
\begin{minipage}[t]{0.3\textwidth}
\textbf{Raw} contribution of up to 400 ppm. Major contributor because of the poor PSF sampling in NIRSpec.\\
\end{minipage} \\
\hline

\begin{minipage}[t]{0.2\textwidth}
Accuracy of the flat-field correction
\end{minipage} &
\begin{minipage}[t]{0.4\textwidth}
Another source of detector response change when the source drifts during a transit observation.\\
\end{minipage} & 
\begin{minipage}[t]{0.3\textwidth}
Raw impact not yet assessed.
\end{minipage} \\
\hline

\begin{minipage}[t]{0.2\textwidth}
PSF variations
\end{minipage} &
\begin{minipage}[t]{0.4\textwidth}
The signal amplitude will change as the footprint of the PSF on the detectors change.\\
\end{minipage} & 
\begin{minipage}[t]{0.3\textwidth}
Raw impact not yet assessed.
\end{minipage} \\
\hline

\begin{minipage}[t]{0.2\textwidth}
Persistence
\end{minipage} &
\begin{minipage}[t]{0.4\textwidth}
Residual signal changing with time.\\
\end{minipage} & 
\begin{minipage}[t]{0.3\textwidth}
Observed in {\it HST}/WFC3 data (e.g. Berta et al. 2012).\\
\end{minipage} \\
\hline

\end{tabular*}
\caption{List of various noise contributions for the NIRSpec transit spectroscopy.
\label{tab:noiseNIRSpec} }
\end{table}

There are several sources of systematic noise that may impact NIRSpec observations of transiting planets. We have listed the major ones in Table~\ref{tab:noiseNIRSpec}. It can be seen that the \textbf{raw} contribution of one of these, intra-pixel sensitivity variations, is large compared to the targeted noise floor and will need to be removed with careful calibration. The good news is that these systematics are similar to those affecting observations with {\it HST} and {\it Spitzer}, so that the analysis techniques honed for current data will provide a good running start for approaching the noise floor for NIRSpec observations. In many cases the sources of systematic error are expected to be reduced relative to earlier observatories, e.g. a more stable orbit  relative to {\it HST} resulting in smaller focus drift drifts and greatly improved detector performance relative to {\it Spitzer} ($\S\ref{testbed}$).

Finally, there is a instrumental limitation on the duration of an uninterrupted transit or phase curve exposure with NIRSpec. A counter in the instrument hardware limits the number of integrations ($RESET-READ-RESET$) to a maximum of 65535. For the shortest integration time and baseline window size, this limits a single exposure to 12.5 hours (for the prisms) and 49.2 hr (for the gratings).

\subsection{\href{http://nexsci.caltech.edu/committees/JWST/NIRISS-JPL-March14v2.pdf}{NIRISS}\label{niriss}}

\subsubsection{Instrument Overview}

The Near Infrared Imager and Slitless Spectrograph (NIRISS) is the Canadian built instrument on the back plane of the Fine Guidance Sensor (FGS). Two filter wheels at the pupil plane accomodate three grisms (the GR150C and its orthogonal twin GR150R, and the GR700XD); a subset of NIRCam filters (F090, F115, F140, F150, F158, F200, F277) a Non Redundant Mask (NRM) and its associated filters (F380, F430 and F480). Four operating modes are possible. 1) The Aperture Masking Interferometry (AMI) mode uses the NRM mask (seven holes spanning 21 interferometric baselines) combined with the red filters (F277 and redder) to searh for companions at very close separation. 2) The Wide Field Slitless Spectroscopy (WFSS) mode uses the GR150 grisms. When coupled with one of the blue filters (F200 and bluer), they diffract in orthogonal directions and are destined to study high-z galaxies. They could also be used, for example, for brown dwarf searches in young star clusters. They are not foreseen to be used for transit spectroscopy because of their very low resolving power (R=150), their faint saturation magnitude limit ($\sim 12$) and the narrow spectral width spanned by the filter required to block higher orders. 3) The NIRCam filter subset can in principle be used for direct imaging either as a NIRCam spare or in parallel observing (not yet supported). 4) The last operating mode of NIRISS was specifically designed for transit spectroscopy. Its description follows.

\subsubsection{Suggested modes for transit work}

The Single Object Slitless Spectroscopy (SOSS) mode uses the GR700XD optics element --- a crossed-dispersed grism whose first surface is a weak cylindrical lens to project defocussed traces in orders 1 and 2, insuring a simultaneous spectral coverage between 0.6 and 2.8 microns at a resolving power of about 700. The trace width is about 20 pixels in the spatial direction and Nyquist-limited along the spectral direction with the goals of  minimizing flat fielding uncertainties and allowing bright targets to be observed without saturating. 

\subsubsection{Bright Star Limit}

\begin{figure}[t!]
\begin{center}  
\includegraphics[width=0.5\textwidth]{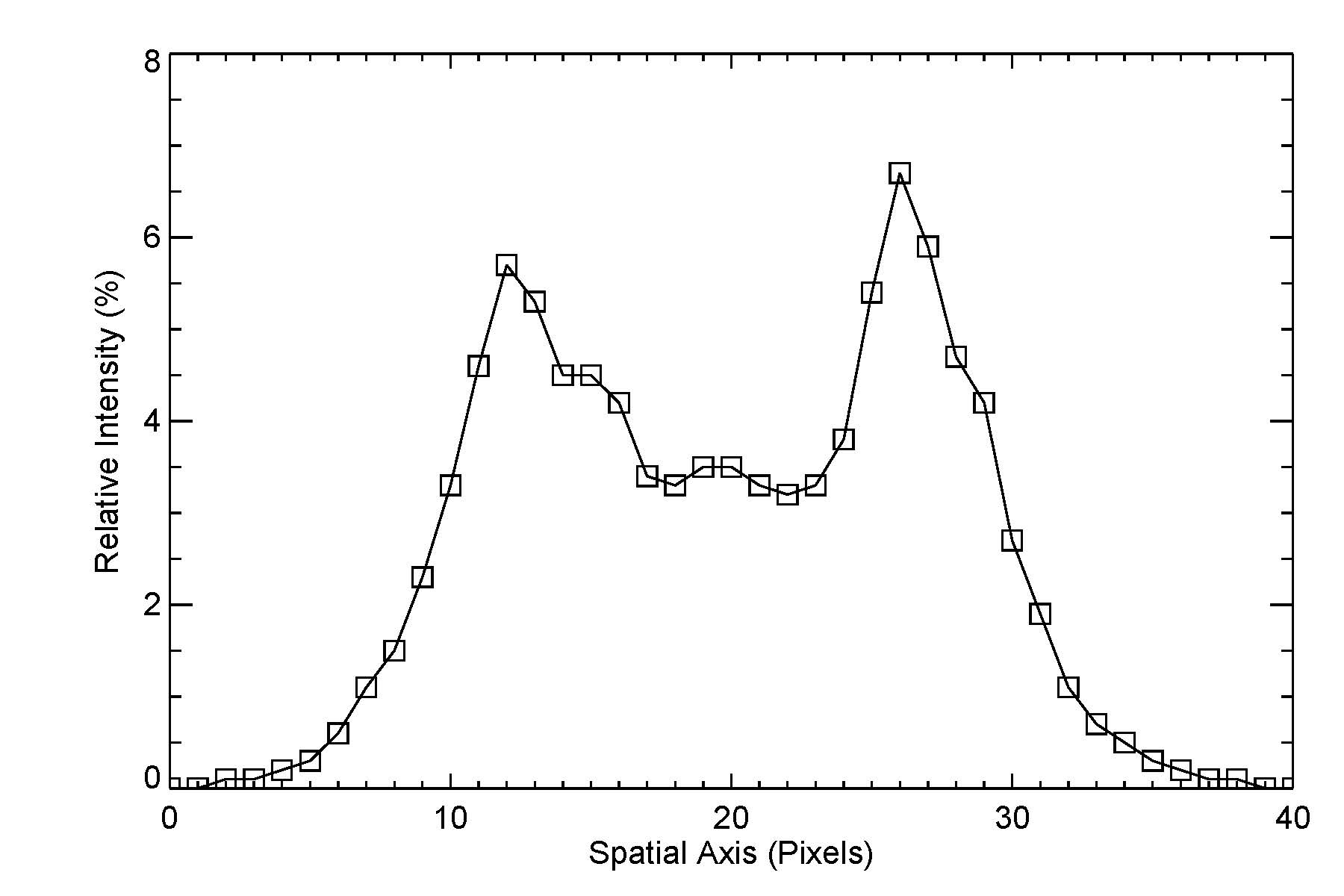}  
\caption{\small\it A cut through the spectral trace along the spatial direction near 1.5 microns as measured at cryo vacuum 1 in 2013. The trace is approximately 20 pixels wide with the brightest pixel receiving about 7\% of the monochromatic light flux. The trace width {\em does not} vary significantly across the wavelength range of 0.6-2.8 microns. \label{fig_niriss_tracewidth}}  
\end{center}  
\end{figure}

Two readout modes are currently implemented: 1) the standard readout mode reads 256x2048 pixels; 2) the bright readout mode reads 80x2048 pixels centered on the Order-1 trace between 1.1 and 2.8 microns (see figure~\ref{fig_niriss_readout}). Both modes use a 1-amplifier read out with a duty cycle/read time of 16.47 (5.49) seconds and 5.66 (1.89) seconds, respectively. The bright readout mode exists specifically to improve the saturation limit and observe brighter stars. Figure~\ref{fig_niriss_tracewidth} is a cut through the PSF along the cross-dispersed direction to show how light is spread over the $\sim20$ pixel-wide trace. Upon saturation, it is the pixels in each peak that saturate first while pixels in the core of the trace can tolerate $\sim$0.7 magnitudes of further abuse. The peak efficiency of the grism is between 1.0 and 1.5 microns in Order-1 and between 0.6 and 0.8 microns in Order-2 so that is the wavelength range where saturation occurs first. Order-2 has twice the spectral dispersion of Order-1 and also has worse throughput which means that the Order-2 trace can tolerate an additional $\approx$1.0 magnitude before saturating. Assuming that saturation occurs at a detector count 70 000 electrons/pixel, using the beginning of integration throughput curves, then the saturation limit for correlated Double Sampling (CDS - $RESET-READ-READ$) images is at $J_{Vega}=8.1$ in standard read out and $J_{Vega}=6.9$ in bright read out. Approaches to improving these limits are discussed in $\S$~\ref{efficient}

\begin{figure}[t!]
\begin{center}  
\includegraphics[width=0.8\textwidth]{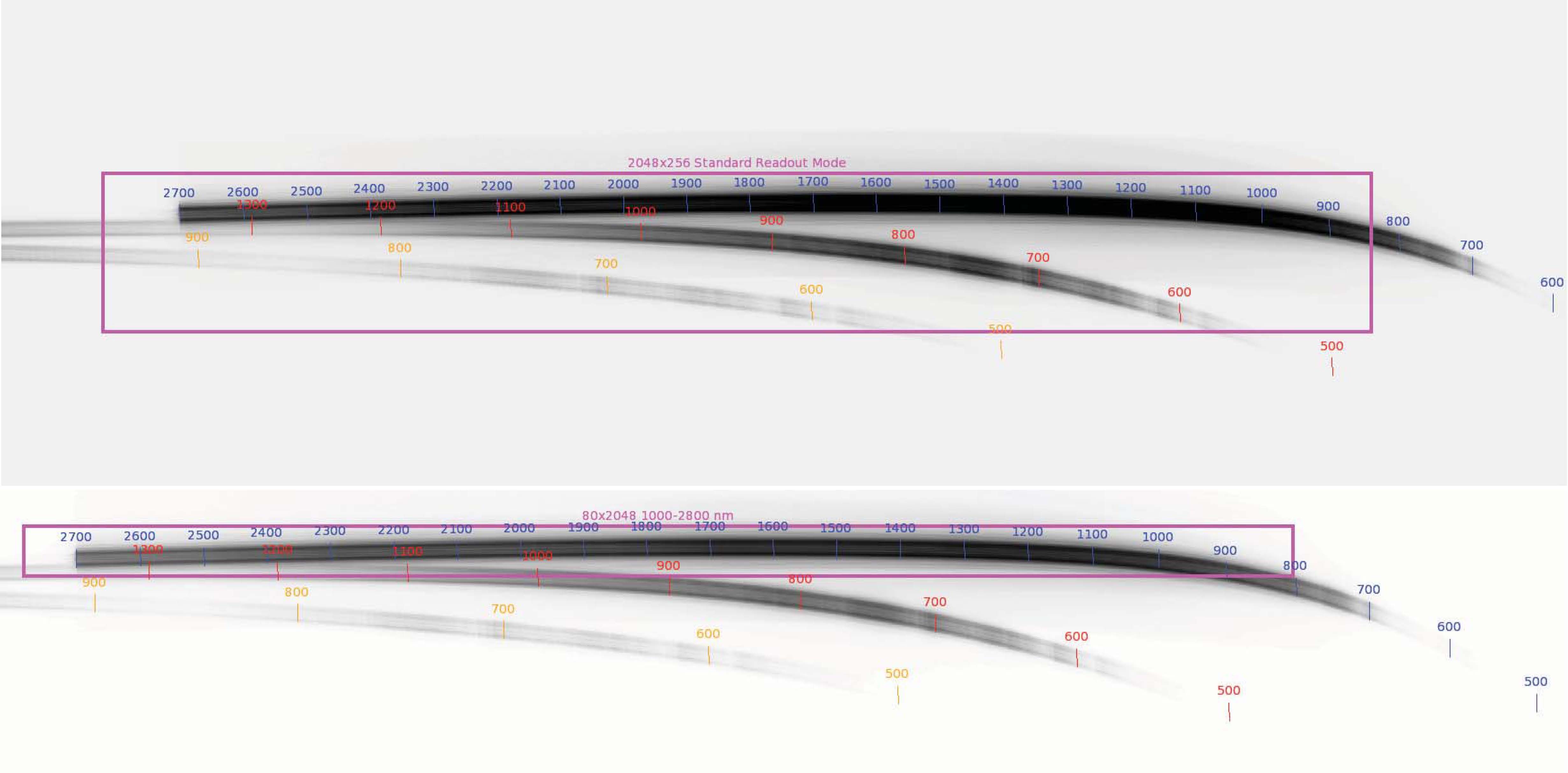}  
\caption{\small\it Simulated traces for the SOSS mode with NIRISS in Orders-1 to 3. The zeroth Order is off the detector on the right side. The magenta rectangles represent the $2048\times256$ (top - standard mode) and $2048\times80$ (bottom - bright mode) detector sub-arrays read-out for those modes. The standard mode covers from 600~nm to 2800~nm (850~nm to 2800~nm in Order-1, 600~nm to 1350~nm in Order-2) while the bright mode covers from 1000~nm to 2800~nm (in Order-1 only). The Order-3 trace has very low throughput and is unlikely to be of any use. Note that these sub-array are along the edge of the detector to make use of the reference pixels and that they are along the amplifier long-axis direction, meaing that the readout {\it can not} be multiplexed using 4 amplifiers. The abrupt cut at 2700~nm is an artifact of the simulation.
\label{niriss_simulation}}  
\end{center}  
\end{figure}

\subsubsection{Target Acquisition and Pointing Requirements}

Target acquisition for the SOSS will consist in slewing to the target, configuring the dual wheels in the NRM+F480 filter (allows acquiring targets as bright as magnitude 3 without saturating) and taking a snapshot with a small sub-array. Then the target will be positioned at a preset sweet spot on the detector to  ensure that the spectral traces fall at the desired position when the GR700XD grism is inserted. 

After target acquisition, the GR700XD grism is inserted and exposures can begin. Each exposure will consist in $N_{I}$ separate integrations (an integration starts with a detector reset) with each integration being as small as  $RESET-READ$ or as large as $RESET$+$N_{read}$ reads where $N_{read}$ can be as large as 88. We expect to operate at small $N_{read}$ for bright targets. The maximum number of integrations allowed is 65536. As an example, in standard mode, the duty cycle for CDS images is 16.47~sec so an exposure can last up to 300 hours. In bright mode, this duration is $\sim 100$ hours. 

\subsubsection{Operational Limitations \& Efficiency}

Wheel repeatability will be the main limitation on the spectral traces position. The encoder precision is equivalent to $0.15^{\circ}$ on the wheel which results in a rotation of $\sim$4 pixels at each end of the Order-1 trace. Contamination by the Order-2 trace increases from  less than 10$^{-3}$ shorter than 2.5 $\mu$m to $\sim10^{-2}$ at 2.8 $\mu$m. Otherwise,  instrumental scattering was modelled to be at a level smaller than 10$^{-3}$ and is a smooth background contamination. Because the SOSS mode is slitless, another limitation is  trace contamination by other stars in the field. For a given target, contamination by other stars will remain fixed and only change with  field rotation as the spacecraft orbits the Sun. So, in a few instances, it may be necessary to schedule the observation to minimize potential contaminants.  

\begin{figure}[t!]
\begin{center}  
\includegraphics[width=.5\textwidth]{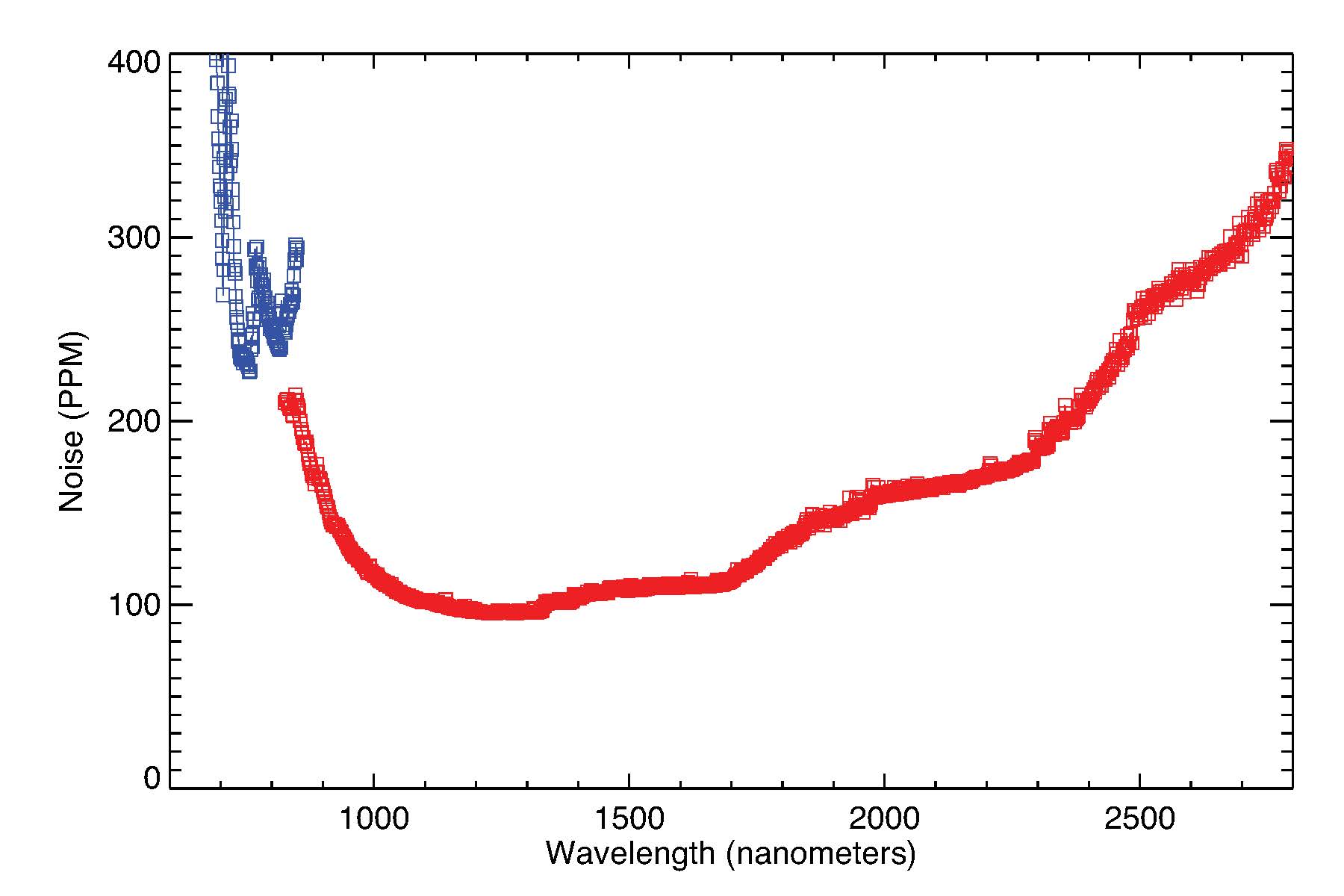}  \\
\caption{\small\it 
Typical curve of the noise level achieved with NIRISS in its full resolving power on a J=10 star for 6 hours of observation.
\label{fig_niriss_snr}}  
\end{center}  
\end{figure} 



\subsubsection{Data Simulations}

Our simulations were developped for exoplanet transit spectroscopy and secondary eclipse spectroscopy. They are photon-noise limited simulations but include an arbitrary noise floor that we currently set at 20 ppm per bin. They are still 1-D simulations, i.e. the flux is not projected onto a 2-D detector. The instrument throughput is that measured or expected at the beginning of integration (BOI) and includes the optical telescope elements (OTE). No conservative fudge factor is applied to the throughput. The grism in NIRISS will be swapped in the fall of 2014 for one that has a $2\times \,(4\times)$ better throughput in order 1(2). We use this new grism for our simulations. The blaze function used for the new GR700XD grism is that measured in the lab before AR coatings were applied but scaled to account for the presence of an AR coating when the grism is swapped in NIRISS. The main efficiency loss comes from the readout overhead incurred (50\% for a reset+read, 33\% for a reset+read+read). In addition, an overhead of 10 minutes is assumed for a telescope slew and 20 minutes for the detector to stabilize for a total of 30 minutes wall clock time loss per visit. Figure~\ref{fig_niriss_snr} presents a noise vs. wavelength curve typical of NIRISS, here a J=8.0 star (T=3200~K) observed for 6 hours with a noise floor of 20 ppm.

At the time of this writing, the effect of the intra pixel response (IPR) on the achieved accuracy has not been modelled. NIRISS was designed such that the monochromatic PSF is tilted by $\sim$2 degrees (see figure~\ref{fig_niriss_tracewidth}), with a spectral trace width of about 20 pixels, that ensures that a spectral line be sampled at about 20 intra-pixel positions and across a full pixel.

\begin{figure}[h!]
\begin{center}  
\begin{tabular}{cc}
\includegraphics[width=0.45\textwidth]{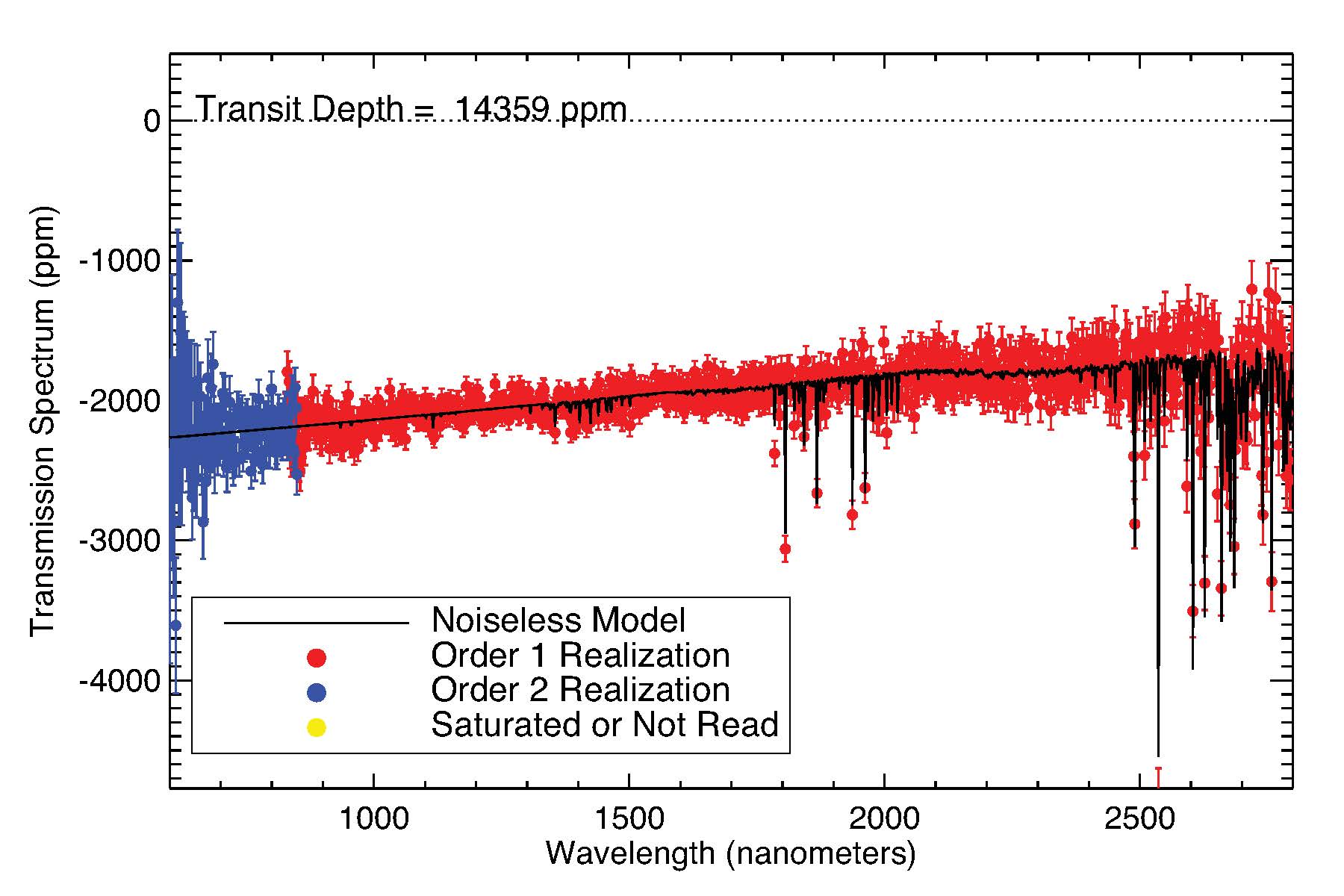}  &\includegraphics[width=0.45\textwidth]{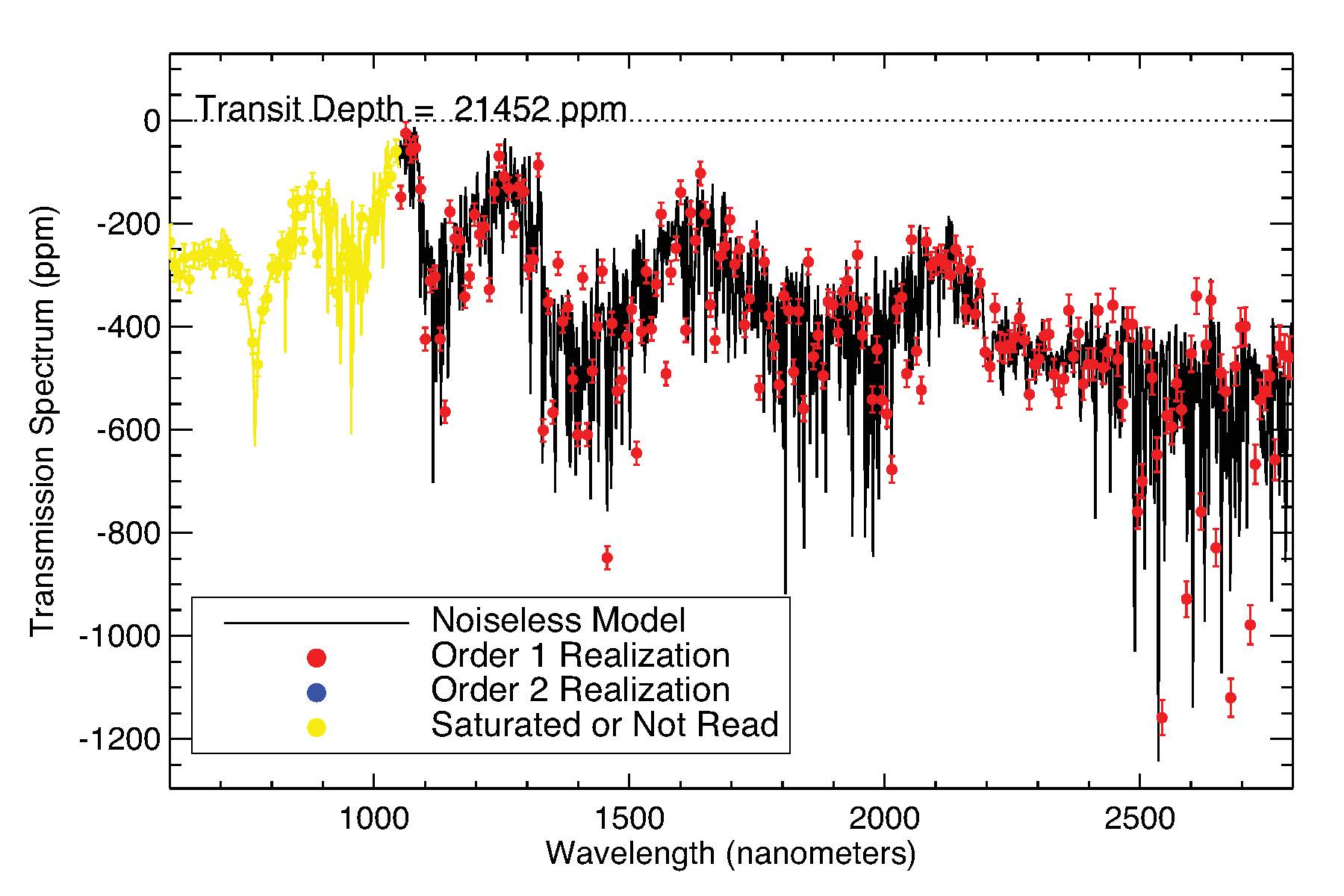}  \\
\includegraphics[width=0.45\textwidth]{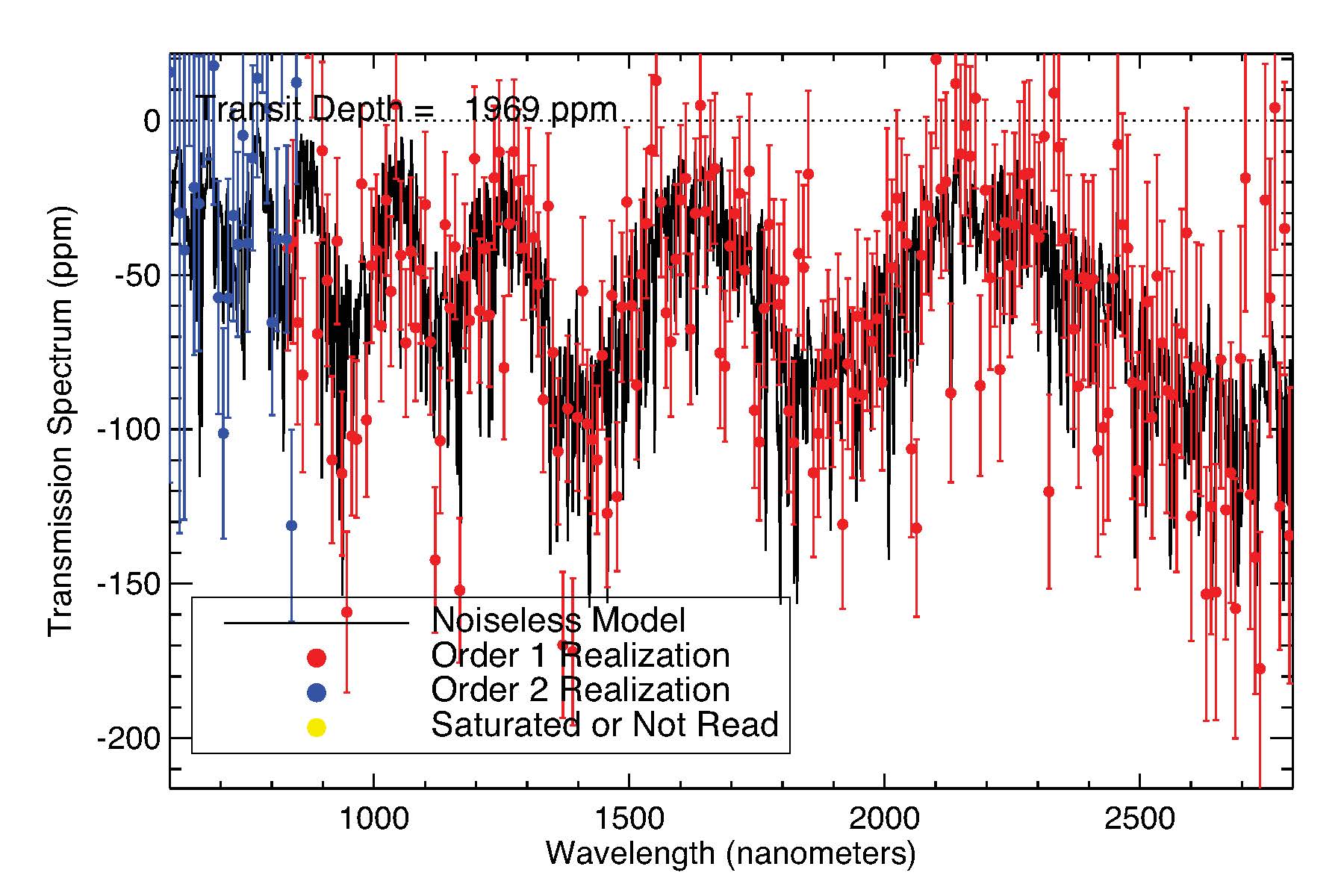}  &\includegraphics[width=0.45\textwidth]{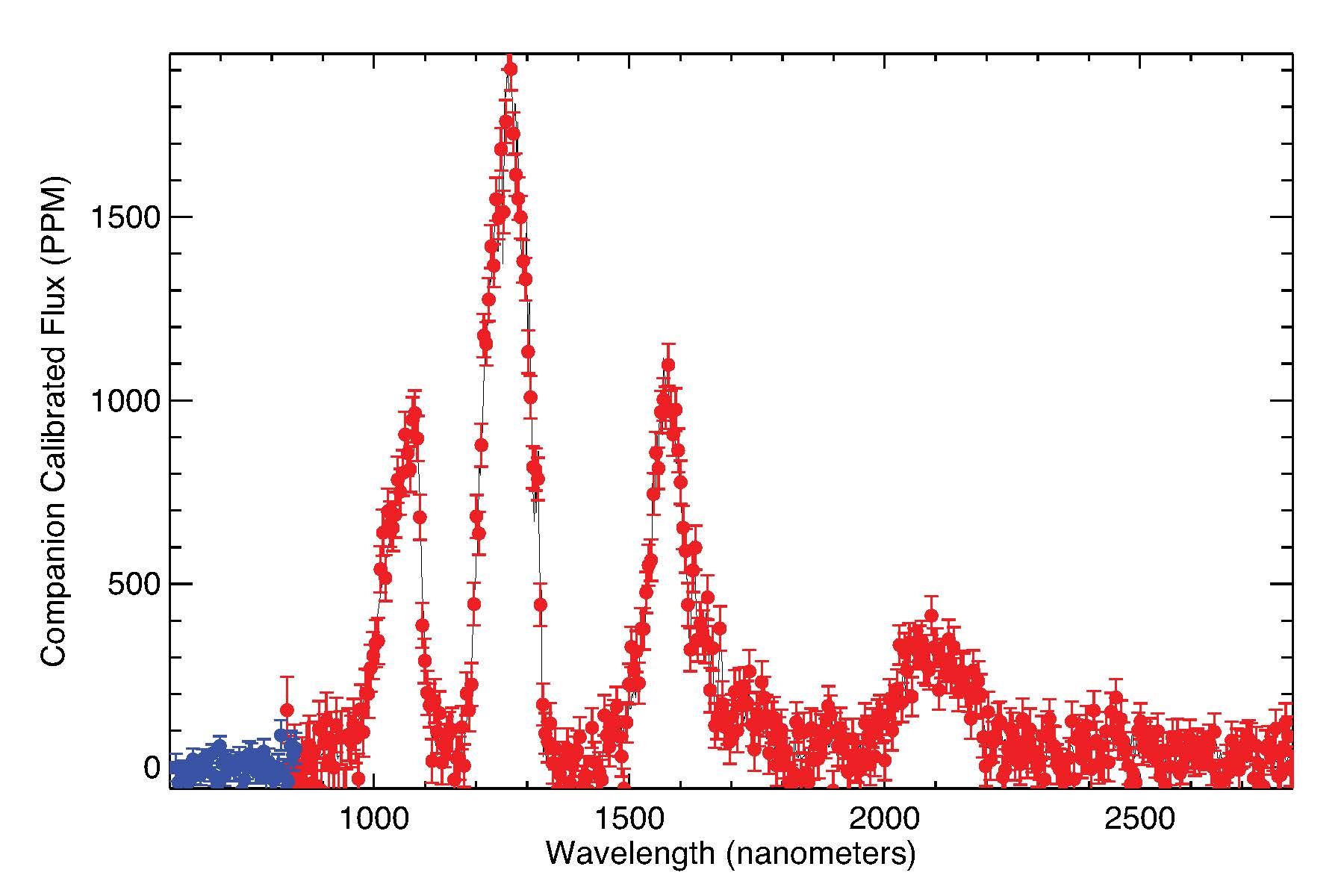}  \\
\end{tabular}
\caption{\small\it top,left) Simulation of GJ~1214b observed with NIRISS for a 50$\times$ solar atmosphere with haze (model from J. Fortney). This simulation assumes 12 hours of clock time spent over 4 transits. At the native NIRISS resolution, individual lines penetrate the haze despite the very flat spectrum up to 1.7$\mu$m. top,right) Simulation of HD189733b observed with NIRISS. This target requires 6.5 hours of clock time if a slew+detector stabilization overhead of 30 minutes is assumed and twice the amount of time is spent out of transit as in-transit. To prevent saturation, a reset+read mode is used rather than a reset+read+read mode. bottom, left) Simulation of an Earth-size water world planet with half the Earth's density observed with NIRISS. We stack 5 transits and assume J=8.0 for a late M (3200~K) star of 0.2 Sun radii orbiting in the habitable zone. This target requires 32 hours of clock time if a slew+detector stabilization overhead of 30 minutes is assumed and twice the amount of time is spent out of transit as in-transit. bottom,right) Secondary eclipse simulation of the brown dwarf LHS~6343~C observed with NIRISS. A single secondary eclipse (5 hours of clock time) delivers a high quality spectrum allowing precise spectral typing to determine the brown dwarf temperature and thus constrain observationally the radius, mass, temperature and luminosity. \label{fig_niriss_sims}}  
\end{center}  
\end{figure}

Figure~\ref{fig_niriss_sims}a presents a transmission spectrum simulation of what we would expect with NIRISS for GJ~1214b, a super-Earth around a star of magnitude J=9.75. The simulation assumes 12 hours of clock time spread over 4 transits, a 20 ppm noise floor. Here, a model (Fortney, private communication) at 50$\times$ solar metallicity and incorporating photochemical hazes dominated by $0.01~\mu$m-sized particles is used. This is a model probably not excluded by the Kreidberg et al. (2014) observations (in the spectral range between 1.1 and 1.7~$\mu$m, the atmosphere signature is almost flat). With NIRISS at the native resolution, individual lines penetrate through the haze. That simulation is of equivalent signal-to-noise ratio as the Kreidberg (2014) observations when binned to the same resolving power. Figure~\ref{fig_niriss_sims}b is a simulation of a hot Jupiter, HD189733b for one transit at its actual magnitude of J=6.1. Part of the spectrum is missing ($\lambda \leq 1.05\mu$m) because of the Bright readout mode used to prevent saturation. The third simulation (Figure~\ref{fig_niriss_sims}c) is that of an Earth-size planet with {\em half} its density harbouring a water vapor atmosphere, orbiting a late M star of 3200~K and 0.2 solar radii in its habitable zone. The brightest M star  {\it TESS} will find harbouring an Earth-size transiting planet will likely be $J=7-8$. We set the magnitude to $J=8.0$ where the standard 256$\times$2048 readout sub-array can be used. Using 5 transits at a resolution binned to R=70, our simulation shows that the main $\sim100$ ppm features can be detected. For an Earth twin (i.e. with the same density), atmospheric features skrink by a factor of 2 for the same noise level such that detection is at the $\sim3-\sigma$ noise threshold. Finally, Figure~\ref{fig_niriss_sims}d gives an example of {\em Secondary eclipse} spectroscopy of LHS~6343~C, an old brown dwarf orbiting an M3 star in the  {\it Kepler}  field. NIRISS will easily allow precise spectral typing of this unique brown dwarf for which 4 physical parameters (radius, mass, temperature and luminosity) can be observationally determined.

What can eventually be achieved with NIRISS - be it a low-density Earth-size water world or an actual Earth-twin - is strongly dependent on the noise floor that the instrument will reach. If NIRISS hits 20 ppm (30 ppm is demonstrated with HST) then planets with $\sim75-100$~ppm atmospheric features will be within reach. An exact Earth-twin would require achieveing a floor of better than $\sim5-10$~ppm to detect features at the $3-\sigma$ level. Work on a NIRISS Optical Simulator (NOS) to better understand the instrumental limitations is starting in the physics laboratory of the Universit\'e de Montr\'eal.

\subsection{\href{http://nexsci.caltech.edu/committees/JWST/Greene\_NIRCam\_transits\_2014March.pdf}{NIRCam}}

%

\begin{deluxetable}{lcccl}
\tablecaption{NIRCam Subarray Modes for Transits}
\tabletypesize{\small}
\tablehead{Size	&	FOV ($\arcsec$)	&	FOV ($\arcsec$)	&	Frame	&		\\
        (pixels)	&	Short Wave	&	Long Wave	& Time	&	Comment	}
\startdata
  64$\times$64	&	2.05	&	4.16	&	49.4ms	& in-focus photometry	\\
  160$\times$160	&  5.12	&	10.4	&	277ms	&	defocused SW photometry	\\
  400$\times$400    & 12.8  &   26.0    &  1.652s &   defocused SW + long events \\
  2048$\times$32	&	N/A	&  133$\times$2.1&	173ms	&	(TBC) LW grism spectroscopy \\
  2048$\times$320	&	N/A	&  133$\times$10.2&	1.682s&	(TBC) LW grism + defocused SW \\
  2048$\times$2048  &  65.3 &  133      & 10.74s & Faint host stars; long events \\

\enddata
\label{table:NIRCamSubarray} 
\end{deluxetable}

NIRCam consists of 2 identical modules with adjacent fields of view that each cover approximately $2\farcm2' \times 2\farcm2$. Each module views the same field through both short wave (SW) and long wave (LW) cameras that are fed via a dichroic beamsplitter with a transition wavelength of 2.4 $\mu$m. The SW cameras are Nyquist sampled at 2.0 $\mu$m and the LW ones are Nyquist sampled at 4.0 $\mu$m.

NIRCam offers narrow (R $=$ 100), medium (R $=$ 10), wide (R $=$ 4) and double wide (R $=$ 2) filters across the $0.7 - 5 \mu$m wavelength range. It also features grisms that provide R $\sim$ 1700 spectroscopy over the LW $2.4 - 5 \mu$m wavelength range with 2 orthogonal dispersion directions per module. These grisms are operated slitlessly, so they are immune to slit losses and any resultant signal modulation during transit observations. The grisms must also be used in series with a filter which selects a portion of the LW spectral region for observation.

NIRCam has a variety of subarray modes to enable observations of bright sources (Table~\ref{table:NIRCamSubarray}). Table~\ref{tab:brightlimits} gives the L band ($\sim 3.8\, \mu$m) magnitudes for the brightest sources that can be observed with either NIRCam's the grism or a 32$\times$32 subarray in the filters listed. These saturation limits assume operation up to 80\% of full well, or approximately 60,000 $e^-$. While subarrays are read out using a single amplifier, the H2RGs can also be configured to read out ``stripes'' of $Nrows\times2048$ using all four readout amplifiers for still faster frame times.

Three NIRCam modes are likely to be most used for transit, eclipse, and phase curve observations. In-focus imaging will allow photometric measurements of planets with host stars as bright as K $\sim 6 -8$ mag (Vega system) with a small (32 $\times$ 32 pixel) detector subarray (Table~\ref{tab:brightlimits}). It is possible to image stars K $\sim$ 6 mag (Vega) or even brighter when using the +8 wave weak lens (WLP8) in series with a SW filter. The WLP8 lens spreads the light of a star over $\sim$ 130 pixels diameter, reducing the maximum flux per pixel as well as reducing total photometric fluctuations due to intrapixel response and flat field variations.

Long wavelength (LW) spectroscopy will be conducted using a grism with either the F322W2 (2.4 - 4.0 $\mu$m) or F444W (3.9 - 5.0 $\mu$m) filter in series; these 2 combinations span the complete NIRCam LW wavelength range. K $\simeq$ 4 mag or brighter stars can be observed with the grism when using a 2048 $\times$ 64 detector subarray in striped mode using all 4 output amplifiers.

NIRCam SW and LW arms can observe the same object simultaneously by virtue of the dichroic beamsplitter that divides the light within each module. It is worth noting that current operations plans call for using identical subarrays and identical integration times in the active SW and LW detectors in each module. This should not be a significant issue when observing in basic imaging modes, but there may be some dynamic range issues when imaging in the complementary arm when imaging with the WLP8 lens in the SW side or obtaining grism spectra in the LW side. These issues can be mitigated by selecting modes (filters, weak lens, and/or grism) that produce similar photon fluxes in each wavelength arm. This versatility allows simultaneous SW and LW photometry or SW photometry plus LW spectroscopy. In addition to providing scientific data at 2 wavelengths, this always provides at least one pointing reference. For example, any photometric variations seen in SW images or simultaneous grism spectroscopy could be decorrelated using any measured SW source motion.

The relatively good spatial sampling of the NIRCam optics and the good noise performance of its detectors (similar for NIRISS and NIRSpec as well) will help NIRCam achieve high spectrophotometric precision. The precision of its SW photometry of bright stars can be improved by using the WLP8 lens. The precision of the grism observations may be improved by binning in the dispersion direction, also reducing the effective resolving power below their native R $\sim$ 1700 (dispersion $\simeq$ 1000 pixels per $\mu$m). Thus we expect that NIRCam will achieve spectrophotometric precision similar to that of {\it HST} WFC3 G141 data,$\sim 35$ ppm \citep{2014Natur.505...66K} even though  {\it JWST}  lacks the fast scanning mode that has been implemented for the {\it HST} observations.

After the discussion of instrument constraints on transit observations at the workshop, the NIRCam team discovered a software limitation to the duration of a single exposure of just 9.1 hours. Once identified, this limit is now being increased to a value of at least 48 hr.

\subsection{\href{http://nexsci.caltech.edu/committees/JWST/Greene\_MIRI\_transits\_2014March.pdf}{MIRI}\label{sec:mirimodes}}

MIRI provides modes for imaging, coronagraphic imaging, low resolution (R
$\sim$ 100), and integral field moderate resolution (R $\sim$ 2000) spectroscopy using three Si:As 1024 $\times$ 1024 pixel detector arrays. The MIRI imaging module (MIRIM) optics provide a $74\arcsec \times 113\arcsec$ field of view  with a scale of 0\farcs11 per pixel. The MIRIM has 10 selectable filters with central wavelengths 5.6 -- 25.5$\mu$m and bandwidths of 0.7 -- 4.6 $\mu$m \citep{bouchet_et_al_2015}. The low resolution spectrograph (LRS) provides slit or slitless R $\sim$ 100 spectra from 5 to $\sim$12 $\mu$m via a double (compound) prism in the MIRIM filter wheel \citep{kendrew_et_al_2015}. MIRIM also provides 3 four quadrant phase
mask coronagraphic fields spanning central wavelengths from 10.65 -- 15.5 $\mu$m and a Lyot coronagraph that operates at 23 $\mu$m \citep{boccaletti_et_al_2015}. A single 1024$\times$ 1024 pixel Si:As IBC detector array collects photons from all of the MIRIM modes, and two similar detectors are used in the MIRI medium resolution spectrograph (MRS) \citep[see][]{rieke_et_al_2015}. The MRS fields are as large as $7\farcs7 \times 7\farcs7$, and they are located adjacent to the MIRM field. The MRS operates at R = 1300 -- 3700, and its two detector arrays see approximately one third of its $\lambda$ = 5 -- 28 $\mu$m spectral coverage at any given time \citep{wells_et_al_2015}. MIRIM imaging and slitless LRS photon conversion efficiencies are on the order of 0.3, and detailed MIRI sensitivity calculations are presented in \citet{glasse_et_al_2015}.

 Three MIRI modes will likely be most useful for transit, eclipse, and phase curve science: MIRIM photometry, low-resolution slitless spectroscopy with the LRS, and medium-resolution integral-field spectroscopy with the MRS. Spectra obtained in MRS mode are interleaved, meaning that 3 separate observations are required to obtain a continuous spectrum over its $5 - 28\, \mu$m range (Fig.~\ref{fig:JWSTmodes}). Given the complexities of diffraction, potential slicer and slit losses, detector undersampling \&\ fringing, and considering the difficulties experienced in past transit observations with integral field instruments \citep[e.g.,][]{angerhausen_et_al_2006}, the MRS mode is unlikely to be useful for obtaining exoplanet spectra over a broad wavelength range with high spectrophotometric precision. However, it may be that focused observations of narrow lines using only a single dichroic/grating setting are tractable. Furthermore, MRS is the only option for spectroscopy beyond $\sim$12\,\micron.

\begin{deluxetable}{lcccl}
\tablecaption{Properties of MIRI Subarray for Transits}
\tablehead{Subarray	&Subarray Size&FAST Frame Time	&	Max Flux 	&	Max Flux	\\
Size & Columns$\times$Rows & (sec) &F560W [mJy] & F2550W [mJy] }
\startdata
FULL	&	1032x1024	&	2.775	&	17	&	360	\\
BRIGHTSKY$^*$	&	968x512	&	1.326	&	34	&	780	\\
SUB256$^*$	&	668x256	&	0.507	&	90	&	2150	\\
SUB128	&	136x128	&	0.119	&	370	&	8400	\\
SUB64	&	72x64	&	0.085	&	520	&	12000	\\
SLITLESSPRISM&72x416&0.159	&\multicolumn{2}{l}{3000 using P750L at 7.5 $\mu$m}		\\
MASK1065&288x224&0.24	&	\multicolumn{2}{l}{3000 using P750L at 7.5 $\mu$m}	\\
\enddata
\label{table:MIRISaturation} 
\tablecomments{$^*$Only available if Burst mode provides acceptable operation}
\end{deluxetable} 

MIRI has a variety of sub-array modes that enable a bright sources to be observed in a variety of instrument configurations (Table~\ref{table:MIRISaturation}) with saturation magnitudes given in Table~\ref{tab:brightlimits}. The bright limits for imaging, slitless LRS, and MRS modes are in the $K$ = 4 -- 6 mag range for late-type host stars. These values are calculated for using subarrays with 2 frames per integration (RESET -- READ -- READ sequence). There is a chance that more frames will need to be acquired in each MIRI integration in order to minimize systematic detector noise, and these bright limits would be impacted if this occurs (e.g., 0.75 mag fainter if 4 frames are required per integration).

The LRS uses a 0.5'' slit to provide $R\sim70$ spectroscopy from 5--12\,\micron. However, it can also be operated in a slitless mode by positioning the source in the field of the Lyot coronagraph region with the LRS double prism selected in the MIRI filter wheel. This will prevent slit losses and their resultant time-correlated variability in time-series spectroscopy. This spatial location of the star also allows use of the SLITLESSPRISM detector subarray, shortening the minimum integration time to provide the high flux bright limit given in Table~\ref{tab:brightlimits}. Like the MIRI Imager, the LRS PSF is spatially undersampled at wavelengths $\lambda \lesssim 7$\,\micron, so there may be variations in the signal correlated with pointing motions due to intrapixel sensitivity variations (as were seen in many {\it Spitzer}/IRAC observations). This challenge can likely be addressed to ensure that MIRI exhibits a noise floor at least as good as that of {\it Spitzer} ($\leq$ 50 ppm, $\S$\ref{IRAC}) given the similarity of their Si:As detectors. Considerable work remains to be done on the planning of MIRI observations and in the development of the MIRI detector pipeline, and hopefully this work will improve the actual noise floor achieved.

Finally, MIRI provides photometry via a set of 10 non-coronagraphic filters over the $5 - 28\, \mu$m range with bandwiths in the range R = 3.5 -- 10. As for the LRS, the MIRI imager PSF is undersampled at $\lambda \lesssim 7 \mu$m, so intrapixel sensitivity variations may impact the spectrophotometric precision of data at those wavelengths. The PSF shows a characteristic ``cross''-like shape due to regularities in the array substrate, but standard aperture and/or PSF-fitting photometry should adequately deal with this. MIRI may be the best instrument for detecting the thermal emissions of cool planets due to its long wavelength coverage.

\section{Illustrative Science Programs and Cross Instrument Opportunities\label{science}}

The Workshop presenters discussed a large number of potential programs involving observations with more than one instrument, including possible early-science and/or Guaranteed Time projects. For gas giants (hot Jupiters down to sub-Neptunes), there was general agreement that early observations should focus on full wavelength coverage, with low- to medium-resolution reconnaissance spectroscopy (R=100--1000) of a modest subset of targets plus a smaller number of targets to be studied at higher spectral resolution (R$>$1000). Illustrative targets (including a nearby  super-Earth that might be discovred by TESS) and observing setups are described in detail in Appendix 2. \ref{tab:stars} and \ref{tab:program} give the SNR  at  representative wavelengths and resolutions after a single transit or phase curve measurement. For objects like GJ1214b for which the single transit SNR will be low at some spectral resolutions, it will be necessary to increase the SNR by either degrading the spectral resolution and/or coadding multiple transits.

 {\it JWST}  generally requires $\sim$4 transits or occultations to assemble a continuous, moderate-resolution spectrum from 0.6--12\,\micron. Such a program would likely use NIRISS (0.6--2.5\,\micron), NIRCam (2.4--5.0\,\micron), and MIRI (5--12\,\micron) and would be limited to targets with $J> 6-8$~mag (see Table~\ref{tab:brightlimits}). Some of the best-known targets like HD~209458b, HD~189733b, and GJ~436b are right at the saturation boundary necessitating careful assessment of saturation limits at each wavelength, but many interesting targets such as GJ~3470b, WASP-43b, WASP-12b are still adequately bright for easy observation and will not present any saturation problems. For stars fainter than about J$>$11 mag it would be possible to use just two modes for full wavelength coverage NIRSpec/CLEAR/PRISM and MIRI/LRS. By using its 4 amplifier, striping readout mode, NIRCam spectroscopy is possible for stars as bright as L$\sim$ 3 mag, thereby recovering some of the brightest targets (Table~\ref{tab:brightlimits}).

Once the limiting capabilities of JWST's instruments are understood after the first year of operation, there will be time in General Observer Cycles 2 and beyond to undertake more ambitious programs. JWST's large collecting area means that many large planets, e.g. gas and ice giants, can be measured with medium to high resolution spectroscopy in just a single transit, and observed with complete wavelength coverage, in just 2-4 transits, depending on the selection of instrument modes and primary vs. secondary transits, etc. A spectroscopic survey of 100-200 planets, covering masses from 0.05 to 5 M$_{Jup}$, metallicities, [Fe/H], from -0.5 to 0.5, and host masses from M5V-F5V would be possible, albeit with a major commitment of observing time. Assuming a sample of 150 planets with an average observation duration of 6 hours (2 hr for the transit plus 4 hr for stellar baseline) and each system observed in 4 different instrument modes, then a survey would require roughly 3600 hr. But by determining elemental abundances and other properties across such a broad range of stellar properties, we would obtain a dataset which would revolutionize our understanding of planet formation. Mini-Neptunes and Super-Earths ($0.015-0.05 M_{Jup}$ or $5-15M_\oplus$) could be addressed in this same way around lower mass stars. A sample of 50 suitable targets is predicted to come from  {\it TESS} ($\S$\ref{targets}) and might require an additional 1200 hr of  {\it JWST}  spectroscopy.

Programs challenging JWST's ultimate levels of precision should also be carried out, e.g. detection H$_2$O, CO$_2$, or O$_3$ via transmission or emission spectroscopy from potentially (or nearly) habitable super-Earths \citep{kaltenegger_et_al_2009,deming_et_al_2009}. The participants agreed that attempts to predict the feasibility of these observations are highly dependent on the systematic--noise floors of the instruments. Early observations of bright objects are best-suited to characterize these floors -- until then, simple extrapolations from the instrument Exposure Time Calculators should be treated with caution. If such programs are feasible it is clear that they will require a significant amount of telescope time because many transits must be observed to build up SNR.  {\it TESS} targets will be critical for this program which will require discovery of small planets orbiting bright stars.

 {\it JWST} 's enhanced sensitivity also renders feasible with one or two transits projects that on {\it Spitzer} required the stacking of many such events. For example, the first 2D map of a hot Jupiter's day side required seven secondary eclipses with {\it Spitzer}/IRAC at 8\,\micron\ \citep{majeau_et_al_2012,dewit_et_al_2012}.  {\it JWST} 's collecting area is $\sim$56$\times$ that of {\it Spitzer} for an SNR increase of 7-8$\times$. In principle MIRI photometry of HD~189733b or a comparable target could generate a dayside map with just a single secondary eclipse. If correlated noise can be sufficiently controlled,  {\it JWST}  may enable eclipse maps at shorter wavelengths (for hot planets) for cooler planets (at longer wavelengths), or perhaps even spectroscopic eclipse maps. Moreover, the smaller number of eclipses required to satisfactorily observe an object makes it more likely that time will be available to use multiple instruments and multiple modes on a single target.

\section{\href{http://nexsci.caltech.edu/committees/JWST/agenda.shtml\#community}{Engage the Community}\label{engage}}

\subsection{\href{http://nexsci.caltech.edu/committees/JWST/pipeline\_deroo.pptx}{Data Processing Challenges and Simulations}}

The reliable reduction of transit, eclipse or phase curve observations is an enormously challenging task. Transit observations involve doing exactly what one's college physics teacher said never to do, namely subtract two big numbers and look for small differences. In the case of transits, we are looking for differences as small as 10 ppm with observations spanning many hours or even days. Every aspect of the stellar host (starspots, variability, confusing effects of its spectrum), the telescope (pointing and focus drifts), the instrument and detectors (numerous bad habits as described in $\S$\ref{detectors}), and post-processing software (over or under-correcting for instrumental effects, developing appropriate measurement uncertainties) must be well characterized and validated.

JWST will provide both planning tools and data processing tools optimized for transit observations. To plan transit observations,  specific observing templates will be available for  each of the science instruments. These templates are being designed
to yield the highest-possible photometric stability by minimizing disturbances
to detector readout operations and to the various observatory systems.

Once the data have been collected, the raw data will be available in the archive and the  JWST pipeline will provide processed single-integration images. It is expected that the science community will play a key role in extracting the ultimate precision from transit data, as it has for data acquired using Hubble and Spitzer. The science operations center will work with the community to incorporate best practices into the pipeline over time, and to  provide any additional information that may be needed to maximize the science  return from transit observations.

To enable rapid and reliable reduction of transit observations, observers must have access to a broad range of  {\it JWST}  data to be able to track the state of the instrument over long periods and to be able to place each observation in context. For example, a large amount of data and detailed analysis was required to characterize {\it HST}/WFC3 instrumental effects and to construct the {\it Spitzer}/IRAC pixel phase maps. It will be essential to be able to download large amounts of  {\it JWST}  data, something for which existing interfaces to mission archives are often not optimized, e.g. extract all NIRSpec observations of stars of some magnitude and spectral type or to track the evolution of the flat-field response. 

The community should be encouraged, perhaps in a coordinated way through the STScI, to participate in end-to-end simulations to quantify the impact of instrument, pipeline and analysis tools on retrieved exoplanet characteristics. Once  {\it JWST}  is in orbit, null-test observations should be carried out very early on for each instrument/mode and made available to the community for detailed analysis. e.g., an exoplanet transit ``null-test" over 8 hours or a phase-curve ``null-test" over 4 days would be of great value in developing post-processing tools.

In developing observational sequences and pipeline tools optimized for transits, STScI should incorporate ``lessons-learned" from current instruments, e.g. the ramp effect, detector noise, inter- and intra-pixel sensitivity variations, and be ready to calibrate and validate reduction and post-processing tools. It will be important to develop a consensus on the best methods to remove these effects. Ground testing can help identify specific detector or instrument issues early. By engaging the community in activities before and immediately after launch, the community will be ready to submit well-founded proposals and to optimize the early return of exciting science from  {\it JWST} .

Once reliable data are in hand, a second level of challenges awaits, namely the retrieval of physical parameters such as abundances of various atomic and molecular species, vertical and horizontal temperature structure in the atmosphere. As noted by \citet{Hansen2014}, a proper treatment of the observational uncertainties in {\it Spitzer} photometry make it impossible to distinguish in a Bayesian sense between a complex atmospheric model and a simpler blackbody model.  {\it JWST} 's higher SNR photometry or, better yet,  multi-wavelength spectroscopic data should greatly improve the interpretative power of the models, but the caution remains. A community effort to agree on a set of test observations to analyze and compare the derived parameters for consistency and accuracy would give the community confidence that reliable results can be obtained from  {\it JWST} 's observations.

\subsection{\href{http://nexsci.caltech.edu/committees/JWST/lee\_jwst\_transit\_2014\_v3.pptx}{Obtaining Transit Observations Soon After Launch}}

As noted in the previous section, workshop participants (both instrument team members and potential General Observers) felt strongly that early access to  {\it JWST}  data would be very important for learning how to work with transit observations.  Analysis of this early data would inform modifications to data acquisition processes, the development of pipelines and post-processing tools, and improve the quality of proposals for Cycle 2 and beyond. 

Such initial transit observations may potentially come from Early Release Observations (ERO), Science Verification (SV) datasets, and/or the Early Release Science Program (ERS).  However, the objectives of each of these three classes of observation are distinct:

\begin{itemize}
  \item {\it Early Release Observations}. As  defined by NASA-SMD policy,  ``Early Release Observations (EROs) will be taken by  {\it JWST}  during both the commissioning and post-commissioning phases of operation. These observations will be chosen to have wide public appeal and are designed to demonstrate the capabilities of the  {\it JWST}  instruments. Publication or reporting in any form of results of these observations is embargoed until the EROs are released. The ERO data become publicly available at that time'' (NASA-SMD Policies and Guidelines for the Operations of the James Webb Space Telescope at the Space Telescope Science Institute, Policy 4).  Thus, as noted by the  {\it JWST}  Advisory Committee (JSTAC\footnote{http://www.stsci.edu/jwst/advisory-committee}), the ERO program ``is a public media activity to demonstrate early mission success. The ERO data will be useful for science, but their primary purpose is a media demonstration that the mission is operational and on track to begin its science program" (JSTAC, March 2014).
  \item {\it Science Verification (SV)} datasets will be taken during the six month commissioning period after launch and are designed to confirm the basic capabilities of the science instruments in the context of observatory operations. They are likely to offer limited opportunities for science observations. The SV datasets will also be embargoed from public access until the release of the EROs.
  \item {\it The Early Release Science (ERS) Program } was recommended by the JSTAC to ``to enable the community to understand the performance of  {\it JWST}  prior to the submission of the first post-launch Cycle 2 proposals that will be submitted just months after the end of commissioning. To meet this goal, science data need to be released as soon as commissioning activities allow. The data would complement the Early Release Observations (ERO) and the Science Verification (SV) datasets [and] should have no proprietary period. The JSTAC recommends that these data be released both in raw form and with any initial calibrations as soon as possible; the key aspect is speed" (JSTAC, June 2010).  While the size and scope of this program are still undefined, the program will be supported by Director's Discretionary Time.  Over the next two years, STScI will work with Guaranteed Time Observers and the community to plan for implementation of ERS observations, ``so that the community involvement in the selection of targets, science objectives and modes is carried out efficiently and is completed when needed before the Cycle 1 Call for Proposals"  (JSTAC, March 2014).  
 \end{itemize}

Given the scope of each of these programs, the ERS program likely provides the best chance for the community to secure the early transit observations that it needs to maximize exoplanet science with JWST.  Workshop participants were greatly in favor of such an approach, and were eager to learn more about how to help define the ERS to ensure that key transit observations are represented in its portfolio of observations.  Participants also emphasized that transit-related observations should be taken as early as possible in the mission to quickly determine instrument performance in a few key modes.

{\it The participants urge the project, the instrument teams, and STScI (along with relevant advisory bodies) to take these urgent recommendations into consideration as they plan various early observations with JWST}.

\section{Conclusions\label{conclusions}}

The  {\it JWST}  Transit Workshop drew together exoplanet observers, modelers and theoreticians, experts in instruments currently making transit observations, and experts in the  {\it JWST}  observatory, its instruments and science operations. The goals of the meeting were to provide a forum where the exoplanet community could come together with representatives of  {\it JWST}  to learn about the capabilities of  {\it JWST}  for transit science, to form ideas for ambitious, yet feasible, observational programs, and to pass on lessons-learned from previous missions. Some of the most important conclusions are listed below. More detailed information is available in individual presentations available at the Workshop website\footnote{http://nexsci.caltech.edu/committees/JWST/agenda.shtml}.

\begin{itemize}
\item All the workshop participants were excited at the prospect of bringing the dramatic new capabilities of JWST, in particular its large collecting area, stable orbit, broad wavelength coverage and spectral resolution, to bear on a wide variety of exoplanet topics.
\item A large sample of exoplanets orbiting bright stars will be available for study from space-based (Kepler, Kepler/K2, TESS, and CHEOPS) as well as ground-based surveys.
\item The validation of these planets, the characterization of their bulk properties, and the accurate determination of their orbital parameters will require a major commitment by ground-based community, supported by NASA, to obtain low and high precision radial velocity as well as adaptive optics imaging.
\item A complete science program over the lifetime of  {\it JWST}  will involve observing wide variety of planets.
\begin{itemize}
\item A survey of 100-200 gas and ice giants with a range masses (0.05-5 M$_{Jup}$) orbiting stars with a broad range of spectral types, metallicity would lead to a transformational breakthrough in our understanding of the formation and evolution of planets. Transit signals are strong enough that a single transit/instrument mode should provide adequate sensitivity in almost all cases.
\item The study of a few 10s of mini-Neptunes or super-Earths (5-15 M$_{Jup}$) would explore a species of planet not found in our own solar system in a variety of stellar environments. In many cases, observations of these smaller planets could be accomplished in a single transit/per mode for planets orbiting M stars.
\item Intensive observations of one or two terrestrial-sized planets (1-5 M$_\oplus$), preferably located in the Habitable Zone of their host stars, might only be possible with filter photometry and might require coadding many 10s of transit observations.  But such observations would offer the promise of characterizing the atmospheres of a planet much like our own.
\item Many individual planets will have properties worthy of careful study, e.g following the phase curve of a highly eccentric planet through periastron passage to explore atmospheric chemistry dynamics under drastic changes in insolation, making two dimensional tomographic maps of planetary atmospheres using high cadence, high SNR observations, etc.
\end{itemize}
\item A complete multi-wavelength  dataset for each object in either primary transit or secondary eclipse will require between 2 and 4 separate observations using various instrument modes. Each transit observation will require between 5 and 20 hours while full phase curve coverage will require between 2-4 days of uninterrupted observation.
\item Experience from {\it Kepler, Spitzer, {\rm and} HST} shows that the acquisition of transit data requires careful operational design to ensure maximum stability of every aspect of  {\it JWST}  for many hours or even days. The project is urged to work with the instrument teams and other experts to understand proposed instrument sequences in great detail and to develop on-orbit tests as early as possible in the mission. Among the lessons-learned from {\it HST}, {\it Kepler} and {\it Spitzer} are the following:
\begin{itemize}
\item Strive to reduce spacecraft interruptions to an absolute minimum during hours-long, or even days-long, transit or phase curve observations. Even small disturbances from heaters, reaction wheels, or discontinuous pointing shifts can have significant effects at the level of a $>>$ 10s of ppm.
\item Recognize the complexity of scheduling a large number of highly time constrained events.
\item Identify detector ``sweet spots" early in the mission to make it possible to characterize a small number of pixels in exquisite detail for all transit observations in a particular instrument mode. Ensure that appropriate operating modes are available to take advantage of these ``sweet spots".
\end{itemize}
\item Given the large amount of effort required to develop on-orbit data acquisition procedures, calibration techniques and post-processing tools for even a few instrument modes, STScI should encourage early identification of a few preferred modes for each instrument. As experience is gained, additional configurations could be supported as well. While this recommendation should not be construed as restricting the community from developing innovative new techniques, e.g. scanning a source with HST/WFC3,  {\it JWST}  has a limited lifetime and diluting the effort with a plethora of modes could prove detrimental.
\item Experience from {\it Kepler, Spitzer, {\rm and} HST} shows that the pipeline and post-processing of transit data require careful calibration and algorithm development. STScI is urged to work with the instrument teams and transit experts to understand the pipelines and requisite calibration parameters in great detail.
\item Prelaunch measurements using either the flight instruments themselves or specialized testbeds can provide important insights to detector and instrument performance.
\item Engage the exoplanet community in the years before launch and soon after launch with data challenges and access to a small amount of transit data taken early in the mission to enable rapid development of optimized observing practices and data reduction tools. Many of the highest precision observing sequences and tools used with {\it HST} and {\it Spitzer} data were developed by the observer community working in close collaboration with science center personnel.
\end{itemize}

JWST will provide dramatic new capabilities for improving our understanding of planetary systems, including  gas and ice giants, mini-Neptunes  and  Super-Earths, and even Earth-sized planets in the habitable zones of their host stars. Careful preparation in advance of launch by the instrument teams, the project, and STScI, coupled with a deepened awareness of JWST's capabilities within the exoplanet community, should bring about a dramatic new era of exoplanet exploration.

\section{Acknowledgements}
The Principal Investigators and members of each instrument team, as well as STScI and  {\it JWST}  Project personnel all gave generously of their time to make this workshop a success. CAB would like to thank Ms. Ellen O'Leary and other NExScI personnel for their invaluable support for all aspects of the workshop. People wishing to cite aspects of this Workshop are encouraged to cite individual presentations (Table~\ref{tab:Agenda}) as given at the ``Caltech  {\it JWST}  Transit Workshop, March 2014". Some of the research described in this publication was carried out in part at the Jet Propulsion Laboratory, California Institute of Technology, under a contract with the National Aeronautics and Space Administration. Copyright 2014 California Inst of Technology. All rights reserved.

\clearpage
\appendix
\section{Meeting Agenda and Attendees} \label{App:AppendixA}

\begin{deluxetable}{ll}
\tablenum{Appendix  1}
\tablecaption{Meeting Agenda\label{tab:Agenda}}
\tablehead{Topic& Speaker}
\tabletypesize{\scriptsize}
\startdata
\multicolumn{2}{l}{\bf DAY 1: March 11, 2014}\\
Topic&Speaker\\
{\it \href{http://nexsci.caltech.edu/committees/JWST/Goals\_for\_Transit\_Meeting2.pptx} {Goals of Meeting}}&C. Beichman\& J. Lunine\\
{\it  Key Science Opportunities} &\\
 \hspace{1 cm}\href{}{}\href{http://nexsci.caltech.edu/committees/JWST/Fortney\_JWST\_giantplanets\_2014.ppt}{Spectroscopy of Giant Planets} &J. Fortney\\
 \hspace{1 cm}\href{http://nexsci.caltech.edu/committees/JWST/JWST\_Kempton.pdf}{Spectroscopy of Super Earths}  &E. Kempton\\
 \hspace{1 cm} \href{http://nexsci.caltech.edu/committees/JWST/knutson\_jwst\_workshop1.pptx}{Atmospheric Dynamics and weather.  Full phase coverage} &H. Knutson\\
 \hspace{1 cm}\href{http://nexsci.caltech.edu/committees/JWST/Deming.pptx}{Transit Photometry} &D. Deming\\
 \hspace{1 cm} \href{http://nexsci.caltech.edu/committees/JWST/kreidberg\_JWST0314\_public.pdf}{Frontiers of Precision Exoplanet Atmosphere}&L.	 Kreidberg\\
 \hspace{2 cm} Characterization with HST&\\
{\it   Transit Best Practices} &\\
 \hspace{1 cm}\href{http://nexsci.caltech.edu/committees/JWST/JWST\_meeting\_Sing.pdf}{HST Best Performance and Best Practices}&D. Sing\\
 \hspace{1 cm}\href{http://nexsci.caltech.edu/committees/JWST/Mandell\_JWST\_Workshop.pptx}{Hot Jupiter Spectroscopy wth HST/WFC3} & A. Mandell\\
 \hspace{1 cm}\href{http://nexsci.caltech.edu/committees/JWST/keplerbestpractices.pptx}{Kepler Best Performance and Best Practices}  & J. Christiansen\\
  \hspace{1 cm}\href{http://nexsci.caltech.edu/committees/JWST/combined\_IRAC-MIPS\_talk\_updated.pptx}{Spitzer Best Performance and Best Practices}  &S. Carey \& Ian Crossfield\\
  \hspace{1 cm}\href{http://nexsci.caltech.edu/committees/JWST/slides\_pasadena\_meeting\_jwst\_2014.pdf}{Lessons from Spitzer Spectroscopy } & J. Bouwman\\
{\it  \href{http://nexsci.caltech.edu/committees/JWST/JWST\_OPS\_MC\_JAS.ppt}{JWST Operations Issues for Transits} }& M. Clampin \& J. Stansberry\\
{\it   Detector Problems and Features}&\\
  \hspace{1 cm}\href{http://nexsci.caltech.edu/committees/JWST/NIRCam\_Detectors\_transit\_mtg.pptx}{ HgCdTe/ASIC} HgCdTe Detectors& Marcia Rieke\\
  \hspace{1 cm}\href{http://nexsci.caltech.edu/committees/JWST/Ressler\_MIRI\_Detectors\_for\_exoplanets.pdf}{Silicon detectors} &M. Ressler\\
 \hspace{1 cm}\href{http://nexsci.caltech.edu/committees/JWST/JWST\_transitworkshop\_IPAC\_swain.pptx}{Challenges in Measurement Repeatability} &M. Swain\\
\multicolumn{2}{l}{\bf DAY 2: March 12, 2014}\\
{\it   Targets for JWST}&\\
 \hspace{1 cm}\href{http://nexsci.caltech.edu/committees/JWST/bean.pptx}{ Ground RV/transits}  &J. Bean\\
 \hspace{1 cm}\href{http://nexsci.caltech.edu/committees/JWST/Howell\_JWST\_Workshop.pptx}{Kepler and K2}  &S. Howell\\
 \hspace{1 cm}\href{http://nexsci.caltech.edu/committees/JWST/Ricker\_TESS\_Pasadena\_JWST\_v2.pdf}{TESS}  &G. Ricker \& D. Latham\\
 \hspace{1 cm}\href{http://nexsci.caltech.edu/committees/JWST/dressing\_jwst\_mdwarfs.pptx}{M Stars as JWST Targets} &C. Dressing\\
 \hspace{1 cm}\href{http://nexsci.caltech.edu/committees/JWST/CHEOPS\_JWST\_transits.pdf}{CHEOPS} &D. Ehrenreich\\
 \hspace{1 cm}\href{http://nexsci.caltech.edu/committees/JWST/JWST\_Gaia\_Sozzetti.pdf}{GAIA} & 	A. Sozzetti\\
  \hspace{1 cm}\href{http://nexsci.caltech.edu/committees/JWST/precursor\_JWST\_ciardi\_20140312a.pptx}{Precursor Data Needs} &D. Ciardi\\
  \hspace{1 cm}\href{http://nexsci.caltech.edu/committees/JWST/McCullough\_JWST\_transits.pptx}{ Challenge of Stellar Variability}&P. McCullough \\
{\it  Instrument modes for transits}&\\
 \hspace{1 cm}\href{http://nexsci.caltech.edu/committees/JWST/20140311-NIRSpec-transit-spectroscopy\_v4.pdf}{NIRSPEC}  &P. Ferruit \& S. Birkmann\\
 \hspace{1 cm}\href{http://nexsci.caltech.edu/committees/JWST/NIRISS-JPL-March14v2.pdf}{NIRISS}   &R. Doyon \& L. Albert\\
 \hspace{1 cm}\href{http://nexsci.caltech.edu/committees/JWST/Greene\_NIRCam\_transits\_2014March.pdf}{NIRCam}   &T. Greene  \\
  \hspace{1 cm}\href{http://nexsci.caltech.edu/committees/JWST/Greene\_MIRI\_transits\_2014March.pdf}{MIRI}  &T. Greene \& P.-O. Lagage\\
\multicolumn{2}{l}{\bf DAY 3: March 13, 2014}\\
{\it  Data Processing challenges and requirements}&\\
 \hspace{1 cm}\href{http://nexsci.caltech.edu/committees/JWST/smallest.pptx}{What is the Smallest Planetary Atmosphere} &D. Deming \\
  \hspace{2 cm}JWST Will Characterize?&\\
  \hspace{1 cm}\href{http://nexsci.caltech.edu/committees/JWST/csb\_TransitExptJWSTMtg3.pptx}{Laboratory Testbeds}  & Gautam Vasisht \& C. Beichman\\
  \hspace{1 cm}\href{http://nexsci.caltech.edu/committees/JWST/pipeline\_deroo.pptx}{Pipeline Data Processing Challenges} &P. Deroo\\
{\it Engage the Community} &\\
  \hspace{1 cm}\href{http://nexsci.caltech.edu/committees/JWST/lee\_jwst\_transit\_2014\_v3.pptx}{Science Timeline for JWST}  &J.Lee\\
  \hspace{1 cm}\href{http://nexsci.caltech.edu/committees/JWST/birkmann\_simulations\_challenges.pptx}{Data Simulations} &S. Birkmann\& J. Valenti\\
\enddata

\end{deluxetable}
 
\newpage

\begin{sidewaystable}
\tiny
\centering
\begin{tabular}{|ll|ll|}
\hline
Workshop Attendees \\
\hline
Name & Email Address & Name & Email Address\\
\hline
Loic Albert & \href{mailto:albert@astro.umontreal.ca}{albert@astro.umontreal.ca} & Eliza Kempton & \href{mailto:kemptone@grinnell.edu}{kemptone@grinnell.edu} \\
Jacob Bean & \href{mailto:jbean@oddjob.uchicago.edu}{jbean@oddjob.uchicago.edu} & Tony Keyes & \href{mailto:keyes@stsci.edu}{keyes@stsci.edu} \\
Charles Beichman & \href{mailto:chas@ipac.caltech.edu}{chas@ipac.caltech.edu} & Heather Knutson & \href{mailto:hknutson@caltech.edu}{hknutson@caltech.edu} \\
Bjoern Benneke & \href{mailto:bbenneke@caltech.edu}{bbenneke@caltech.edu} & Laura Kreidberg & \href{mailto:kreidberg@uchicago.edu}{kreidberg@uchicago.edu} \\
Stephan Birkmann & \href{mailto:Stephan.Birkmann@esa.int}{Stephan.Birkmann@esa.int} & Jessica Krick & \href{mailto:jkrick@caltech.edu}{jkrick@caltech.edu} \\
Gary Blackwood & \href{mailto:gary.h.blackwood@jpl.nasa.gov}{gary.h.blackwood@jpl.nasa.gov}& David Lafreni\`ere & \href{mailto:david@ASTRO.UMontreal.CA}{david@ASTRO.UMontreal.CA} \\
Jeroen Bouwman & \href{mailto:bouwman@mpia-hd.mpg.de}{bouwman@mpia-hd.mpg.de} & Pierre Olivier Lagage & \href{mailto:pierre-olivier.lagage@cea.fr}{pierre-olivier.lagage@cea.fr}\\
Sean Carey & \href{mailto:carey@ipac.caltech.edu}{carey@ipac.caltech.edu} & David Latham & \href{mailto:dlatham@cfa.harvard.edu}{dlatham@cfa.harvard.edu} \\
 Jessie Christiansen & \href{mailto:christia@ipac.caltech.edu}{christia@ipac.caltech.edu} &Janice Lee & \href{mailto:jlee@stsci.edu}{jlee@stsci.edu} \\
 David Ciardi & \href{mailto:ciardi@ipac.caltech.edu}{ciardi@ipac.caltech.edu} & Jonathan Lunine & \href{mailto:jlunine@astro.cornell.edu}{jlunine@astro.cornell.edu} \\
 Mark Clampin & \href{mailto:mark.clampin-1@nasa.gov}{mark.clampin-1@nasa.gov} & Avi Mandell & \href{mailto:Avi.Mandell@nasa.gov}{Avi.Mandell@nasa.gov} \\
Ian Crossfield & \href{mailto:ianc@lpl.arizona.edu}{ianc@lpl.arizona.edu} & Peter McCullough & \href{mailto:pmcc@stsci.edu}{pmcc@stsci.edu} \\
 Drake Deming & \href{mailto:lddeming@gmail.com}{lddeming@gmail.com} & Michael E. Ressler & \href{mailto:michael.e.ressler@jpl.nasa.gov}{michael.e.ressler@jpl.nasa.gov}\\
Pieter D. Deroo & \href{mailto:pieter.d.deroo@jpl.nasa.gov}{pieter.d.deroo@jpl.nasa.gov}& George Ricker & \href{mailto:grr@space.mit.edu}{grr@space.mit.edu} \\
Jean-Michel Desert & \href{mailto:desert@colorado.edu}{desert@colorado.edu} & Marcia Rieke & \href{mailto:mrieke@as.arizona.edu}{mrieke@as.arizona.edu} \\
Rene Doyon & \href{mailto:doyon@ASTRO.UMontreal.CA}{doyon@ASTRO.UMontreal.CA} & Massimo Robberto & \href{mailto:robberto@stsci.edu}{robberto@stsci.edu} \\
Courtney Dressing & \href{mailto:cdressing@cfa.harvard.edu}{cdressing@cfa.harvard.edu} & Avi Shporer & \href{mailto:shporer@gps.caltech.edu}{shporer@gps.caltech.edu} \\
Pierre Ferruit & \href{mailto:pferruit@rssd.esa.int}{pferruit@rssd.esa.int} & David Sing & \href{mailto:sing@astro.ex.ac.uk}{sing@astro.ex.ac.uk} \\
Jonathan Fortney & \href{mailto:jfortney@ucolick.org}{jfortney@ucolick.org} & Roger Smith & \href{mailto:rsmith@astro.caltech.edu}{rsmith@astro.caltech.edu} \\
Paul Goudfrooij & \href{mailto:goudfroo@stsci.edu}{goudfroo@stsci.edu} & Alessandro Sozzetti & \href{mailto:sozzetti@oato.inaf.it}{sozzetti@oato.inaf.it} \\
Tom Greene & \href{mailto:tom.greene@nasa.gov}{tom.greene@nasa.gov} & John Stansberry & \href{mailto:jstans@stsci.edu}{jstans@stsci.edu} \\
Matthew Greenhouse & \href{mailto:matthew.a.greenhouse@nasa.gov}{matthew.a.greenhouse@nasa.gov} & Mark Swain & \href{mailto:mark.r.swain@jpl.nasa.gov}{mark.r.swain@jpl.nasa.gov} \\
Dean Hines & \href{mailto:hines@stsci.edu}{hines@stsci.edu} & Jeff Valenti & \href{mailto:valenti@stsci.edu}{valenti@stsci.edu} \\
Steve Howell & \href{mailto:steve.b.howell@nasa.gov}{steve.b.howell@nasa.gov} & Gautam Vasisht & \href{mailto:gv@s383.jpl.nasa.gov}{gv@s383.jpl.nasa.gov} \\
\hline
\end{tabular}
\tablenum{Appendix  2}
\caption[Appendix:  2]{Workshop Attendees}
\label{tab:Attendees} 
\end{sidewaystable}

\clearpage

\section{Detector Operations} \label{App:AppendixC}

The  {\it JWST}  standard for specifying and acquiring science exposures is called ''MULTIACCUM''. MULTIACCUM exposures consist of one or more sample-up-the-ramp \emph{integrations}, each integration is composed of one or more \emph{groups}, while each group is made up of one or more \emph{frames} (i.e. non-destructive reads of the signal on each pixel). In order to limit data rates, the frames from within a group can be co-added before being stored on the solid-state recorder (SSR). The overall science exposure scheme is illustrated for a simple example in Figure~\ref{fig:multiaccum}.  Terms and definitions are given in Appendix~\ref{App:AppendixC}, Table~\ref{tab:multiaccum}.

Exposure commands are issued by the on-board scripts subsystem (OSS), are received and translated by the flight software (FSW) for the specified science instrument (SI). For the three instruments (NIRCam, NIRISS and NIRSpec) using the 2k $\times$ 2k HgCdTe  (H2RG) detectors, the exposure parameters are then sent to ASIICs (application specific integrated circuits) that are co-located with the detectors. Each ASIC controls, and digitize the data from, a single H2RG detector.  Once an ASIC receives the exposure parameters and the signal to start an exposure, it executes that exposure autonomously without further intervention from SI FSW or from OSS. The 3 MIRI Si:As detectors are interfaced to warm focal plane electronics (FPE) boards that control the detectors' biases, operate their clocks, and also amplify and digitize their signals.

\begin{figure}[h!]
\centering 
\includegraphics[width=0.75\textwidth]{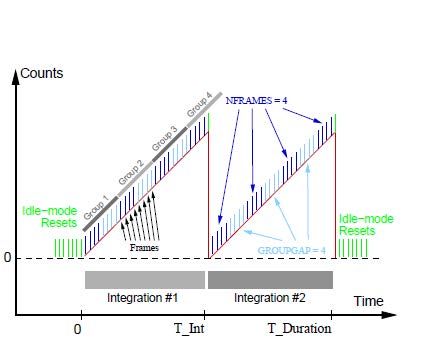}
\caption{\small\it Schematic showing an  example of a possible MULTIACCUM exposure consisting of two integrations (NINTS=2). In this example, each integration has 4 groups (NGROUPS=4), and each group is composed of 8 frames (i.e. non-destructive reads). Within all but the last group in an integrations, 4 frames are marked for storage (NFRAMES=4) and 4 frames are discarded (GROUPGAP=4). In the final group in an integration, the detector is reset rather than performing the final GROUPGAP samples of the detector (resulting in higher efficiency and reducing possible saturation and resulting latent images).
\label{fig:multiaccum}} 
\end{figure}

As of this writing, both reads and resets (during which accumulated signal on the detectors is cleared out) are performed on a pixel-by-pixel basis. The time to acquire a read, or to reset all the pixels, depends primarily on the number of pixels configured to be addressed. Small overheads are incurred when telemetry from the ASIC is inserted into the data stream. The time to acquire one frame of data is given by:
$$
TFrame = ({{NCOL}\over{Noutput}} + 12) \times (NROW + 1) \times 10 \mu\hbox{sec}
$$
where $NCOL$ and $NROW$ define the dimensions of the subarray that is configured (including full array), and 10 $\mu$sec is the pixel addressing clock period. $Noutput$ refers to the number of output channels used, and is 4 in full-frame mode (i.e. $NCOLS = NROWS = 2048$), and is generally 1 for subarrays. (Subarrays with $NCOLS = 2048$ can be configured to use 4 output channels (this is known as \emph{stripe-mode}), and will be supported, \emph{e.g.} for NIRCam grism observations of point sources.) These complexities aside, frame times can be approximated as the total number of pixels being used ($NCOLS \times NROWS$) times 10 $\mu$sec. In full-frame mode this gives a frame time of 10.49 seconds (compared to the actual value of 10.74 seconds); for $320^2$ and $64^2$ subarrays the approximate and actual frame times are 1.02 and 1.07 seconds ($320^2$) and 41 and 49 msec ($64^2$). 

MIRI subarrays work differently. All four outputs are always used, including for  subarrays. However, the the lack of a hardware window mode (which the H2RG detectors do have) means that time must be spent clocking past the unused rows. As a result, the frame time for MIRI subarrays depend both on the size and location of the subarray, as follows:
$$
TFrame = (({{SCOL+NCOL-1}\over{4}} + 10) \times
NROW+(1024-NROW)\times7) \times 10 \mu\hbox{sec}
$$
where $SCOL$ is the starting column of the subarray. The overhead requred
to clock past the unused rows leads to a lower
limit of 70 ms for a 1x1 subarray.

\begin{table}[t!]
\centering
\small
\begin{tabular*}{\textwidth}{|l|l|}
\hline
\bf{Term} & \bf{Definition}\\
\hline
\begin{minipage}[t]{0.2\textwidth}
Frame \\
\end{minipage} &
\begin{minipage}[t]{0.75\textwidth}
Single non-destructive read performed during an integration. \\
\end{minipage} \\

\hline
Tframe & \begin{minipage}[t]{0.75\textwidth}
The time required to read all of the pixels on a detector (or within a subarray). \\
\end{minipage} \\ 

\hline
Group & \begin{minipage}[t]{0.75\textwidth}
A series of \emph{frames}, some of which are stored, and some of which may be discarded. \\
\end{minipage} \\

\hline
NFRAMES & \begin{minipage}[t]{0.75\textwidth}
Number of \emph{frames} in a \emph{group} to store and downlink. NFRAMES must be a power of 2. If NFRAMES $>$ 1 the frames are coadded before being stored. \\
\end{minipage} \\

\hline
GROUPGAP & \begin{minipage}[t]{0.75\textwidth}
Number of \emph{frames} in a \emph{group} to discard. The total number of \emph{frames} within a \emph{group} is NFRAMES $+$ GROUPGAP. \\
\end{minipage} \\

\hline
NGROUPS & \begin{minipage}[t]{0.75\textwidth}
The number of \emph{groups} within a single \emph{integration}. \\
\end{minipage} \\ 

\hline
Tgroup & \begin{minipage}[t]{0.75\textwidth}
 The time to perform all of the reads in a single \emph{group}, equal to Tframe $\times$ (NFRAMES$+$GROUPGAP). \\
\end{minipage} \\ 

\hline
Integration & \begin{minipage}[t]{0.75\textwidth}
Single up-the-ramp set of samples of the pixels of a detector or within a subarray. Each integration is preceded by a reset. NGROUPS samples are stored and downlink for conversion to slope images by the ground system. \\
\end{minipage} \\ 

\hline
\end{tabular*}
\caption{MULTIACCUM exposure parameters (ALL CAPS) and related quantities.
\label{tab:multiaccum} }
\end{table}

\begin{table}[t!]
\centering
\tablenum{\ref{tab:multiaccum}}
\begin{tabular*}{\textwidth}{|l|l|}
\hline
\bf{Term} & \bf{Definition}\\

\hline
NINTS & \begin{minipage}[t]{0.75\textwidth}
The number of consecutive \emph{integrations} within an \emph{exposure}. For NIRCam, NIRISS, and NIRSpec, \textbf{NINTS can not exceed 65535.} \\
\end{minipage} \\ 

\hline
\begin{minipage}[t]{0.2\textwidth}
Integration Time \end{minipage} & 
\begin{minipage}[t]{0.75\textwidth}
The duration of a single up-the-ramp \emph{integration}, equal to the time between the \emph{reset} at the beginning of the integration and the time of the final (saved) \emph{frame} in the last \emph{group} of the \emph{integration}. \\
\end{minipage} \\ 

\hline
Exposure & \begin{minipage}[t]{0.75\textwidth}
A series of one or more \emph{integrations} acquired without interruption. \\
\end{minipage} \\ 

\hline
Exposure Time & \begin{minipage}[t]{0.75\textwidth}
The total time spent collecting signal during an exposure, equal to NINT $\times$ \emph{Integration Time} \\
\end{minipage} \\ 

\hline
 Exposure Duration & 
 \begin{minipage}[t]{0.75\textwidth}
 The total time spent executing and exposure, including resets and GROUPGAP frames. Equal to \emph{tframe} $\times$ (NFRAME$+$GROUPGAP$+$1) $\times$ NINTS. \\
\end{minipage} \\ 

\hline
\end{tabular*}
\caption{JWST MULTIACCUM exposure definitions (continued).}
\end{table}


\clearpage

\section{Illustrative Science Programs} \label{App:AppendixB}
To illustrate the power of JWST's instruments we have selected potential targets for  observation, choosing either a primary transit, a secondary eclipse, or a full phase curve. Table~\ref{tab:stars}) lists some of the properties of the planetary systems and Table~\ref{tab:program} gives a representative observing program including integration times. To these times must be added a significant amount of overhead for slew taxes, telescope acquisition and instrument setup times.

\begin{deluxetable}{lccccccc}
\tablenum{Appendix  3}
\tablecaption{Illustrative  {\it JWST}  Target Stars\label{tab:stars} }
\tablehead{Star &Spec &	K &	Rp & Depth & Period & Transit Dur$^1$ & Teff \\
Name &Type &	 (mag)&	 (R$_{Jup}$)& (ppm)& (day)& (hr) & (K)}
\startdata
GJ436		&M1&	6.1&	0.38&	7,000&	2.6&	1&	700\\
Gliese1214	&M4.5&	8.8&	0.20&	13,000&	1.6&	1&	550\\
HD189733	&G0&	5.5&	1.14&	24,000&	2.2&	2&	1200\\
HD209458	&F8&	6.3&	1.38&	14,00&	3.5&	3&	1450\\
HD80606		&G5&	7.3&	0.98&	10,600&	111&	60&	400-1500\\
Kepler-7	&G1&	11.5&	1.50&	6,800&	4.9&	120& 1400\\
KOI-02311.01&G2&	11.0&	0.10&	100&	192&	5&	310\\
Sample TESS Target$^2$ &M2&	7.5&	0.13&	800&	30&	 3&	300\\
\enddata
\tablecomments{$^1$Observation is 3$\times$ the transit duration to account for time to measure the star before and after the transit/eclipse. For HD 80606 and Kepler-7 the duration is that required to observe a phase curve.$^1$ Illustrative  properties of a nearby M dwarf that  {\it TESS} might discover to have a Habitable Zone Super Earth.}
\end{deluxetable}
 
\clearpage
\begin{sidewaystable}[t!]
\tiny
\tablenum{Appendix  4}
\centering
\begin{tabular}{|lccccccc|}
\hline
 Source & Science & Instrument & Disperser/ & $\lambda$ ($\mu$m)& Resln & Event$^1$ & SNR (or Noise)$^2$ \\
 Name & Driver & Mode & Filter & &  & & (1 Event) \\
 \hline
GJ 436 & Hot Neptune Dayside Map & NIRcam Imaging & F356W & 3.11-4& 4 &  SE & 8.9$^3$\\
GJ 1214 & Sub-Neptune Flux & MIRI Imaging & F1000W & 9-11& 5 & SE & 7.0\\
 & Sub-Neptune Spectrum & NIRISS Grism & GR700XD & 0.6-2.3& 300-800& Tr & 75-100$^4$\\
 & Sub-Neptune Spectrum & NIRcam Grism & F322W2 & 2.4-4.0& 1100-1700& Tr & 
55$^5$\\
HD189733 & Emission Spectrum & NIRcam Grism & F356W & 3.15-3.97& $\sim$1500 & SE & 21$^6$\\
& Transmission Spectrum & NIRISS Grism &GR700XD & 1.10-2.70& $\sim$700 & Tr & 425$^7$\\
 & Tomographic imaging & MIRI Imaging & F770W & 6.6-8.8& 3.5 & SE & 33\\
 & Tomographic spectro-imaging & NIRSpec & G140M/F170LP & 1.7-3.1 & 700-1300 & PC & 100-250 ppm$^{14}$\\
HD209458 & Tomographic imaging & NIRcam Imaging & F200W+WL& 1.75-2.25& 4 & SE & 80\\
HD80606 & Periastron light curve & NIRcam Imaging & F444W & 3.89-5& 4 & PC & 22$^8$\\
 & Emission Spectrum & MIRI Imager & Prism & 5-12& 40-200 & SE & 12$^9$\\
 & Emission Spectrum (high res)& NIRSpec & G140H/F070LP & 0.7-1.2& 1300-2300& SE & 100-20 ppm$^{14}$\\
 & & & G140H/F100LP & 1.0-1.8& 1900-3600& SE & 20-35 ppm$^{14}$\\
 & & & G235M/F170LP & 1.7-3.1& 1900-3600& SE & 20-50 ppm$^{14}$\\
 & & & G395H/F290LP & 2.9-5.2 & 1900-3600& SE & 30-80 ppm$^{14}$\\
Kepler-7 & Silicate Cloud Phase Curve & NIRcam Imaging & F444W & 3.89 - 5& 4 & PC & 9.5\\
 & Silicate Cloud Phase curve & MIRI MRS/Ch2 & Ch2 & 7- 12& 3000 & PC & 0.3$^{10}$\\
KOI-02311.01& Confirmation& NIRcam Imaging & F200W &1.75-2.25& 4& Tr & 5\\
TESS-001& Super Earth& NIRcam Grism & F322W2 & 2.42-4.01& 1100-1700& Tr & 35$^{11}$\\
 & Super Earth& MIRI Imager& F1280W & 11.6-14.0& 5.3 & SE &  0.1\\
 & Super Earth& NIRSpec & G140M/F070LP & 0.7-1.2& 500-850& Tr & 1000-80 ppm$^{14}$\\
 & & & G140M/F100LP & 1.0-1.8& 700-1300 & Tr & 60-100 ppm$^{14}$\\
 & & & G235M/F170LP & 1.7-3.1& 700-1300 & Tr & 70-160 ppm$^{14}$\\
 & & & G395M/F290LP & 2.9-5.2& 700-1300 & Tr & 90-200 ppm$^{14}$\\
\hline
\end{tabular}
\caption[Illustrative Science Programs]{Description of illustrative programs \label{tab:program} }

\end{sidewaystable} 

The following notes pertain to Table~\ref{tab:program} and give details of the instrument setup and SNR calculations.

\begin{enumerate}
\item Observed events are either a transit (Tr), secondary eclipse (SE), or a partial phase curve (PC).

\item SNR values given are for measuring either a single transit, the continuum day-side emission at eclipse, or the continuum emission for 1 hour of a phase curve using the system parameters in Table~\ref{tab:stars}  unless otherwise specified. Actual time spent observing a transit or eclipse will be $\sim3\times$ the transit duration for sufficient time on the star+planet as well as the star alone. Spectral SNRs are for the indicated event at the indicated resolution. SNRs for expected spectral or temporal features must be scaled from these single event values.

\item  GJ 436 exceeds the NIRCam imaging bright limit. SNR calculated for a NIRCam grism + F356W filter observation of 1 secondary eclipse with all spectral information co-added. 

\item  NIRISS GJ 1214 transit spectrum SNR calculated for $\lambda = 0.6 to 2.7 \mu$m and R=700. See Figure~\ref{fig_niriss_sims}a.

\item NIRCam GJ 1214 transit spectrum SNR calculated for $\lambda = 3.22 \mu$m and R=1000. SNR improves with R$^-0.5$ if binned down further.

\item NIRCam HD 189733 secondary eclipse spectrum SNR calculated for $\lambda = 3.56 \mu$m and R=1000. SNR improves with R$^{-0.5}$ if binned down further.

\item NIRISS HD 189733 primary transit SNR calculated for $\lambda = 3.6 \mu$m and R=700. See Figure~\ref{fig_niriss_sims}b.

\item NIRCam HD 80606 b phase curve SNR calculated for planet temperature of 1000 K. HD 80606 exceeds the NIRCam imaging bright limit. SNR calculated for a NIRCam grism + F444W filter with all spectral information co-added. 

\item MIRI HD 80606 b secondary eclipse SNR calculated for planet temperature of 1400 K at $\lambda = 8.0 \mu$m and R=100 for 1 hour before eclipse.

\item Kepler-7b MIRI Silicate phase curve SNR calculated for $\lambda = 10.0 \mu$m and R=3000. SNR may improve as much as R$^-0.5$ if binned down to lower R. Beware that full spectral coverage requires 3 exposures with different dichroic / grating settings.

\item KOI-02311.01 NIRCam transit spectrum SNR calculated for $\lambda = 3.22 \mu$m and R=1000. SNR improves with R$^-0.5$ if binned down further (SNR = 110 at R=100).

\item All NIRCam calculations assume a systematic noise floor of 10 ppm per spectral / spatial resolution element per transit or eclipse event. NIRCam phase curve observation SNRs are calculated for 1 hr during a phase curve that also covers 1 hr of secondary eclipse.

\item All MIRI calculations assume a systematic noise floor of 40 ppm per spectral / spatial resolution element per transit or eclipse event. MIRI phase curve observations SNRs are calculated for 1 hr during a phase curve that also covers 1 hr of secondary eclipse.

\item SNR expressed as a relative noise level in parts-per-million (ppm of the signal from the parent star). All calculations assume a minimum achievable noise level (systematic noise floor) of 20 ppm similar to what has been achieved in the best cases with HST. Note that the capability of NIRSpec to reach such low levels of systematics will only be demonstrated once in orbit. All numbers are given per spectral pixel (no spectral binning; see the spectral resolution curves in Fig.~\ref{fig:Rlambda}) and assuming the transit durations listed in Table~3 (with 2 times the transit duration spent out of transit or eclipse). For (spectroscopic) phase curve observations, we have assumed that up to 1 hour of observation could be added for each point of the curve. To maximise the efficiency of the observations, we have allowed the use of integrations with a single frame readout ({\it "RESET-READ"} pattern, see discussion in Section~6.3) whenever relevant.

 \end{enumerate}

 
\newpage


 \end{document}